\documentclass{aa}  

\usepackage{natbib}
\usepackage{graphicx}
\usepackage{txfonts}
\usepackage[normalem]{ulem}
\usepackage{xcolor}

\newcommand{\beforeReferee}[1]{}
\newcommand{\afterReferee}[1]{{#1}}

\newcommand{\ignoreThis}[1]{ }

\providecommand{\dt}[1]{{\tt #1}} 

\newcommand\gaia{\textit{Gaia}}
\newcommand\gdrone{\gaia~DR1}
\newcommand\gdrtwo{\gaia~DR2}
\newcommand\hip{\textsc{Hipparcos}}
\newcommand\tyc{\textit{Tycho}}
\newcommand\tyctwo{\textit{Tycho}-2}

\newcommand\secref[1]{Sect.~\ref{#1}}
\newcommand\figref[1]{Fig.~\ref{#1}}
\newcommand\figsref[1]{Figs.~\ref{#1}}

\bibpunct{(}{)}{;}{a}{}{,} 

\newcommand\bpminrp{\ensuremath{G_\mathrm{BP}-G_\mathrm{RP}}}
\newcommand\gminbp{\ensuremath{G-G_\mathrm{BP}}}
\newcommand\ebpminrp{\ensuremath{E(G_\mathrm{BP}-G_\mathrm{RP})}}
\newcommand\gminrp{\ensuremath{G-G_\mathrm{RP}}}

\def\teff{$T_{\rm eff}$}
\def\a0{$A_{\rm 0}$}

\def\gmag{$G$}
\def\gbp{$G_{\rm BP}$}
\def\grp{$G_{\rm RP}$}

\def\kms{\,km\,s$^{-1}$}
\def\masyr{\,mas\,yr$^{-1}$}

\def\logg{$\log g$}
\def\feh{[Fe/H]}
\def\parallax{$\varpi$}

\begin{document} 

   \title{{\gdrtwo}}

   \subtitle{Catalogue validation 
   }

\author{
F.~Arenou\inst{\ref{inst:gepi}}, 
X.~Luri\inst{\ref{inst:ieec}}, 
C.~Babusiaux\inst{\ref{inst:IPAG},\ref{inst:gepi}}, 
C.~Fabricius\inst{\ref{inst:ieec}}, 
A.~Helmi\inst{\ref{inst:kapteyn}}, 
T.~Muraveva\inst{\ref{inst:bologna}}, 
A.~C.~Robin\inst{\ref{inst:utinam}}, 
F.~Spoto\inst{\ref{inst:nice},\ref{inst:imcce}}, 
A.~Vallenari\inst{\ref{inst:padova}}, 
T.~Antoja\inst{\ref{inst:ieec}},
T.~Cantat-Gaudin\inst{\ref{inst:padova},\ref{inst:ieec}},
C.~Jordi\inst{\ref{inst:ieec}}, 
N.~Leclerc\inst{\ref{inst:gepi}}, 
C.~Reyl\'e\inst{\ref{inst:utinam}}, 
M.~Romero-G\'omez\inst{\ref{inst:ieec}},
I-C.~Shih\inst{\ref{inst:gepi}},
S.~Soria\inst{\ref{inst:ieec}},
C.~Barache\inst{\ref{inst:SYRTE}}, 
D.~Bossini\inst{\ref{inst:padova}}, 
A.~Bragaglia\inst{\ref{inst:bologna}},
M.~A.~Breddels\inst{\ref{inst:kapteyn}},  
M.~Fabrizio\inst{\ref{inst:roma},\ref{inst:asdc}},
S.~Lambert\inst{\ref{inst:SYRTE}}, 
P.~M.~Marrese\inst{\ref{inst:roma},\ref{inst:asdc}},
D.~Massari\inst{\ref{inst:kapteyn}},
A.~Moitinho\inst{\ref{inst:FCUL}}, 
N.~Robichon\inst{\ref{inst:gepi}}, 
L.~Ruiz-Dern\inst{\ref{inst:gepi}}, 
R.~Sordo\inst{\ref{inst:padova}}, 
J.~Veljanoski\inst{\ref{inst:kapteyn}}, 
P.~Di Matteo\inst{\ref{inst:gepi}}, 
L.~Eyer\inst{\ref{inst:Geneva}}, 
G. Jasniewicz\inst{\ref{inst:montpellier}}, 
E.~Pancino\inst{\ref{inst:0024}},
C.~Soubiran\inst{\ref{inst:Bordeaux}}, 
A.~Spagna\inst{\ref{inst:torino}}, 
P.~Tanga\inst{\ref{inst:nice}},
C.~Turon\inst{\ref{inst:gepi}},
C. Zurbach\inst{\ref{inst:montpellier}}
}

\institute{
GEPI, Observatoire de Paris, Universit{\'e} PSL, CNRS, 5 Place Jules Janssen, 92190 Meudon, France\\
\email{Frederic.Arenou@obspm.fr}
\label{inst:gepi}
\and
Dept. FQA, Institut de Ciències del Cosmos (ICCUB), Universitat de Barcelona (IEEC-UB), Martí Franquès 1, E08028 Barcelona, Spain
\label{inst:ieec}
\and
Kapteyn Astronomical Institute, University of Groningen, Landleven 12, 9747 AD Groningen, The Netherlands
\label{inst:kapteyn}
\and
Institut UTINAM, CNRS, OSU THETA Franche-Comt\'e Bourgogne, Univ. Bourgogne Franche-Comt\'e, 25000 Besan\c{c}on,
France\label{inst:utinam}
\and
INAF, Osservatorio Astronomico di Padova, Vicolo Osservatorio, Padova, I-35131, Italy
\label{inst:padova}
\and
Observatoire de Gen\`eve, Universit\'e de Gen\`eve, CH-1290 Versoix, Switzerland
\label{inst:Geneva}
\and
Institute of Astronomy, University of Cambridge, Madingley Road, Cambridge CB30HA, United Kingdom
\label{inst:ioa}
\and 
SYRTE, Observatoire de Paris, Universit\'e PSL, CNRS, Sorbonne Universit\'e, LNE, 61 avenue de l'Observatoire, 75014 Paris, France
\label{inst:SYRTE}
\and
CENTRA, Universidade de Lisboa, FCUL, Campo Grande, Edif. C8, 1749-016 Lisboa, Portugal
\label{inst:FCUL}
\and
Leiden Observatory, Leiden University, Niels Bohrweg 2, 2333 CA Leiden, The Netherlands
\label{inst:leidenO}
\and
Université C\^ote d'Azur, Observatoire de la C\^ote d'Azur, CNRS,
Laboratoire Lagrange, Bd de l'Observatoire, CS 34229, 06304 Nice cedex
4, France\label{inst:nice}
\and
INAF - Osservatorio Astronomico di Roma, Via di Frascati 33, 00078 Monte Porzio Catone (Roma), Italy
\label{inst:roma}
\and
ASI Science Data Center, Via del Politecnico, Roma
\label{inst:asdc}
\and
Laboratoire d'astrophysique de Bordeaux, Univ. de Bordeaux, CNRS, B18N, all{\'e}e Geoffroy Saint-Hilaire, 33615 Pessac, France
\label{inst:Bordeaux}
\and
INAF - Osservatorio Astrofisico di Arcetri, Largo Enrico Fermi 5, I-50125 Firenze, Italy                                \label{inst:0024}
\and 
INAF - Osservatorio di Astrofisica e Scienza dello Spazio di Bologna, via Piero Gobetti 93/3, 40129 Bologna,  Italy                                        \label{inst:bologna}
\and 
INAF - Osservatorio Astronomico di Torino, via osservatorio 20, Pino Torinese, Torino,  Italy                           \label{inst:torino}
\and 
Univ. Grenoble Alpes, CNRS, IPAG, 38000 Grenoble, France
\label{inst:IPAG}
\and
IMCCE, Observatoire de Paris, PSL Research University, CNRS, Sorbonne
Universités, UPMC Univ. Paris 06, Univ. Lille, 77 av.
Denfert-Rochereau, 75014 Paris, France\label{inst:imcce}
\and Laboratoire Univers et Particules de Montpellier, Universit\'{e} Montpellier, CNRS, Place Eug\`{e}ne Bataillon, CC72, F-34095 Montpellier Cedex 05, France\relax \label{inst:montpellier}
}

   \date{ }

\abstract
{
The second {\gaia} data release (DR2), contains very precise astrometric and photometric properties for more than one billion sources, astrophysical parameters for dozens of millions, radial velocities for millions, variability information for half a million of stellar sources and orbits for thousands of solar system objects.
}
{
Before the Catalogue publication, these data have undergone dedicated validation processes. 
The goal of this paper is to describe the validation results in terms of completeness, accuracy and precision of the various {\gdrtwo} data. 
}
{
The validation processes include a systematic analysis of the Catalogue content to detect anomalies, either individual errors or statistical properties, using statistical analysis, and comparisons to external data or to models. 
}
{
Although the astrometric, photometric and spectroscopic data are of unprecedented quality and quantity, it is shown that the data cannot be used without a dedicated attention to the limitations described here, in the Catalogue documentation and in accompanying papers. A particular emphasis is put on the caveats for the statistical use of the data in scientific exploitation.
}
{}

   
   \titlerunning{{\gdrtwo} -- Catalogue validation} 
   \authorrunning{F. Arenou et al.}

   \maketitle
\section{Introduction}
This paper describes the validation of the second data release, {\gdrtwo}, from the European Space Agency mission {\gaia} \citep{2016A&A...595A...1G, DR2-DPACP-36}.  
The approach followed by this catalogue validation is an external, transverse analysis of the properties of the various contents. 

A large variety of the Catalogue properties are described together with their dedicated validation by  \citet{DR2-DPACP-51} for the astrometry, \citet{DR2-DPACP-40} for the photometry, \citet{DR2-DPACP-47} and \citet{DR2-DPACP-54} for the spectroscopic data, \citet{DR2-DPACP-43} for the astrophysical parameters, \citet{DR2-DPACP-49} for the variable stars, \citet{DR2-DPACP-32} for the Solar System objects and \citet{DR2-DPACP-30} for the reference frame. Besides, science demonstration papers such as \citet{DR2-DPACP-31} for the H-R diagram, \citet{DR2-DPACP-33} for the Milky Way disk kinematics or \citet{DR2-DPACP-34} for the Milky Way satellites have also indirectly contributed largely to demonstrate the overall quality of the catalogue and \cite{DR2-DPACP-36} summarises its impressive characteristics. For this reason, a special emphasis is put here on the caveats attached to the data, in order to allow a better exploitation of the Catalogue.

We mention here only what is strictly necessary and invite the reader to refer to the above papers or to the {\gaia} on--line documentation\footnote{\url{http://gea.esac.esa.int/archive/documentation/gdrtwo/index.html}. In this paper, we note the catalogue fields with a special font, e.g. \dt{astrometric\_chi2\_al}. The description of these fields can be found in Chapter 14 of the Catalogue documentation.} for details. As will be evident below, understanding the properties of the Catalogue is mandatory for a proper scientific use of the data; reading of the above papers is important -- and inspiring.

This paper is organised as follows. We first describe the general consistency of the data (\secref{sec:data}) then the completeness of the Catalogue from small to large scale (\secref{sec:comp}). We describe in turn the astrometric properties, systematics and random (\secref{astroqual}), the photometric quality (\secref{photoqual}), the spectroscopic results (\secref{radial}), the astrophysical parameters (\secref{astropar}), and validation of solar system objects (\secref{sso}).

\section{Data and general validation tests}\label{sec:data}

\subsection{Data integrity and consistency}\label{ssec:data_integrity}
The data release consists of several data tables. In most of this paper we focus on the gaia\_source Catalogue with the mean parameters for about 1.7 billion point-like sources. In addition, smaller tables contain the results of the analysis of light curves for variable sources, Sect.~\ref{photovar}, and the results for solar system objects, discussed in Sect.~\ref{sso}.

The gaia\_source Catalogue contains positions and \gmag-band mean photometry, both with several auxiliary parameters, for all sources. For a large subset, 1.3 billion sources, it also gives proper motions and parallaxes, again with many auxiliary parameters; for another large subset, 1.4 billion sources, photometry in the \gbp\ and \grp-bands; for smaller subsets, between 77 and 161 million sources, various astrophysical parameters; and for a more modest seven million sources the radial velocity. Light curves are given for half a million variables and two million individual CCD-observations in 330\,000 transits of fourteen thousand asteroids.

For all preliminary versions of the {\gdrtwo} Catalogue, one of the validation tasks consisted in several basic verification tests in order to check the internal consistency of the data records, e.g.\ that data fields were present when and only when expected, that fluxes were converted consistently to magnitudes, or that positions were expressed equally well in equatorial, ecliptic, and Galactic coordinates. The fields were corrected when needed for the final Catalogue and the results are not reported here.

Beside this, the data in the {\gdrtwo} as a whole generally behaves following expectations. This has been established for example, by comparing the (clustering) behaviour of multi-dimensional distributions of the observables and their errors for different regions on the sky (symmetric with respect to the disk, and with similar number of transits/observations), using the Kullback-Leibler divergence statistic \citep[KLD,][]{kullback1951}.
Furthermore, comparisons to Galactic models confirm that the global behaviour of \afterReferee{most of} the data, at a surface level, is as expected.

\subsection{Duplicate entries}\label{sec:err_dup}\label{sec:dup}
The {\gaia} data processing is complex, cf.\ e.g. \citet[][Sect.~2]{DR2-DPACP-51}, and has still not reached full maturity. It may therefore happen that the same source is processed twice, but based on disjoint sets of observations. In the published Catalogue only one of the solutions has been kept, and the flag \dt{duplicated\_source} has been set, but the removed duplicated solution was made available for validation. Although these duplicates are in themselves relatively harmless, decreasing their number for the next data release would nevertheless allow to increase the number of observations per star.
Turning the weakness into strength, such duplicated sources offered an interesting opportunity for the validation as discussed in the various sections below.  

\begin{figure}\begin{center}
\includegraphics[width=0.8\columnwidth]{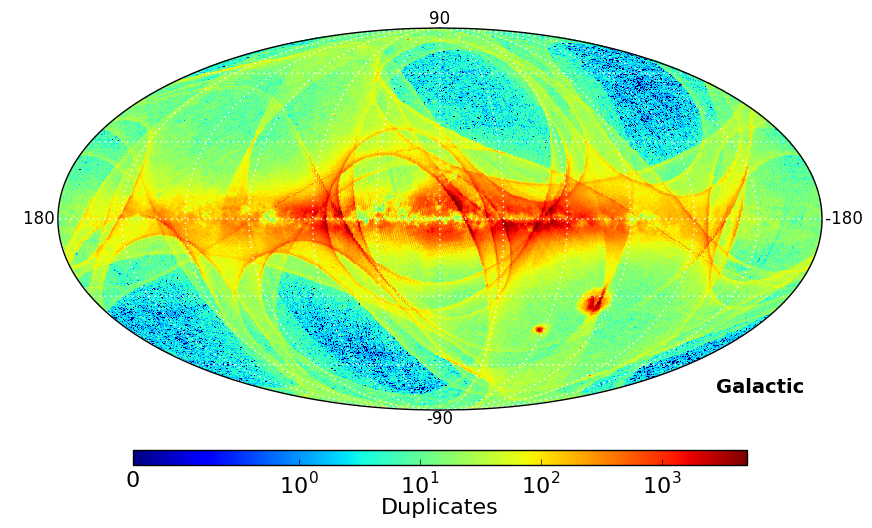}
\caption{Map of duplicated sources.}\label{fig:map-duplicates}
\end{center}\end{figure}

\begin{figure}\begin{center}
\includegraphics[width=0.7\columnwidth]{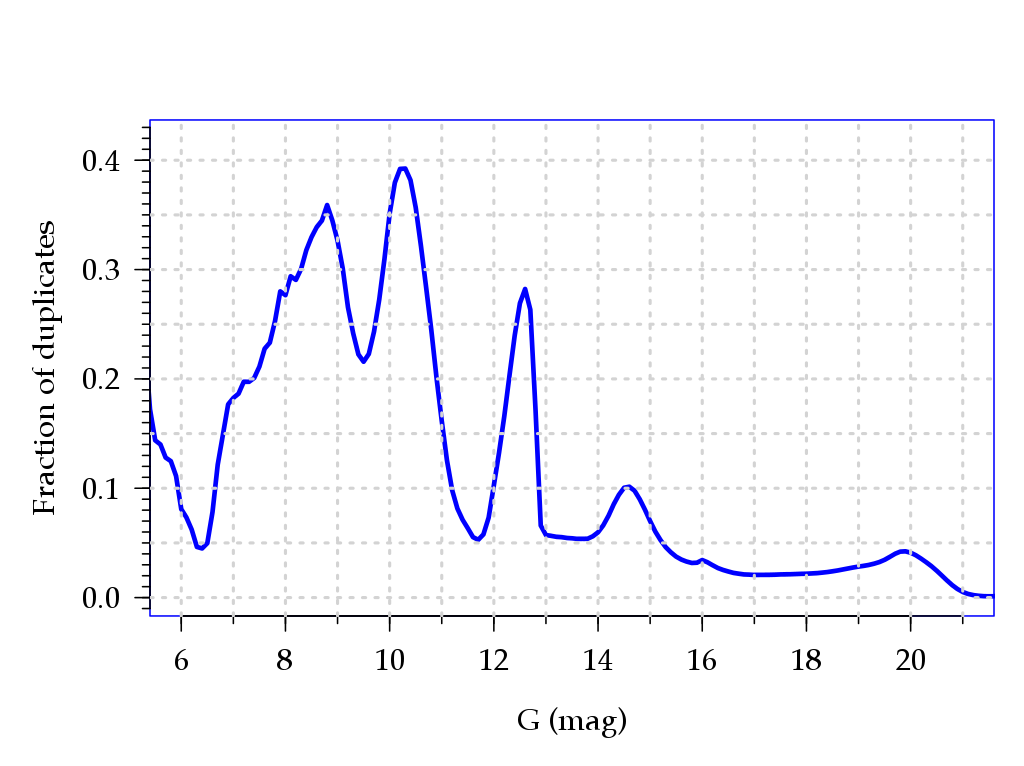}
\caption{Fraction of duplicated sources vs $G$ magnitude. The peaks must be due to a combination of problems in the on-board detections and the cross-match process.} \label{fig:hist-dup}
\end{center}\end{figure}

The duplicated sources, with two independent solutions in the initial versions of {\gdrtwo}, are found all over the sky (Fig.~\ref{fig:map-duplicates}), but because of various details related to on-board as well as on-ground processing, they are not a random subset of the Catalogue and are seen more often on the bright side, reaching 39\% at $G=10.3$ as shown in Fig.~\ref{fig:hist-dup}. Conclusions based on this subset are therefore not necessarily representative for the full Catalogue. This is especially the case for sources without the full astrometric solution, where the quality indicators show poorer results for the duplicate solutions. For sources with full astrometry, on the other hand, the quality is only marginally affected.

Image sizes in {\gaia} in the along-scan direction  (AL) are around 0\farcs2. It follows, however, from the way observations are acquired, that sources separated by less than 0\farcs4--0\farcs5 cannot be resolved without a dedicated process. Such a process is still not in place, and for {\gdrtwo} duplicated sources were therefore simply defined as solutions separated by less than 0\farcs4. The average separation within duplicate pairs is 0\farcs019, so small that it shows that the pairs represented basically the same sources, and that resolved double stars can only represent a very small fraction of them.

Contamination by close-by sources may indeed give erroneous solutions as discussed in \secref{sec:err_ast}. The processing for {\gdrtwo} rests on the assumption that all sources are isolated point sources. When this condition is not fulfilled, the resulting photometry and astrometry may suffer distortions. The {\gbp} and {\grp} photometry is especially vulnerable because it is based on aperture photometry of dispersed spectra.

\section{Sky coverage and completeness}\label{sec:comp}

In this section the completeness of the {\gdrtwo} Catalogue is described with respect to the actual sky content. The situation is obviously more complicated for what concerns the various data which may, or not, be available for each source. In this respect, appendix \ref{chap:selectFunc} details how the satellite observation first, then the various processing steps have built the Catalogue content, i.e. the fraction, for each category of data, of the total number of sources, and we refer to \cite{DR2-DPACP-36} for characteristic figures of the Catalogue.

\subsection{Limiting magnitude}\label{ssec:data_limit}

\begin{figure}\begin{center}
\includegraphics[width=0.8\columnwidth]{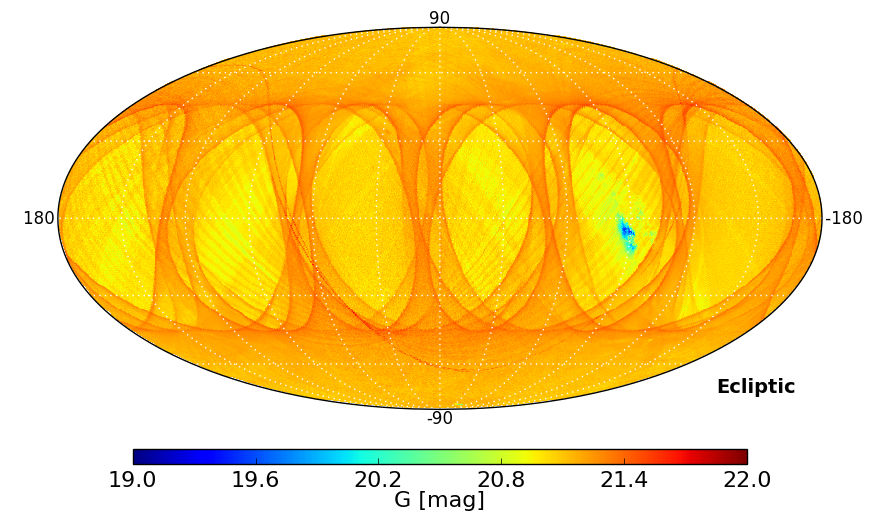}
\caption{Sky map in ecliptic coordinates of limiting magnitude: 99th percentile in \gmag. } \label{fig:g_percentile_99}
\end{center}\end{figure}

Figure~\ref{fig:g_percentile_99} illustrates the variation in limiting magnitude (99th percentile) across the sky. The map is in ecliptic coordinates in order to emphasize the importance of the scanning law. The brightest limit is found near the Galactic centre where the star density is very high and where we have relatively few scans. On the other hand, the faintest limit is achieved near the caustics of the scanning law at ecliptic latitude $\pm45\degr$, where there are more observations.

\subsection{Overall large-scale coverage and completeness}\label{ssec:data_coverage}

\begin{figure}\begin{center}
\includegraphics[width=0.8\columnwidth]{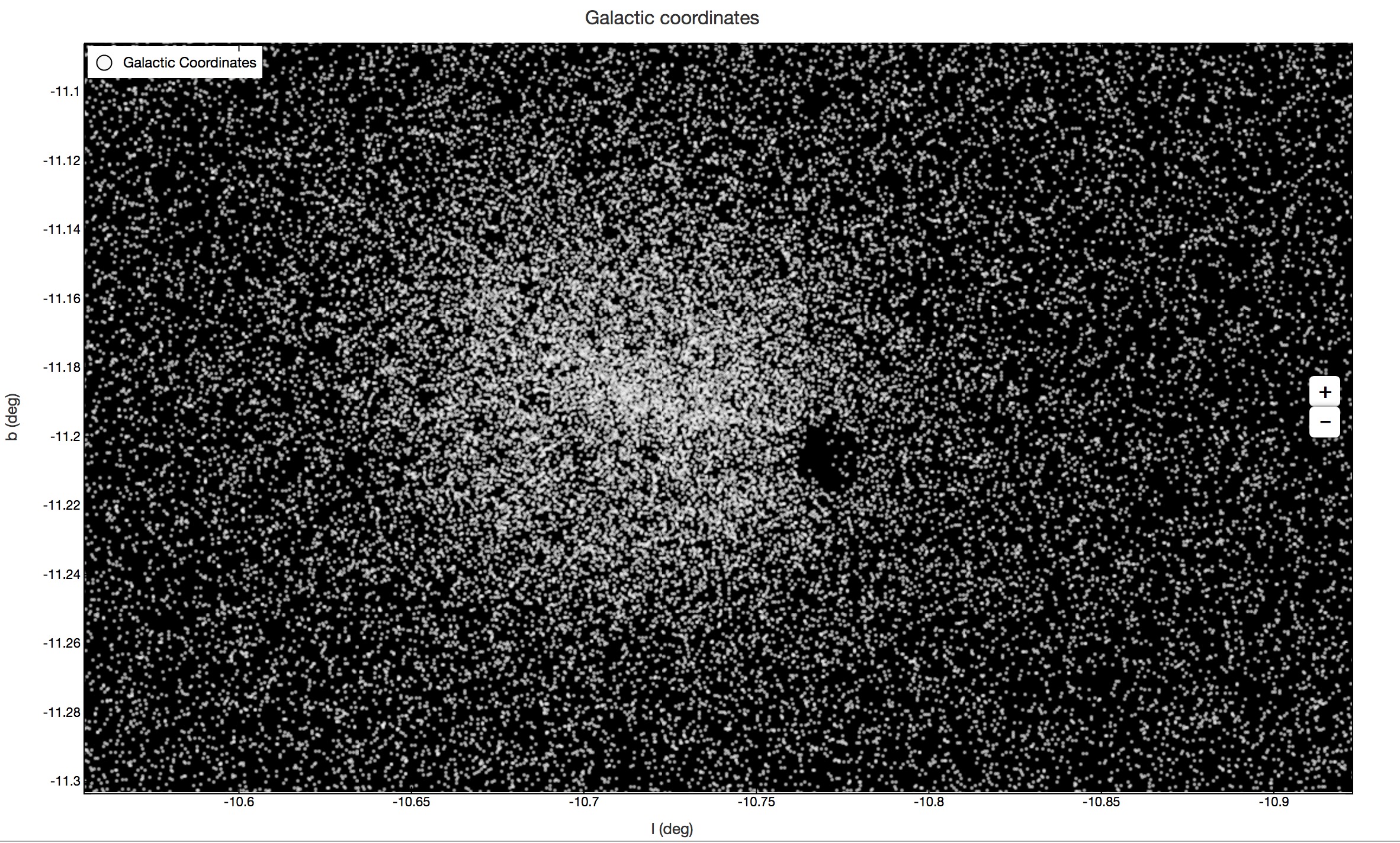}
\caption{There are only a few underscanned regions due to lack of on-board resources, such as here, on the edge of NGC 6541.}\label{fig:NGC6541}
\end{center}\end{figure}

For {\gdrone} several regions suffered from limited on-board resources, which created holes in the sky coverage; these regions are now covered, and only a few remain, such as near NGC 6541 globular cluster, Fig.~\ref{fig:NGC6541}.

\begin{figure}\begin{center}
\includegraphics[width=0.49\columnwidth]{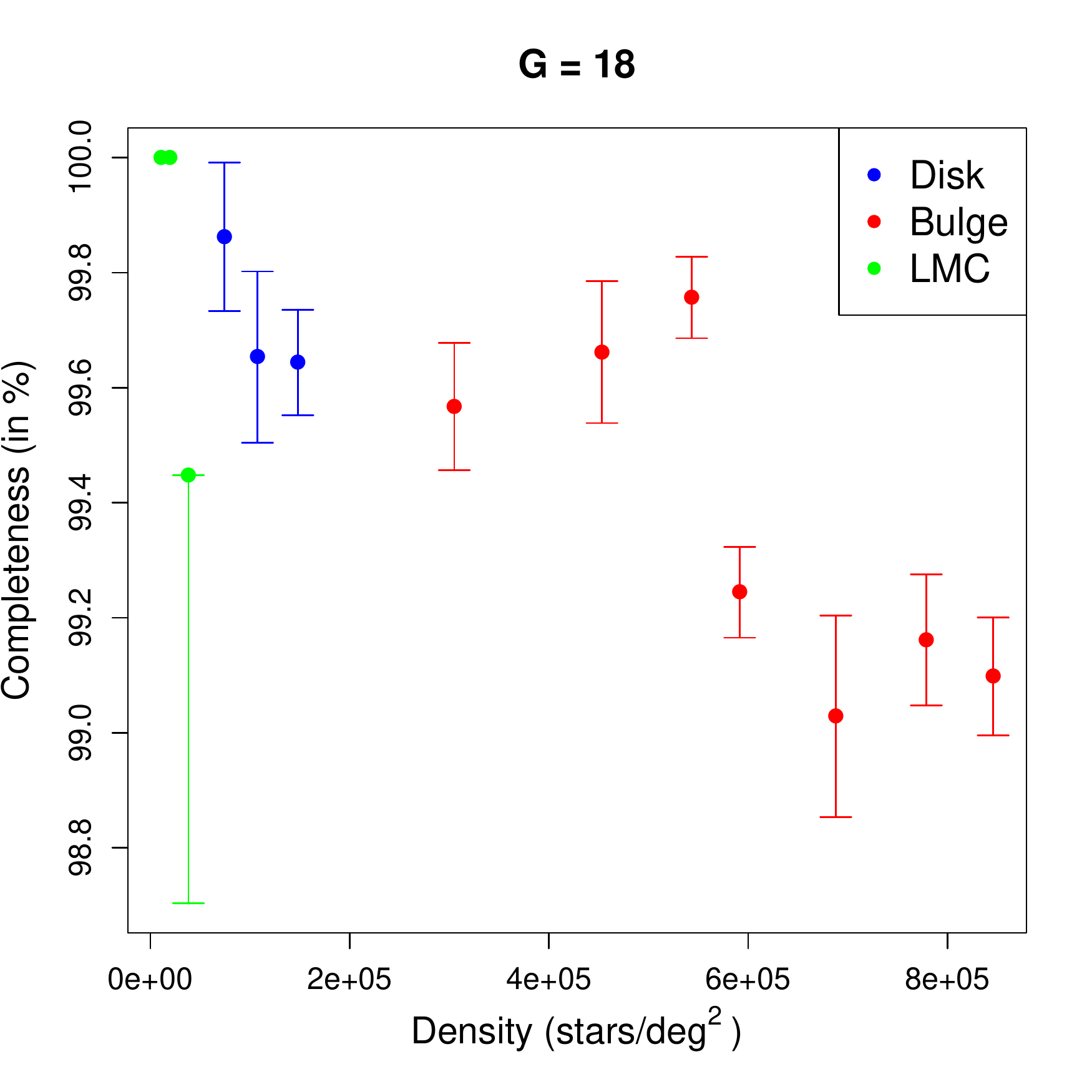}
\includegraphics[width=0.49\columnwidth]{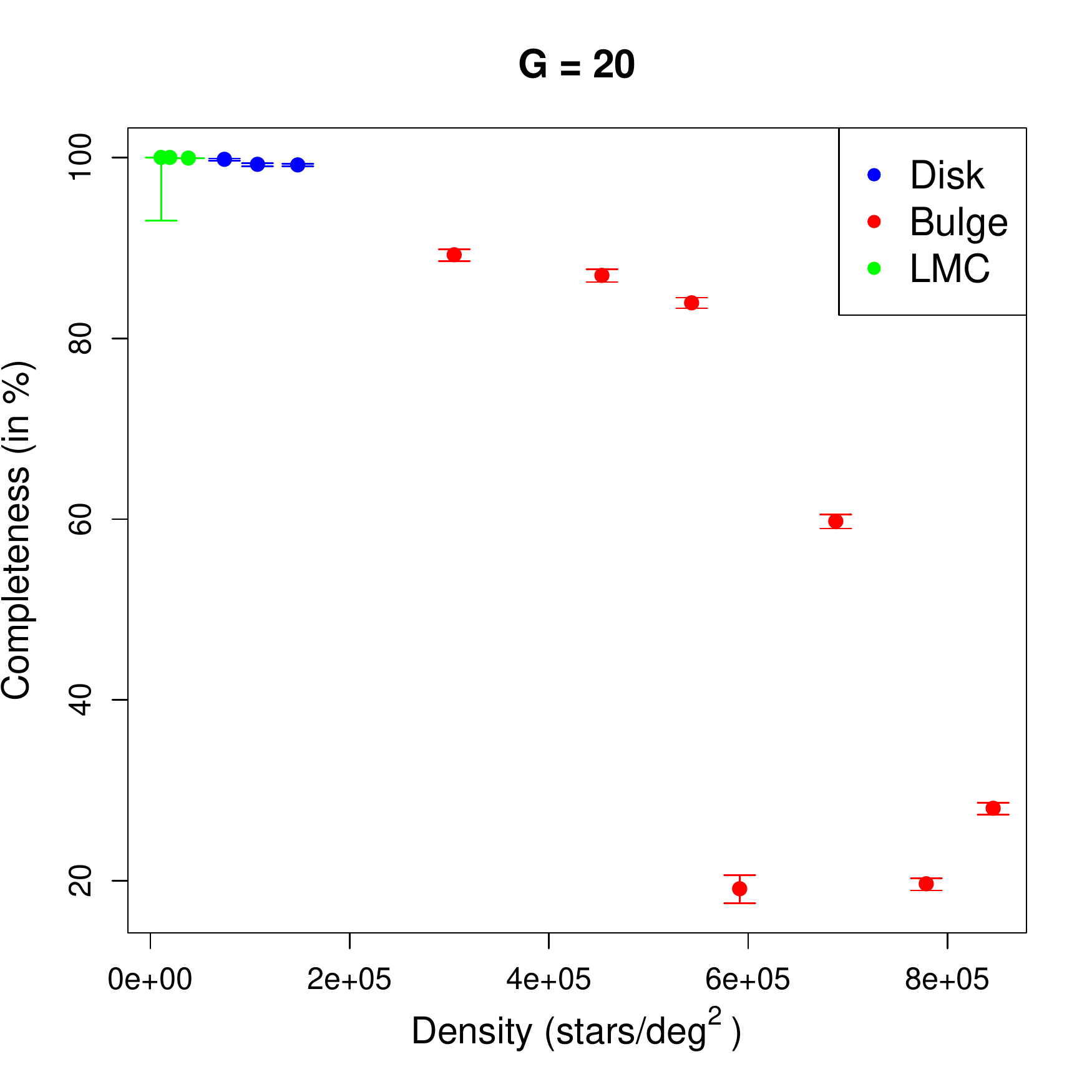}
\caption{{\gdrtwo} completeness vs some OGLE fields at $G=18$ and $G=20$ as a function of the measured density at $G=20$. Note the very different scale between the two plots.}\label{fig:oglecompl}
\end{center}\end{figure}

Figure~\ref{fig:oglecompl} shows the completeness versus OGLE data \citep{2008AcA....58...69U} in some selected fields with different sky density. The OGLE spatial resolution being worse than {\gaia}, comparison with OGLE provide upper limits to the Gaia completeness. Compared to {\gdrone} \citep[Fig. 15 of][]{2017A&A...599A..50A} the coverage is now much better, the {\gdrtwo} Catalogue being almost complete at $G=18$, whereas it was less than 80\% for {\gdrone} as soon as the density was above one hundred thousand stars per square degree. 

\begin{figure}\begin{center}
\includegraphics[width=0.8\columnwidth]{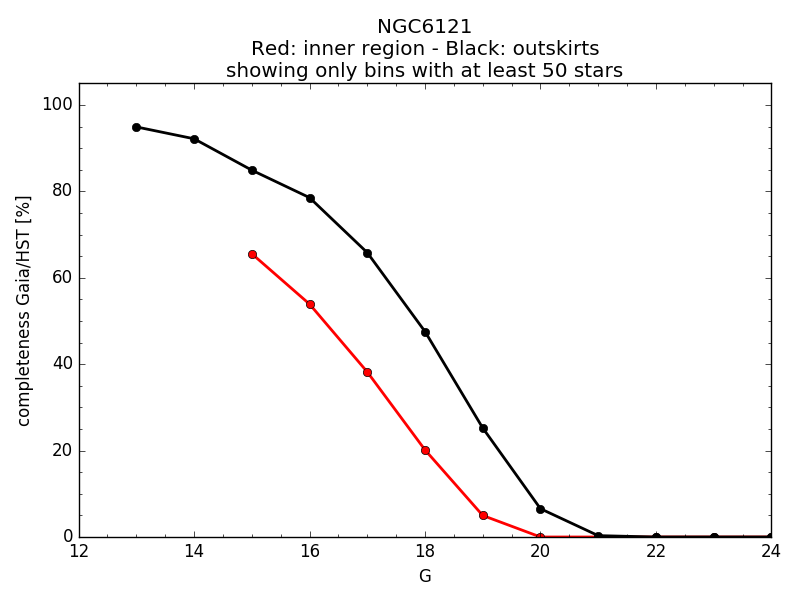}
\caption{Completeness level with respect to HST data in the inner (within 0.5 arcmin, red) and outer region (black) of the cluster NGC 6121.} \label{fig:completeness_NGC6121_coreVSoutskirts}
\end{center}\end{figure}

\begin{figure}\begin{center}
\includegraphics[width=0.9\columnwidth]{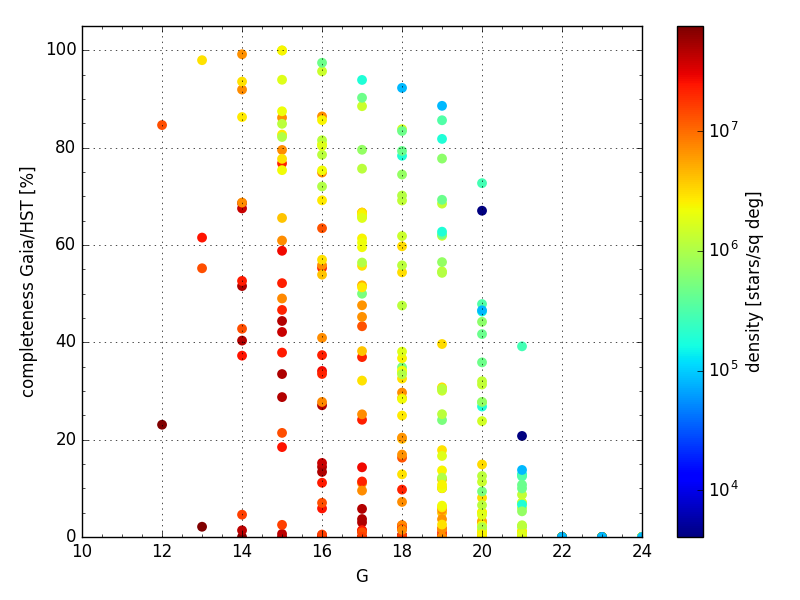}
\caption{Completeness levels with respect to HST data in different regions of 26 globular clusters with various local density, showing the influence of crowding on the completeness.} \label{fig:completeness_allGCs_colourbydensity}
\end{center}\end{figure}

For very crowded regions, we used HST observations of 26 globular clusters, which are expected to be complete down to at least $G\sim24$ and with a spatial resolution comparable to the {\gaia} one. The HST data we employed are the same as were used in \citet{2017A&A...599A..50A}. They were acquired by \citet{2007AJ....133.1658S} with the ACS and contain photometry in F606W and F814W filters, which we transformed to \textit{Gaia} $G$ magnitudes 
through a direct comparison of the magnitudes of the stars in {\gmag}, F606W and F814W passbands. This avoids issues due to variations of metallicity, and interstellar extinction.
For each cluster, we compared the number of sources in various magnitude slices in the inner (inside 0.5 arcmin) and outer (0.5 to 2.2 arcmin) regions. The result of all clusters is given in Table~\ref{tab:completeness26gcs}, and an example for NGC~6121 (M4) is shown in Fig.~\ref{fig:completeness_NGC6121_coreVSoutskirts}. The information contained in Table~\ref{tab:completeness26gcs} is also visually represented in Fig.~\ref{fig:completeness_allGCs_colourbydensity}, where it is clearly visible that the completeness level depends on both magnitude and local density (for $G<20$). Overall the completeness level of {\gdrtwo} is much higher than in {\gdrone}.  

\subsection{Small-scale completeness of Gaia DR2}\label{ssec:data_completeness}

\begin{figure}\begin{center}
\includegraphics[width=0.6\columnwidth, height=0.5\columnwidth]{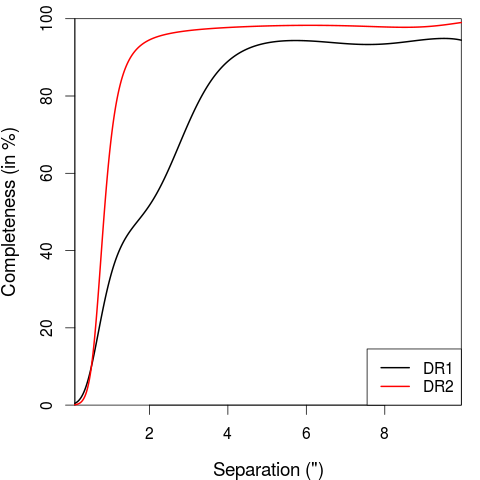}
\caption{Improvement of the completeness (\%) of visual double stars from the WDS Catalogue as a function of the WDS separation between components, from {\gdrone} (black) to {\gdrtwo} (red).}\label{fig:WDScompleteness}
\end{center}\end{figure}

One first way to check the spatial resolution of the {\gaia} Catalogue is to use known double stars, using the Washington Visual Double Star Catalogue \citep[WDS;][]{WDS}. Figure~\ref{fig:WDScompleteness} shows the completeness as a function of the separation between visual double stars as provided by the WDS. This shows that the completeness starts to drop at around 2{\arcsec}, while it was around 4{\arcsec} in {\gdrone}.

\begin{figure}\begin{center}
\includegraphics[width=0.9\columnwidth]{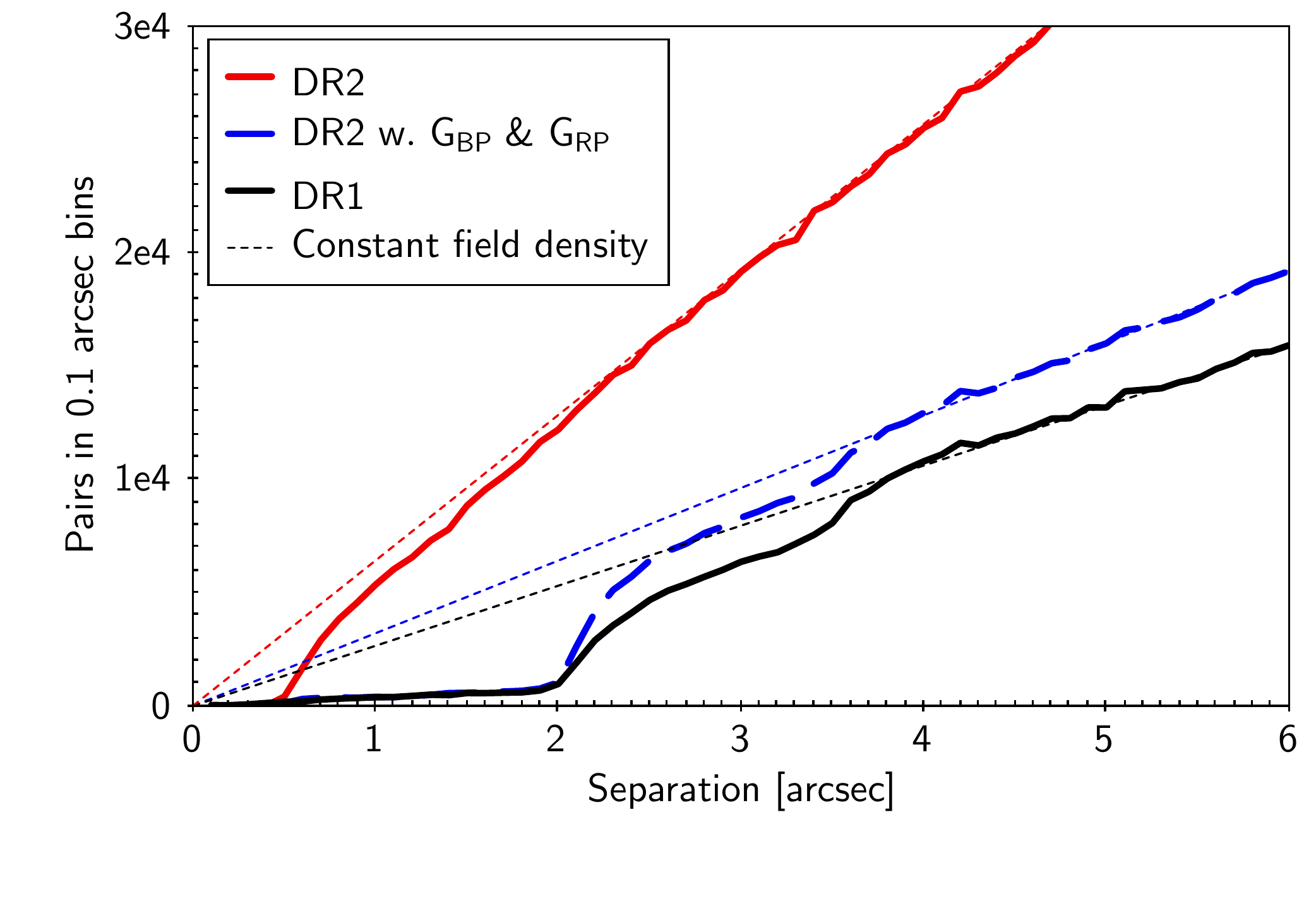}
\includegraphics[width=0.9\columnwidth]{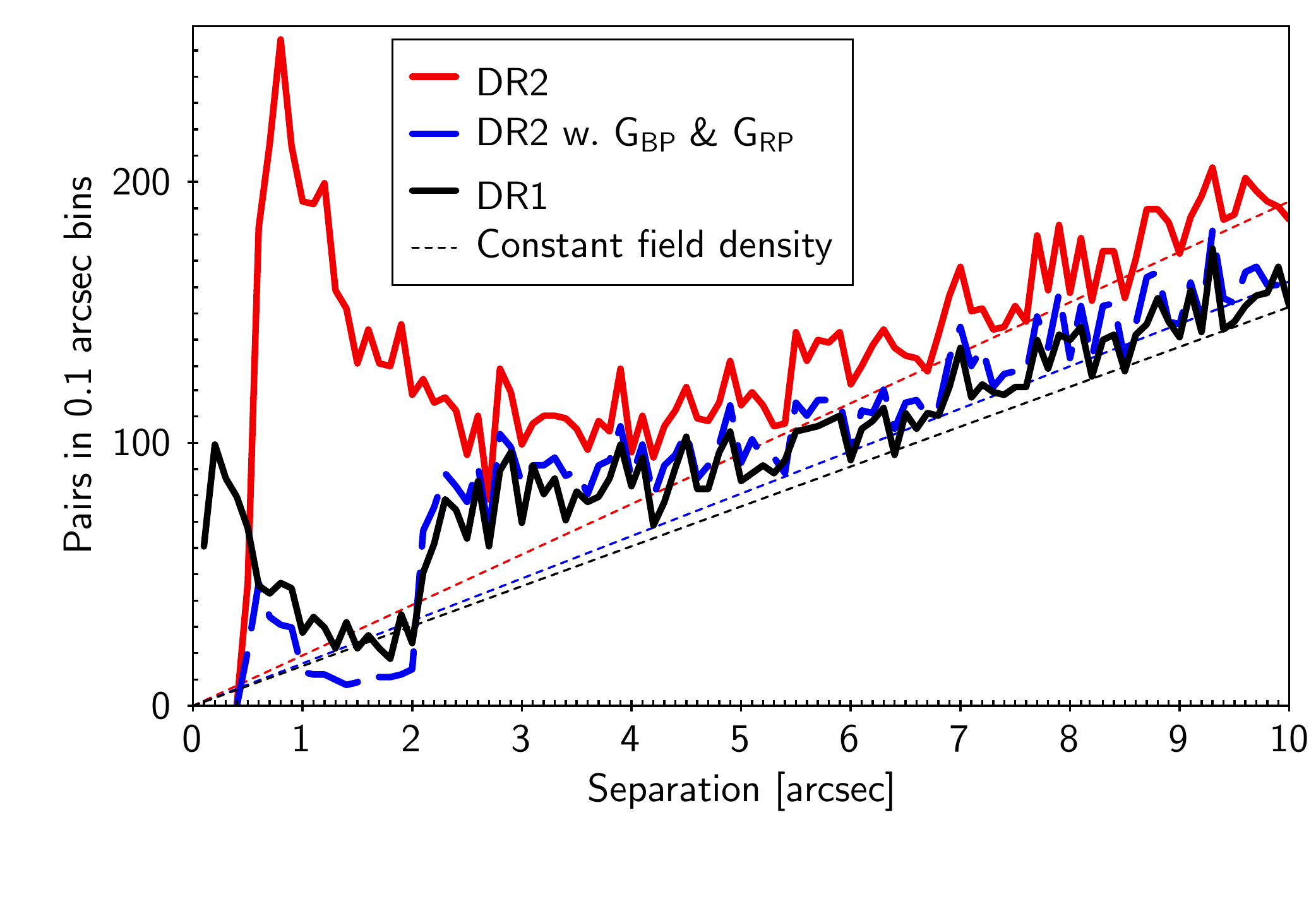}
\caption{Histogram of source pair separations in two test fields
for {\gdrtwo} sources; for {\gdrtwo} sources with \gbp\ and \grp\ photometry; and
for {\gdrone} sources. {\em Top:} a dense field at $(l,b) = (-30\degr,-4\degr)$,
{\em Bottom:} a sparse field at $(l,b) = (-100\degr,-60\degr)$.
The thin, dotted lines show the relation for a random distribution.
\label{fig:pair_stat}}
\end{center}\end{figure}

The small-scale completeness can also be evaluated from the distribution of
distances between source pairs in the whole Catalogue.
Figure~\ref{fig:pair_stat} shows distributions in two test fields, a dense
field near the Galactic plane and a sparse field at $-60\degr$ Galactic
latitude. The dense field contains 456\,142 sources in a circle of radius
0.5\degr, while the sparse has 250\,092 sources within a radius of 5\degr.
The sparse field has therefore 200 times smaller surface density than the
dense one.  From {\gdrone} to {\gdrtwo},
the dense field has obtained 56\% more sources, whereas the sparse field has
only gained 12\%. The top panel shows the distributions for the dense field. In
{\gdrone} (lower, black curve) there is a deficit of pair distances smaller
than 3\farcs7 and extremely few below 2\arcsec. For {\gdrtwo} (upper, red
curve) the deficit does not set in until 2\farcs2 and drops gently to zero
around 0\farcs5.  For {\gdrone} it was required that all sources had a known
colour, but this requirement has been waived for {\gdrtwo}. This difference
explains the gain in angular resolution as illustrated by the middle, blue,
dashed curve showing the distribution for {\gdrtwo} sources with known colour.
This curve shows the same features as the one for {\gdrone}, but lies a bit
higher due to a gain of 10\% in the number of sources.  The specific distances
where the resolution changes are related to the size of the data acquisition
windows, $0\farcs7\times2\farcs1$ for the point images and
$3\farcs5\times2\farcs1$ for the low-dispersion spectra. The situation for the
sparse field is dramatically different for separations below 2\arcsec, where we
now see a strong peak of binaries. A population of binaries must also be
present in the dense field, which however is dominated by several times more
remote sources.

In view of \figref{fig:pair_stat}b, one would have expected the binaries to grow towards low separations, naively down to the angular resolution, $\sim 0\farcs12$. One may thus wonder where did the missing binaries with a $0\farcs12<\rho<0\farcs5$ separation go. Because there was no special treatment for non-single objects, the missing binaries were actually handled as single objects, which may have sometimes corrupted astrometry or colour photometry and produced either the sources with two astrometric parameters only, or the spurious solutions discussed at \secref{sec:err_ast}.

\subsection{High proper motion stars}\label{ssec:hpm}

Looking for known high proper motion (HPM) stars we find that
17\% of the SIMBAD HPM  stars with a proper motion larger than 0.6~arcsec\,yr$^{-1}$ are missing in {\gdrtwo}, preferentially bright stars.  

In {\gdrone} much more HPM stars were missed because the cross match of the observations to the sources relied on ground-based proper motions. For {\gdrtwo} the cross match is much more independent of a star catalogue and this has given a significant improvement, and further improvements are already in place for the future.

\section{Astrometric quality of {\gdrtwo}}\label{astroqual}
We have mainly checked the astrometric quality of {\gdrtwo} for sources with the full, five-parameter astrometric solution and with focus on the parallaxes and proper motions. 
The remaining 360~million sources, with only two published parameters, are either fainter than 21~mag, have only few transits, or gave very bad fits to the five-parameter model (binaries, diffuse objects, etc.). The quality of this group is much lower than for the rest of the sources and it is therefore of limited interest. 

We have also checked the reference frame, which was aligned to the IRCF3-prototype, but we do not mention our results here as they are in full agreement with \cite{DR2-DPACP-30} and \cite{DR2-DPACP-51}, to which we refer.

\subsection{Spurious astrometric solutions}\label{sec:err_ast}

\begin{figure}\begin{center}
\includegraphics[width=0.8\columnwidth]{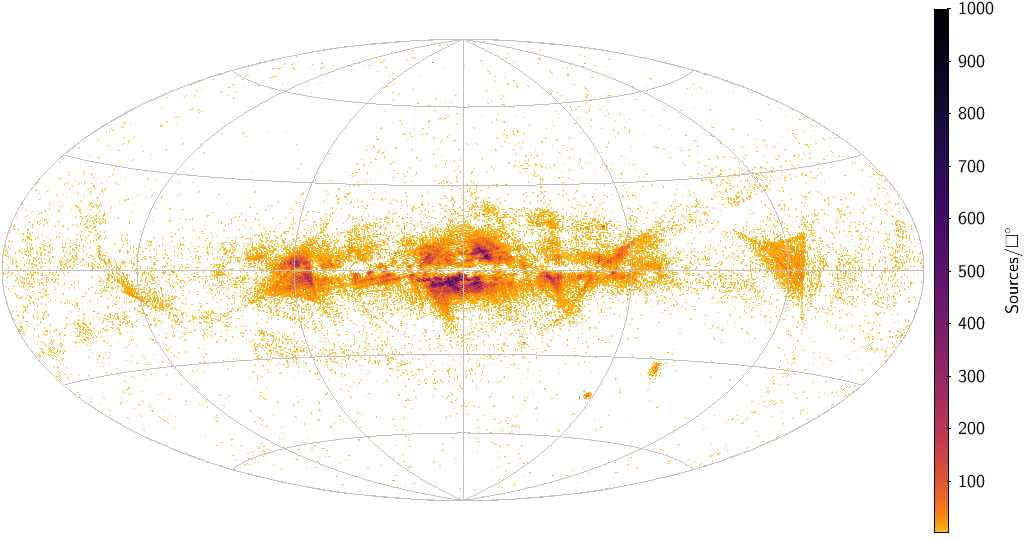}
\includegraphics[width=0.8\columnwidth]{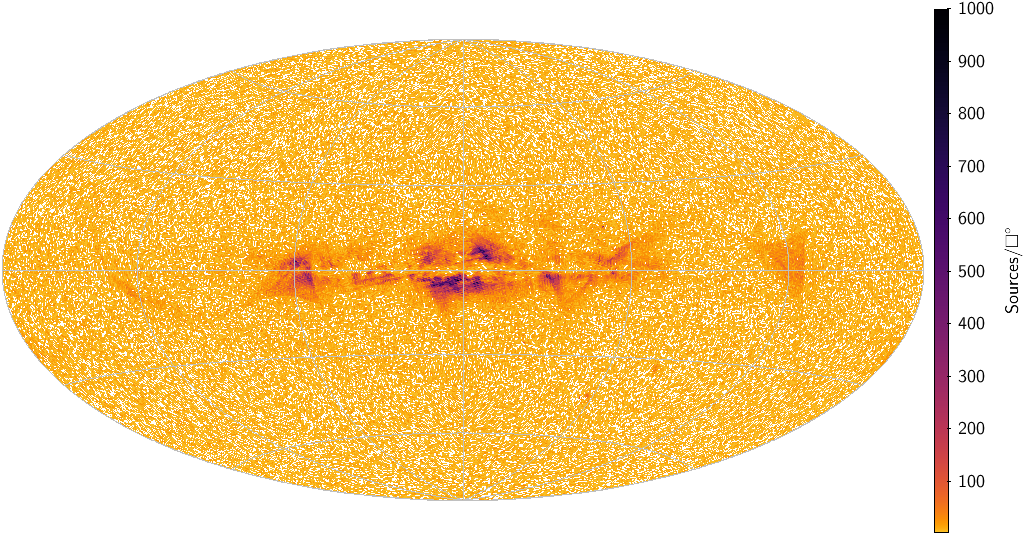}
\includegraphics[width=0.8\columnwidth]{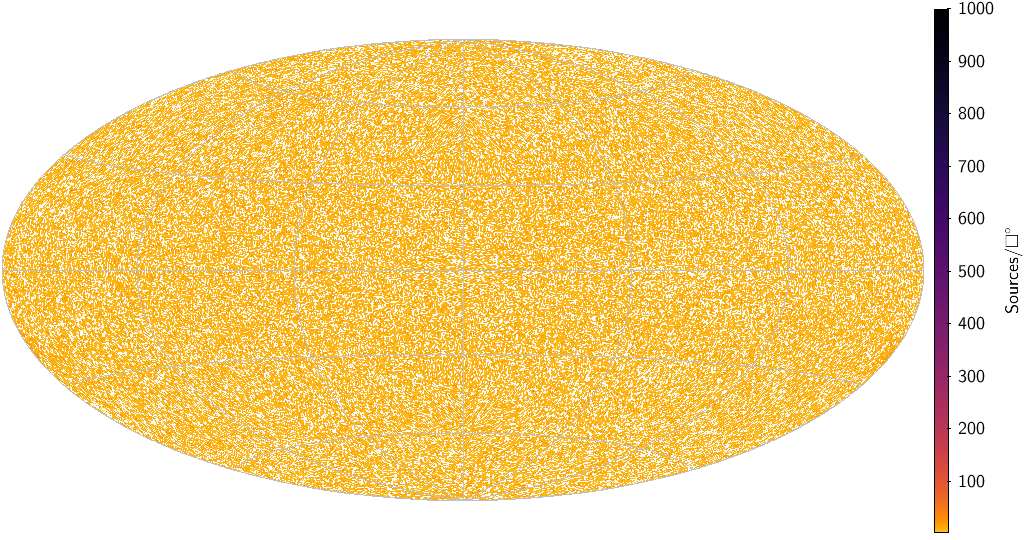}
\caption{Sky maps in Galactic coordinates showing sources with parallaxes with
less than
10\% error and significantly different from zero, {\em Top:} parallaxes below $-10$\,mas;
{\em Centre:} parallaxes larger than $+10$\,mas; and {\em Bottom:}
parallaxes larger than $+10$\,mas after applying the quality filters in Eqs.
\ref{eq1} and \ref{eq2}.
}\label{fig:map_par10}
\end{center}\end{figure}

Good astrometric solutions can only be obtained if there are many scans well spread in scanning angle and in a sufficiently long period of time\footnote{In this respect, the Catalogue field  \dt{visibility\_periods\_used} indicates the number of groups of observations separated from other groups by at least 4 days.} \citep{DR2-DPACP-51}. In some parts of the sky this fundamental requirement was not met during the 21 months of observations used in {\gdrtwo} astrometry. Solutions will in these areas be more susceptible to e.g.\ disturbances introduced by a close-by source. Especially difficult are areas where one or two scan directions dominate and even more so if one of these directions is perpendicular to the direction to the Sun and therefore insensitive to parallax. In future data releases, based on longer time series, this problem will diminish. 

An obvious way to check for problematic astrometric solutions is to look for significantly negative parallaxes. Figure~\ref{fig:map_par10} shows, top panel, the sky distribution of the 113\,393 sources with parallaxes below $-10$\,mas and $\varpi/\sigma_\varpi < -10$. They clearly concentrate in the dense areas of the Galactic plane and the Magellanic clouds, and especially in some areas delineated by scan patterns. In the centre panel, showing the same, but for 439\,020 positive parallaxes, we see the same patterns, but with a uniform background of supposedly well-behaved, astrometric solutions. Finally the bottom panel shows the 254\,007 positive parallaxes after the application of the quality filters defined in Eqs.~\ref{eq1} and \ref{eq2}. The same filters reduce the number of negative parallaxes to just 298. We conclude that some sky areas contain sources with spurious astrometry, and that these poor solutions may equally well contain a negative as a positive parallax. 

Proper motions are as concerned as parallaxes.
For example, 6189 stars have a proper motion larger than 500 {\masyr} in {\gdrtwo}, of which only 70\% are known in SIMBAD. Selecting only the stars with $\dt{visibility\_periods\_used}>8$, i.e.\ with a better astrometric quality, this number raises to 93\%, showing that the non selected are probably spurious.

\begin{figure}\begin{center}
\includegraphics[width=0.43\columnwidth,height=0.5\columnwidth]{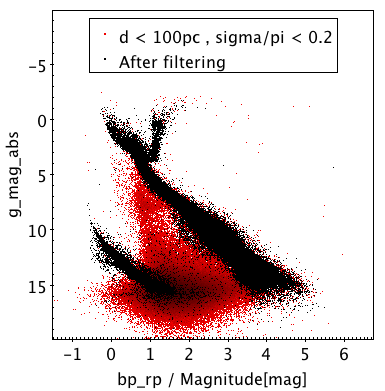}
\includegraphics[width=0.47\columnwidth]{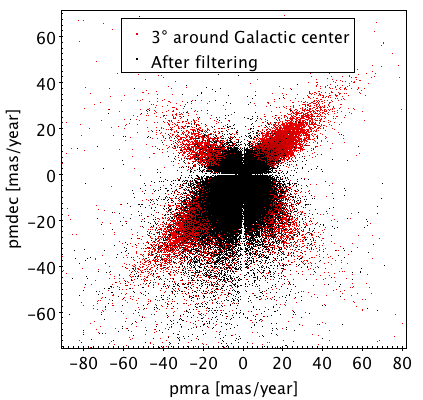}
\caption{H-R diagram of stars closer than 100 pc (left) and proper motion diagram near the Galactic center (right), with (black) or without (red) filtering of spurious solutions with \eqref{eq1}+\eqref{eq2}. In both cases, a 20\% relative uncertainty truncation on the astrometric parameters has also been applied (which generates a ``void cross'' at the origin on the right).}\label{fig:HR-filtered}\end{center}\end{figure}

The detrimental impact of the spurious solutions appears clearly on H-R diagrams (Fig.~\ref{fig:HR-filtered}a) or proper motion diagrams (Fig.~\ref{fig:HR-filtered}b). These figures also show that, fortunately, quality filters can be devised to clean the samples.

The filter used in these plots is the same as defined by \cite{DR2-DPACP-31} for their study of the HR diagram and in appendix~C of \cite{DR2-DPACP-51}, theirs Eqs.\ C-1 and C-2. Defining:
\begin{itemize}
\item $\chi^2$ = \dt{astrometric\_chi2\_al}
\item $\nu$ = \dt{astrometric\_n\_good\_obs\_al -5}
\item $u = \sqrt{\chi^2/\nu}$
\item $E$ = \dt{phot\_bp\_rp\_excess\_factor}\footnote{\dt{phot\_bp\_rp\_excess\_factor} is the ratio of the sum of {\gbp} and {\grp} fluxes over the $G$ flux and should be around one for normal stars.}
\end{itemize}
we accept solutions fulfilling the conditions:
\begin{equation}
    u < 1.2\times\max(1,\exp(-0.2(G-19.5))) \, , \label{eq1}
\end{equation}
and
\begin{equation}
    1.0+0.015(G_\text{BP}-G_\text{RP})^2 < E < 1.3+0.06(G_\text{BP}-G_\text{RP})^2 \, . \label{eq2}
\end{equation}

By rejecting large $\chi^2$, \eqref{eq1} helps filtering contamination from double stars, astrometric effects from binary stars and also from calibration problems.
As surprising as it seems, the photometric filtering defined at \eqref{eq2} cleans even more efficiently the spurious astrometric solutions, by detecting the perturbations due to close-by sources, but it mostly cleans the faint stars (that would have been rejected otherwise by a filtering on the photometric precision). 

To realise that filtering does not come cheap, and why it could not have been applied for the production of the {\gdrtwo} Catalogue, the fraction of remaining sources is illustrative: in Fig.~\ref{fig:HR-filtered}a, filtering keeps 39\% of the sources, and only 26\% survive in Fig.~\ref{fig:HR-filtered}b. The filter \eqref{eq2} has the largest effect; if it had been applied alone, it would have kept only 30\% of the sources. Obviously, the fraction of filtered data depends on magnitude, on parallax and proper motion, and it introduces additional selection effects.

\begin{figure}\begin{center}
\includegraphics[width=0.45\columnwidth]{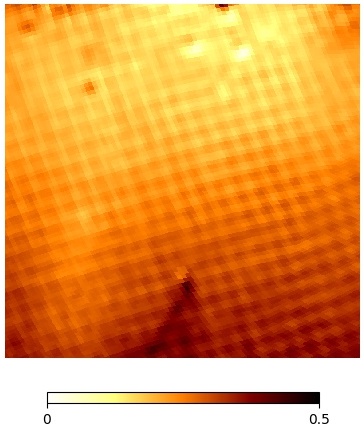}
\includegraphics[width=0.45\columnwidth]{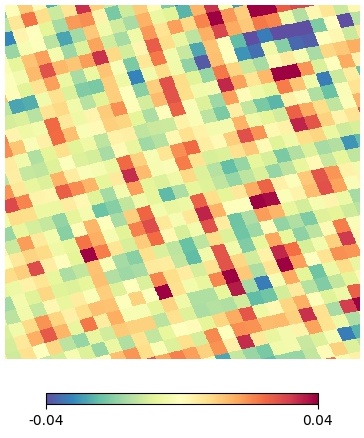}
\caption{Small scale systematics: map of median parallaxes (mas) in a 10{\degr} field centered on $(l,b)=(0\degr,-12\degr)$ (left). Residuals (mas) of median parallaxes in field $(1\degr,-7\degr)$, size 3{\degr} for stars brighter than $G=17$ only, after subtraction of a 0.7{\degr} running median  (right).}\label{fig:astro-syst-bulge}
\end{center}\end{figure}
\begin{figure}\begin{center}
\includegraphics[width=0.45\columnwidth]{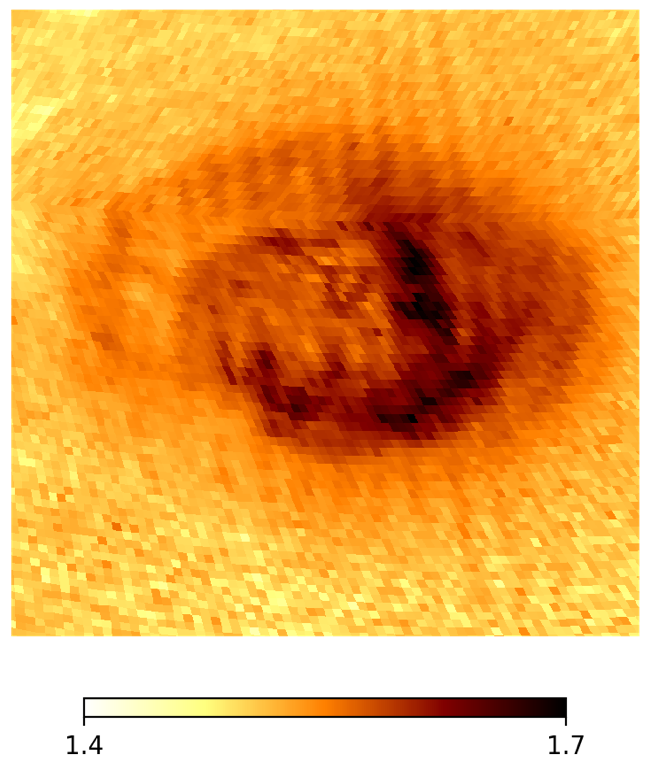}
\includegraphics[width=0.45\columnwidth]{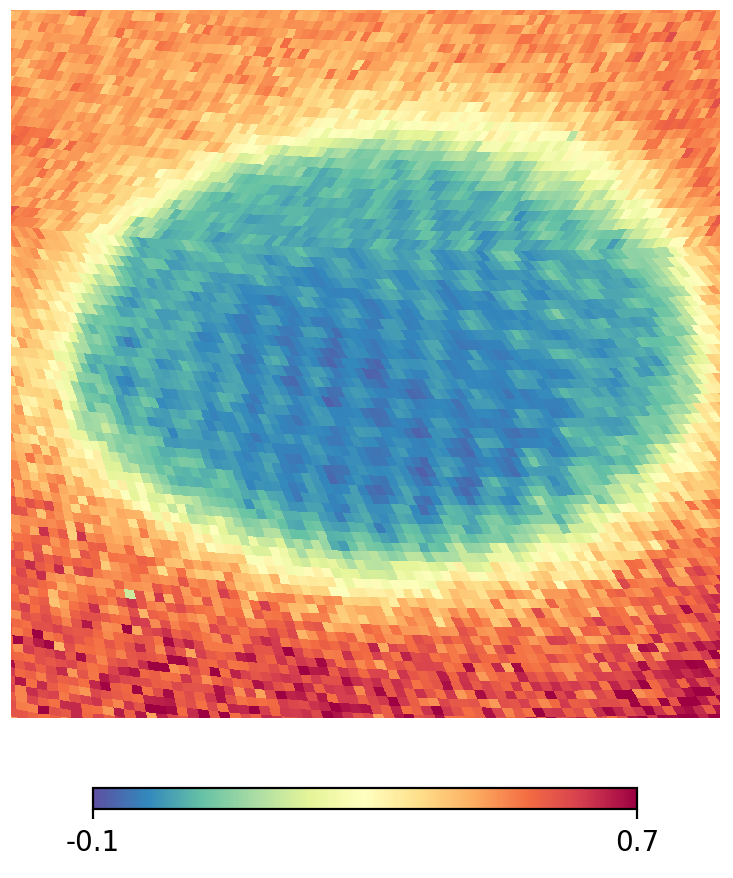}
\caption{In the direction of the LMC, the median of the pseudo colour (left) shows a similar banding effect as for parallaxes (in mas, right).}\label{fig:pseudo-colour-LMC}
\end{center}\end{figure}
\begin{figure}\begin{center}
\includegraphics[width=0.45\columnwidth]{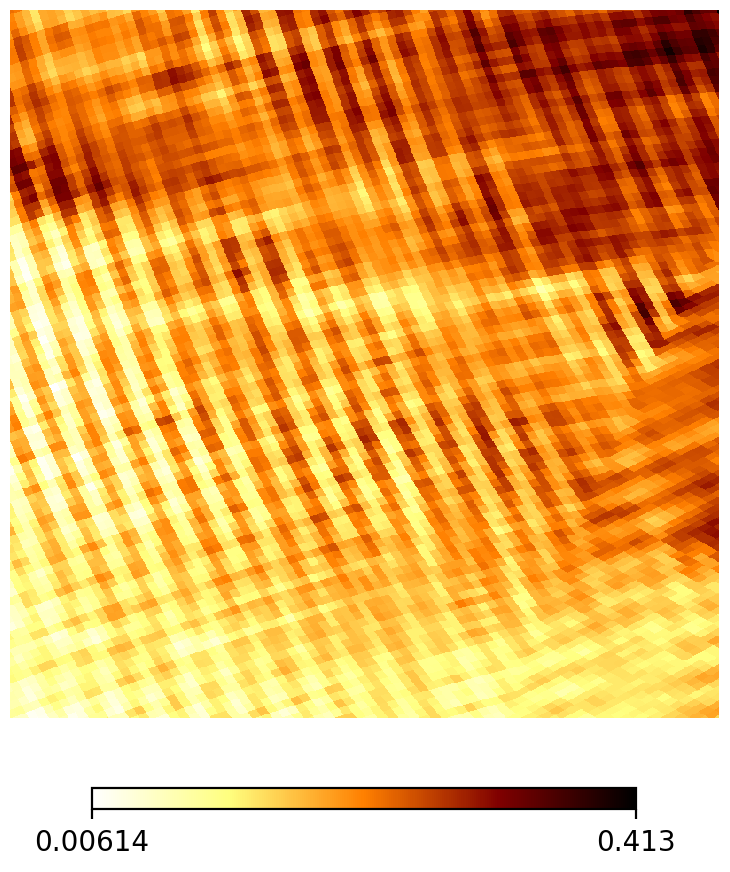}
\includegraphics[width=0.45\columnwidth]{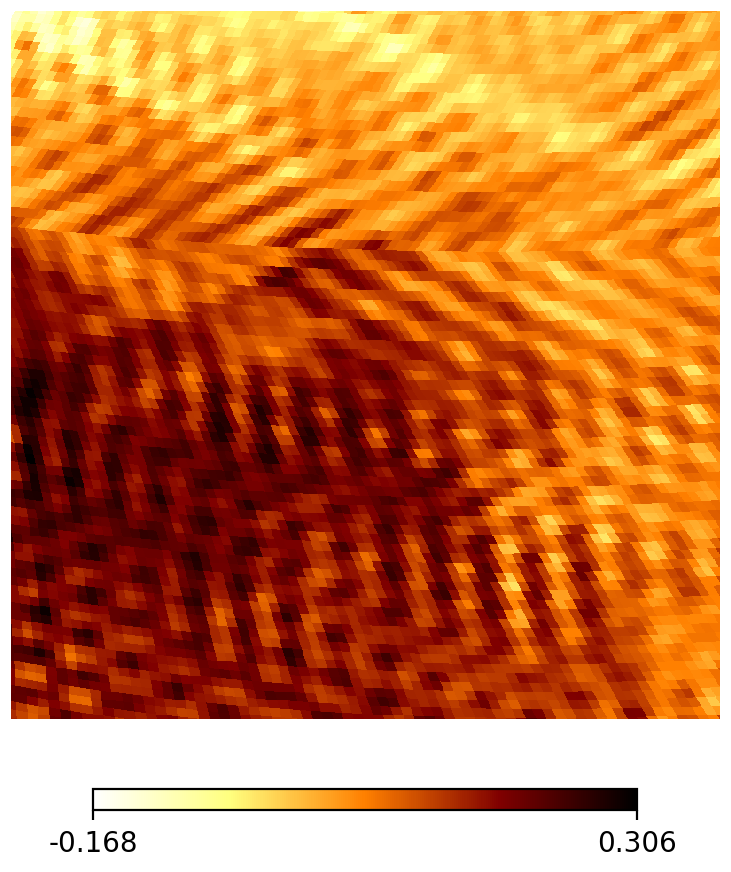}
\caption{Correlation $\rho(\varpi,\mu_\delta)$  towards the bulge (left) and $\rho(\alpha,\delta)$  towards the LMC, same fields as, respectively, \figref{fig:astro-syst-bulge}a and \figref{fig:pseudo-colour-LMC}. }\label{fig:correlations}
\end{center}\end{figure}

Other filters may of course be defined depending on the application. For instance, replacing \eqref{eq2} by 
\begin{equation}
\dt{visibility\_periods\_used}>8  \label{eq3}
\end{equation}
may look at first sight removing many good solutions having not enough observations, but it would actually be preferable for the bulge proper motions as it would increase from 26\% to 72\% the remaining data in Fig.~\ref{fig:HR-filtered}b, while probably cleaning enough the bad solutions as the influence of crowding on photometry is perhaps not an issue here; however this filtering would not be sufficient for the HR diagram,  Fig.~\ref{fig:HR-filtered}a, leaving too many spurious intrinsically faint stars; however, if external colours are being used for the HR diagram, the criterion \eqref{eq2} may not be necessary either.
To summarise, the quality filters to be applied may typically be either \eqref{eq1}+\eqref{eq2} or \eqref{eq1}+\eqref{eq3} depending on whether a photometric filtering is needed or not.

To end on a positive note, if the fraction of rejected source may appear at first sight very high, the probability of a bad solution when taking a star at random is quite low. What happens is that spurious solutions produce large astrometric values: selecting high proper motion stars will preferentially select spurious proper motions; making an HR diagram with nearby stars only will select large parallax values, with a larger fraction of spurious ones. In some other random sample, however, robust statistics may be enough to mitigate their effect.

\subsection{Small scale systematics}\label{ssec:astroacc_small}

As shown in \cite{DR2-DPACP-51}, spatial correlations are present in the astrometry, producing small scale systematic errors. In scientific applications, this means that the average parallax or proper motion in a small field will be biased if the systematic error is not accounted for. In practice, they limit the asymptotic precision gain on samples of stars to $\sqrt\rho$ instead of the expected $1/\sqrt N$, where $\rho$ is the correlation between sources.

Although probably present over the whole sky, these correlations can be more easily seen in fields mostly made of distant stars, where the true parallax is small compared to the parallax error, e.g. in dSphs  \citep{DR2-DPACP-34}, in the direction of the LMC, Fig.~\ref{fig:pseudo-colour-LMC}, or the bulge, Fig.~\ref{fig:astro-syst-bulge}. In the latter field, the scanning law pattern appears clearly, and the systematics have at least a $\approx 0.02$ mas RMS over a $\approx 0.6\degr$ period, and they are present for faint as well as for more brighter stars. This banding pattern producing systematics for parallaxes and proper motions at small angular scale is however more difficult to handle as it changes shape, orientation and amplitude across the sky.

\subsection{Large scale systematics}\label{ssec:astroacc_large}

\begin{figure}\begin{center}
\includegraphics[width=0.8\columnwidth]{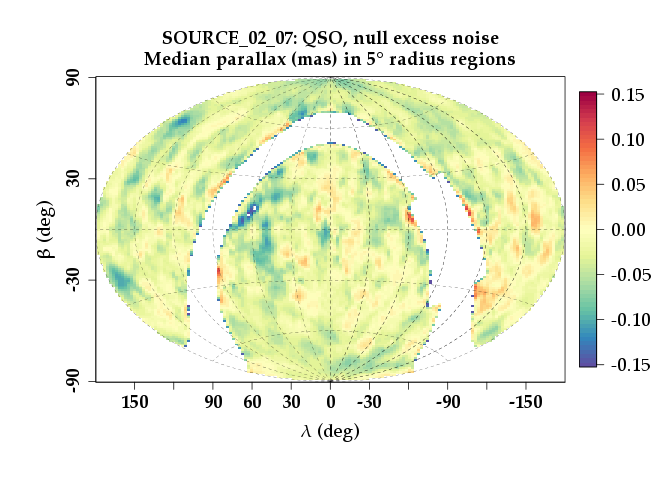}\caption{Variations of the QSO parallaxes (mas) in 5{\degr} radius fields, ecliptic coordinates. Beside a $\approx -0.03$ mas median zero-point, large scale variations also appear with the ecliptic longitude.}\label{fig:MDB02.05-ecl-qso-reg-MedPlx-r6-zoom}
\end{center}\end{figure}

The systematics on a larger scale can be studied using QSOs.
Compared to the thorough QSO selection made in \cite{DR2-DPACP-51}, 
we used a simpler approach using sources identified as QSO in the {\gdrtwo} Catalogue\footnote{Catalogue field {\tt\small frameRotatorObjectType} equal to 2 or 3, i.e. a source assumed to be extragalactic and used to determine the reference frame.}. However, we kept only those with a parallax over error smaller
than 5 in absolute value and a null astrometric excess noise\footnote{The {\tt\small astrometric\_excess\_noise}, cf.\ \citet{DR2-DPACP-51}, expresses the astrometric goodness of fit (GoF) in angular measure. This is the quantity that would be needed to be added to the measurement uncertainties to make the weighted sum of squared residuals equal or smaller than the degree of freedom. It is $>0$ only for poor fits.} to avoid bad astrometric solutions due to e.g. perturbations by nearby sources.
Non-QSO may however remain in the sample and we will not pay
attention to the results near the galactic plane where there are not enough sources. Only 220\,178 QSO remain in this sample.

In order to estimate how the systematics are spatially represented, median of parallaxes have been computed in random regions defined by a given radius and with at least 20 QSOs. Compared with a similar plot done for {\gdrone} \citep[][Fig.~24]{2017A&A...599A..50A}, 
Fig.~\ref{fig:MDB02.05-ecl-qso-reg-MedPlx-r6-zoom} shows an improvement by a factor at least 3 for the amplitude of the systematics. 

There is a significantly negative global zero point ($\approx -0.03$ mas, cf. Table~\ref{tab:cu9val_wp944_summaryplx}), and some variations appear on a larger scale. For example, in a 5{\degr} radius around $(\alpha,\delta)=(191\degr, 50\degr)$ the median parallax is $-0.105\pm 0.031$ mas, to be compared with $+0.028\pm 0.027$ mas in a 5{\degr} radius region around $(\alpha, \delta)=(30\degr, -8\degr)$, i.e., there can be up to a 0.13 mas peak-to-peak 
variation of the parallax systematics over the sky. A statistical study of these angular correlations has been done by \citet[][Sect.~5.4]{DR2-DPACP-51}. Large scale systematics are of smaller, though not negligible amplitude.

\subsection{Global systematics from external comparisons}\label{sect:colourSystematics}

\begin{table*}
\caption[Comparison between the {\gdrtwo} parallaxes and the external catalogues]{Summary of the comparison between the {\gaia} parallaxes and the external catalogues.}
\begin{center}
\begin{tabular}{lcccccc} 
\hline\hline
{ Catalogue} & Nb & { Outliers} & $<G>$ & { \parallax~difference} & { \parallax~uwu}  \\ 
 \hline
  {\hip} & 62484 & 0.1\% & 8.3 & \textcolor{purple}{$-0.118 \pm 0.003$} & \textcolor{purple}{$1.25 \pm 0.003$}  \\ 
 \hline
  VLBI & 40 & 2.5\% & 8.2 & \textcolor{black}{$-0.07 \pm 0.03$} & \textcolor{purple}{$1.9 \pm 0.2$}  \\ 
  HST & 51 & \textcolor{purple}{33\%} & 11.7 & \textcolor{black}{$-0.01  \pm 0.02$} & \textcolor{purple}{$2.1 \pm 0.3$} \\
  RECONS & 432 & 3\% & 12.6 & \textcolor{purple}{$-0.71 \pm 0.06$} & \textcolor{purple}{$1.69 \pm 0.06$}  \\ 
 \hline
  GCVS RR~Lyrae & 197 & 2\% & 14.9 & \textcolor{purple}{$-0.033 \pm 0.009$}  & \textcolor{purple}{$1.51 \pm 0.08$}  \\ 
  Gaia RR~Lyrae & 795 & 3\% & 15.6 & \textcolor{purple}{$-0.056 \pm 0.005$}  & \textcolor{purple}{$1.38 \pm 0.04$}  \\ 
  Gaia Cepheids & 1417 & 2\% & 15.6 & \textcolor{purple}{$-0.0319 \pm 0.0008$}  & \textcolor{purple}{$1.53 \pm 0.03$}  \\ 
 \hline
 APOGEE & 5212 & 2\% & 13.9 & \textcolor{purple}{$-0.048 \pm 0.002$}  & \textcolor{purple}{$1.44\pm0.01$}  \\ 
 LAMOST & 174  & 9\% & 14.9 & \textcolor{purple}{$-0.040 \pm 0.005$}  & \textcolor{purple}{$1.50\pm0.08$}  \\ 
 SEGUE Kg & 3151 & 0.2\% & 16.5 & \textcolor{purple}{$-0.041 \pm 0.002$}  & \textcolor{purple}{$1.10\pm0.01$}  \\ 
 \hline
 LMC & 51162 & 1\% & 19.2 & \textcolor{purple}{$-0.038 \pm 0.0004$} &  \textcolor{purple}{$1.098 \pm 0.004$}  \\ 
 LMC Vr & 319 & 4\% & 12.8 & \textcolor{purple}{$-0.042 \pm 0.001$} &  \textcolor{purple}{$1.34 \pm 0.05$}  \\ 
 SMC & 26404 & 2\% & 16.4 & \textcolor{purple}{$-0.0268 \pm 0.0004$} &  \textcolor{purple}{$1.43 \pm 0.006$}  \\ 
 SMC Vr & 114 & 8\% & 12.5 & \textcolor{purple}{$-0.037 \pm 0.002$} &  \textcolor{purple}{$1.4 \pm 0.1$}  \\ 
 Draco & 427 & 0\% & 19.3 & \textcolor{purple}{ $-0.047 \pm 0.008$} & \textcolor{purple}{$ 1.08 \pm 0.04 $} \\ 
 Ursa Minor & 78 &  0\% & 17.4 & \textcolor{purple}{ $-0.054 \pm 0.008 $} & \textcolor{black}{$ 1.03 \pm 0.08 $}  \\ 
 Sculptor & 1287 & 0.3\% & 19.1 & \textcolor{purple}{ $-0.028 \pm 0.006 $} & \textcolor{purple}{$ 1.11 \pm 0.02 $} \\ 
 Sextans & 375 & 0.3\% & 19.3 & \textcolor{purple}{ $-0.09 \pm 0.02$} & \textcolor{purple}{ $1.07 \pm 0.04 $} \\ 
 Carina & 864 & 0\% & 19.8 & \textcolor{purple}{ $-0.020 \pm 0.007$} & \textcolor{black}{ $1.05 \pm 0.03 $} \\ 
 Crater2 & 63 & 0\% & 19.1 & \textcolor{purple}{ $-0.06 \pm 0.03$} & \textcolor{black}{ $0.96 \pm 0.09 $}  \\ 
 Fornax & 2659 & 0.4\% & 18.8 & \textcolor{purple}{$ -0.052 \pm 0.004$} & \textcolor{purple}{$1.18  \pm  0.02 $}  \\ 
 CVnI &  51 & 0\% & 20.0 & \textcolor{black}{$ -0.030 \pm 0.08$} & \textcolor{black}{$0.91  \pm  0.09 $}  \\ 
 LeoII & 123 & 0\% & 19.5 & \textcolor{black}{ $0.05 \pm 0.05$} & \textcolor{black}{$1.0  \pm  0.06 $}  \\ 
 LeoI & 292 & 0.7\% & 19.6 & \textcolor{purple}{ $-0.23 \pm 0.05$} & \textcolor{purple}{$ 1.30  \pm  0.05 $}  \\ 
 Phoenix &  81 & 0\% & 20.6 & \textcolor{black}{ $0.09 \pm 0.07$ } & \textcolor{black}{$ 1.07  \pm  0.08 $} \\ 
 all dSph & 6300 & 0.3\% & 19.0 & \textcolor{purple}{ $-0.044 \pm 0.002$} & \textcolor{purple}{$ 1.13 \pm 0.01$}  \\ 
 \hline
 ICRF2 & 2347 & 0.3\% & 18.8 & \textcolor{purple}{$ -0.031 \pm 0.003$} & \textcolor{purple}{$1.16 \pm 0.02$}  \\ 
 RFC2016c &  3523 & 0.3\% & 18.9 & \textcolor{purple}{$ -0.031 \pm 0.002$} & \textcolor{purple}{$1.15 \pm 0.01$}  \\ 
 LQRF & 79631 & 0.04\% & 19.1 & \textcolor{purple}{$ -0.0322 \pm 0.0008$} & \textcolor{purple}{$1.088 \pm 0.003$}  \\ 
 \hline
\end{tabular}
\tablefoot{
The total number of stars used in the comparison (Nb) as well as the percentage of outliers excluded (at 5$\sigma$, in purple if larger than 10\%) as well as the median {\gmag} of the sample are presented.  The parallax differences ($\varpi_G-\varpi_E$, in mas) and unit-weight uncertainty (uwu) that needs to be applied to the uncertainties to adjust the differences are indicated in purple when they are significant (p-value limit: 0.01). 
}
\end{center}
\label{tab:cu9val_wp944_summaryplx}
\end{table*}

As in \cite{2017A&A...599A..50A}, we analysed the parallax systematics using a comparison to many external catalogues. Direct comparison of the parallaxes has been done with {\hip} \citep{2007A&A...474..653V}, VLBI \citep{2014ARA&A..52..339R}, HST \citep{2015IAUGA..2257159B, 2007AJ....133.1810B} and RECONS \citep{2015IAUGA..2253773H} parallaxes (using their database as of January 2018). 

Distance moduli were compiled for distant stars, distant enough so that the uncertainty on their parallax is 10 times smaller than the {\gaia} one. Distance moduli from variable period-luminosity relations were obtained for RRab RR~Lyrae and fundamental mode Cepheid stars using both GCVS \citep{GCVS} variables and directly {\gaia} provided ones, using both the supervised classifications and the SOS component of the variability pipeline \citep{DR1-DPACP-15}. We used the 2MASS \citep{2006AJ....131.1163S} magnitude independent of extinction  $K_{J-K} = K - \frac{k_K}{k_J-k_K} (J-K)$, with $k$ the extinction coefficients, and the period luminosity relation of \cite{2015ApJ...807..127M} for RRLyrae (using the metallicity information from the {\gaia} light curve when available, assuming $-1$~dex with a dispersion of 0.6~dex otherwise) and of \cite{2007A&A...476...73F} for Cepheids.

Distance moduli were also compiled from spectroscopic surveys, here APOGEE DR14 \citep{2015AJ....150..148H} and LAMOST DR2 \citep{LamostDR1}, using $K_{J-K}$ and Padova isochrones \citep[CMD 2.7]{Bressan12}. A catalogue of distances of SEGUE K giants \citep{2012AJ....144....4M} was also used. In contrast to the {\gdrone} validation, we do not provide anymore comparisons with asteroseismic distances due to the small number of stars with a distance information significantly smaller than the {\gaia} one. 

Very distant stars, for which the true parallaxes can be considered as almost zero, were also compiled through Milky Way satellites confirmed members, mostly using their radial velocities for dSph. For the LMC and SMC, the bright subset for which we could use the {\gaia} radial velocities to confirm their membership was also tested (called LMC/SMC Vr in Table~\ref{tab:cu9val_wp944_summaryplx}). Finally, parallaxes of confirmed QSOs were tested from the ICRF2 \citep{2015AJ....150...58F}, RFC2016c\footnote{http://astrogeo.org/vlbi/solutions/rfc\_2016c/} and LQRF \citep{2009A&A...505..385A} catalogues.

More details about the construction of those catalogues are provided in the on-line Catalogue Documentation, Section~10.4.
The results of the comparison are summarised in Table~\ref{tab:cu9val_wp944_summaryplx}. All the catalogues point towards a global zero point bias in the parallax of about -0.03~mas, with sky variations illustrated by the dSph members \citep[see also][]{DR2-DPACP-34}. 

For most of the tests, variations with magnitude, colour and pseudo-colours\footnote{The {\tt\small astrometric\_pseudo\_colour} is an astrometrically determined effective wavenumber given in $\mu{\text m}^{-1}$, see \citet[][Sect.~3.1]{DR2-DPACP-51}.} have been found, depending whether we look at the weighted mean differences or at the normalised differences,
indicating correlations with the uncertainty estimates (see \secref{ssec:astrocal_2}). 
The strongest correlation of the differences with colour and magnitude is seen with APOGEE, the difference being larger for the redder sources which are also the faintest, which may be due to systematics linked to the isochrones used.  
For the Cepheids, variations with the astrometric excess noise and GoF are present, indicating possible contamination with binaries.

\begin{figure*}[ht]\begin{center}
\includegraphics[width=1.6\columnwidth, height=0.8\columnwidth]{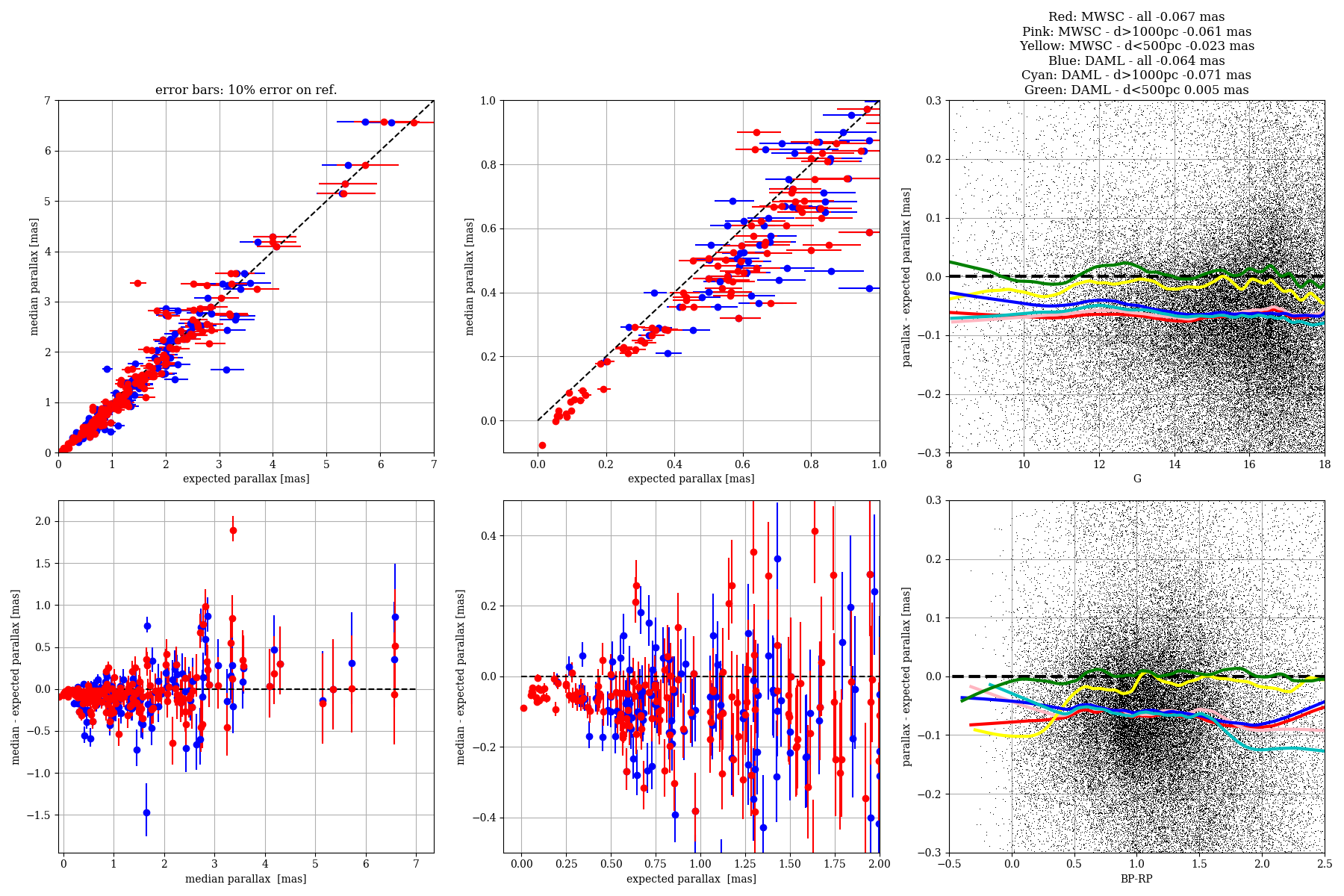}
\caption{Upper left panel: cluster median DR2 parallaxes compared with literature: MWSC (red dots) and DAML (blue dots). Upper central panel: same as left panel, restricted to $\varpi < 1$ mas. Upper right panel: parallax differences for the cluster star sample as a function of $G$ (black dots). Lower left panel: parallax differences for the whole sample. Lower central panel: same as left panel, for $\varpi < 2$ mas. The right panels are the analogous for the whole star sample (black dots). Lower right panel: parallax differences as a function of the colour $(G_{BP}-G_{RP})$). In the right panels, lines show the smoothing for both reference catalogues for different distances. Red line indicate  the whole MWSC, pink is MWSC OCs with distance $d > 1000$ pc; yellow is MWSC OCs with $d < 500$ pc; blue is DAML, all OCs; cyan is DAML with $d > 1000$ pc; green is DAML with $d < 500$ pc.}\label{fig:allmembers_parzeropoint}
\end{center}\end{figure*}
\begin{figure*}\begin{center}
\includegraphics[width=1.3\columnwidth]{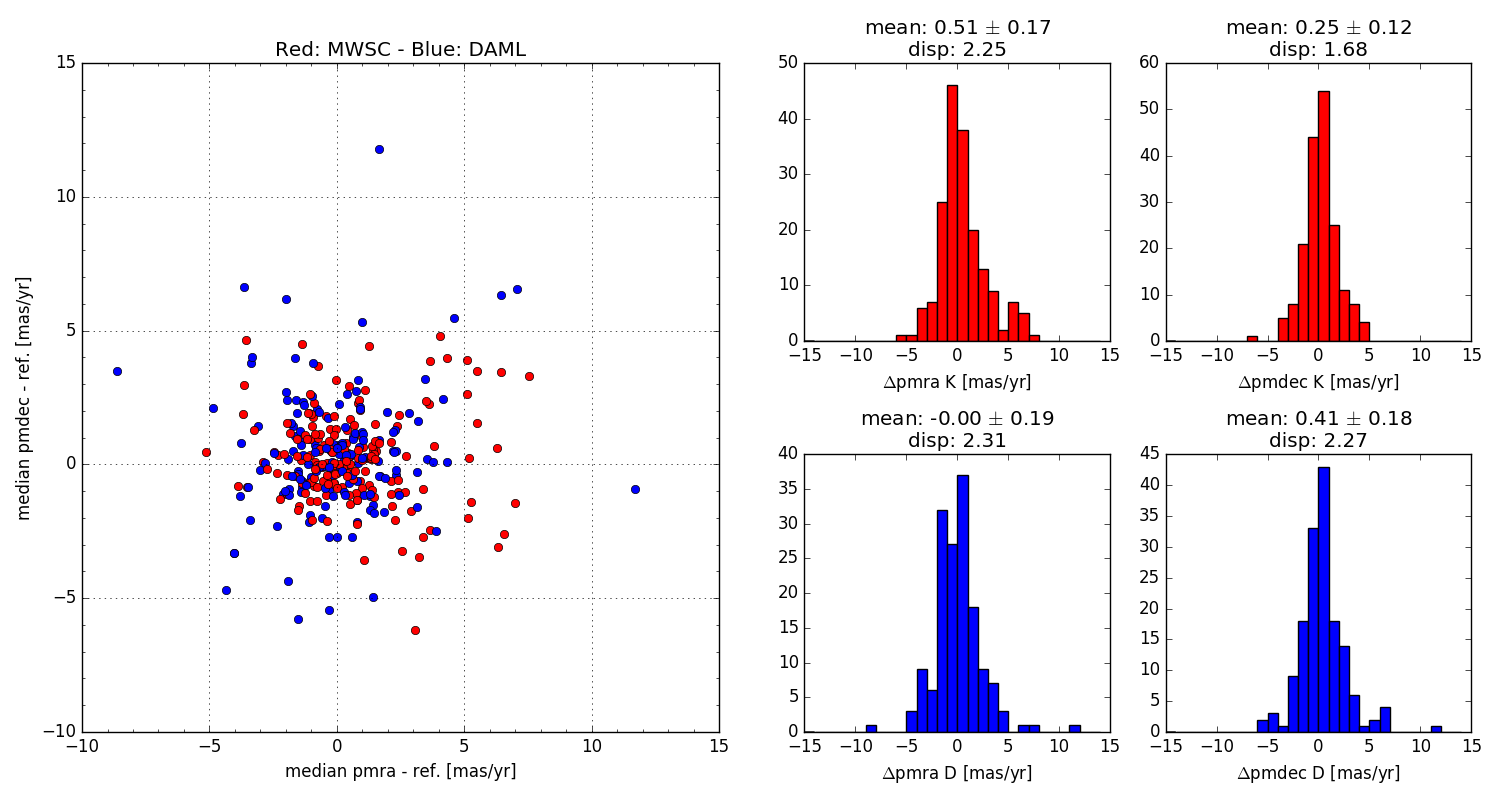}
\caption{Difference between the median DR2 proper motion of the clusters in $\mu_{\delta}$ vs  $\mu_{\alpha}\cos\delta$   for DAML (blue) and MWSC (red) sample (left panel). Distribution of the differences in $\mu_{\alpha}\cos\delta$  and  $\mu_{\delta}$ for MWSC (right upper panel) and the analogous for DAML (right lower  panel) }\label{fig:propermotion_literature}
\end{center}\end{figure*}

The analysis of about 200 clusters, including open (OCs) and globular (GCs) clusters, also shows a residual zero point in  parallaxes. We computed the differences between the actual DR2 value and the reference value for all the stars in the clusters using the DAML \citep{2014A&A...564A..79D} and MWSC \citep{2013A&A...558A..53K} catalogues. The difference depends on the reference catalogue, on the distance of the cluster, and on the colour of the stars. On the average, for the whole cluster sample, the residual zero point is $\varpi_{\rm Gaia}-\varpi_{\rm reference}=-0.067 \pm 0.12$ mas for MWSC and $-0.064 \pm 0.17$ mas for DAML. 
Fig.~\ref{fig:allmembers_parzeropoint} shows the comparison for both catalogues.

Concerning proper motions, a comparison with literature values for clusters is presented in Fig.~\ref{fig:propermotion_literature}. DAML and MWSC proper motion catalogues present significant differences for the same clusters. While average zero point differences are less than 1 {\masyr}, the dispersion around this value can be of the order of 3-4 {\masyr} \citep[see][for a discussion]{2014A&A...564A..79D}. This is reflected in the comparison with {\gdrtwo} proper motions.  We find that the residual zero point is $\mu_{\rm Gaia}-\mu_{\rm reference} = 0.51 \pm0.17$ \masyr, and 0.25$\pm 0.12$ \masyr\ for $\mu_{\alpha} \cos\delta$ and $\mu_{\delta}$ respectively  for the MWSC, while the analogous quantities for DAML are  $0.0 \pm 0.19$ \masyr, and $0.41 \pm 0.18$ \masyr. These values are consistent with the differences between the two catalogues. On the basis of this comparison we have no evidence for the presence of a significant residual proper motion zero point in the {\gdrtwo}.

\subsection{Managing systematics}\label{sect:systSystematics}
For samples on a small spatial scale, one first concern is how to evaluate the presence of the systematics. Figure~\ref{fig:correlations}a compared to \figref{fig:astro-syst-bulge}a and \figref{fig:correlations}b compared to \figref{fig:pseudo-colour-LMC}b show that, locally, some hint of astrometric non-uniformity may perhaps be indicated by local variations of the correlations. 

\begin{figure}\begin{center}
\includegraphics[width=0.49\columnwidth, height=0.45\columnwidth]{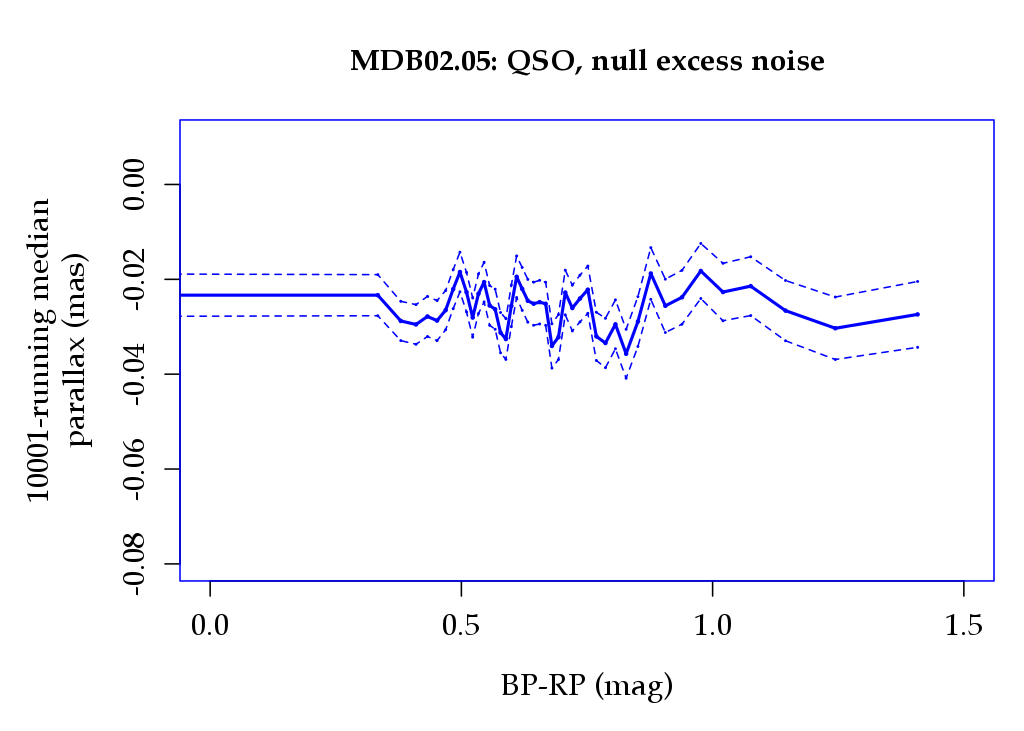}
\includegraphics[width=0.49\columnwidth, height=0.45\columnwidth]{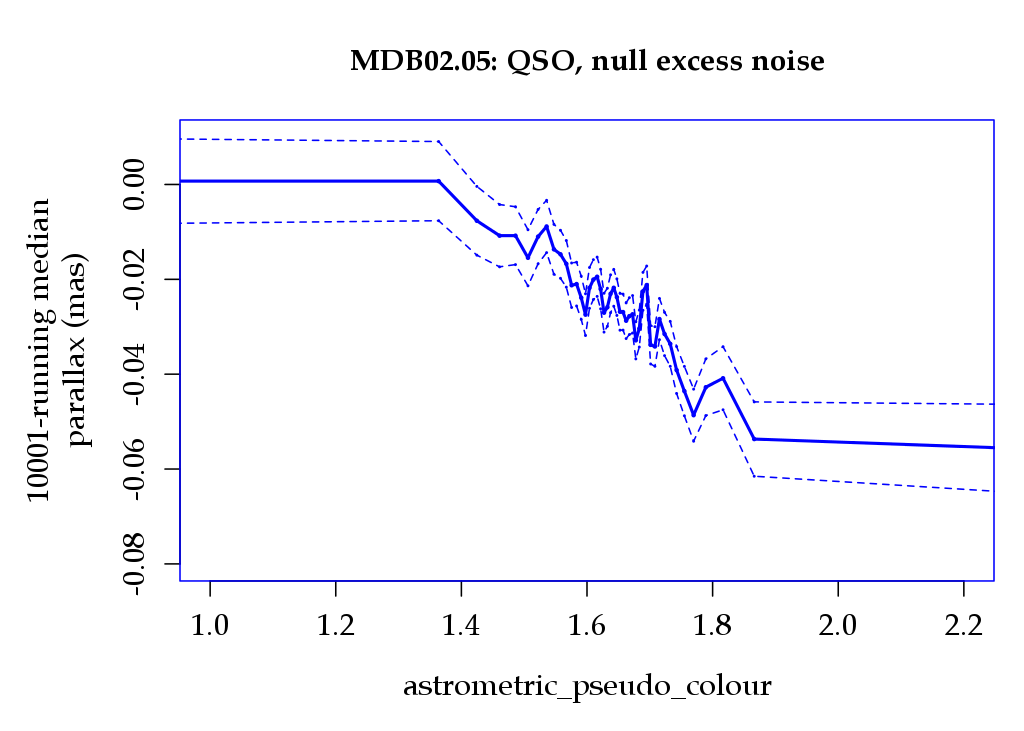}
\includegraphics[width=0.49\columnwidth, height=0.45\columnwidth]{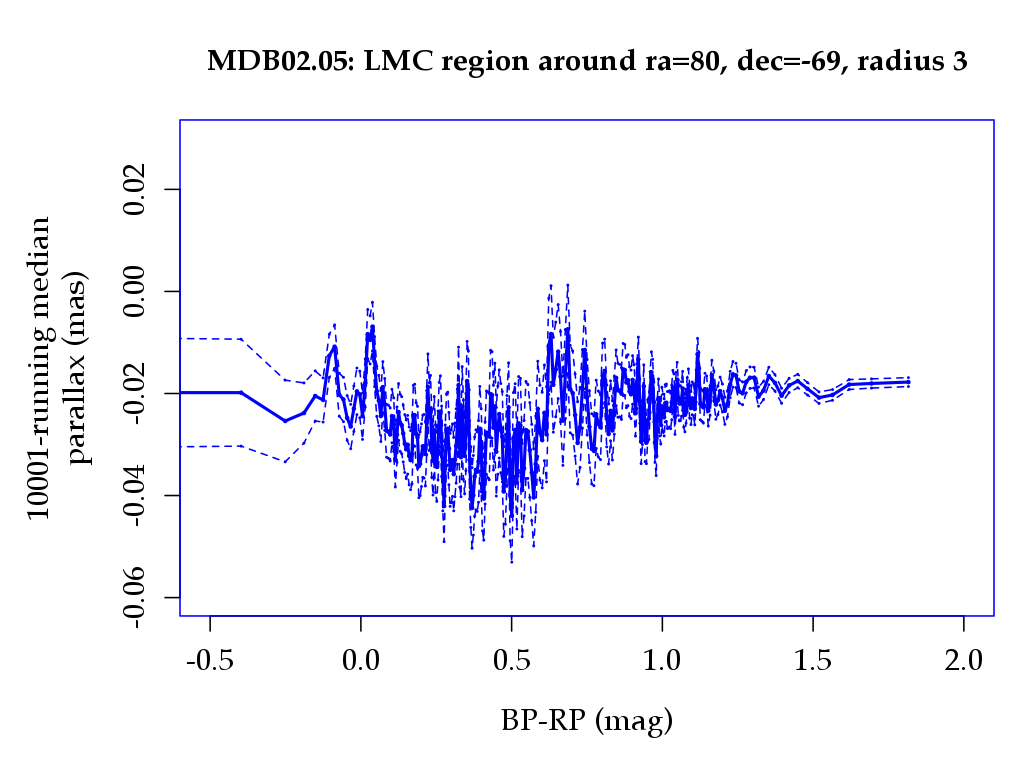}
\includegraphics[width=0.49\columnwidth, height=0.45\columnwidth]{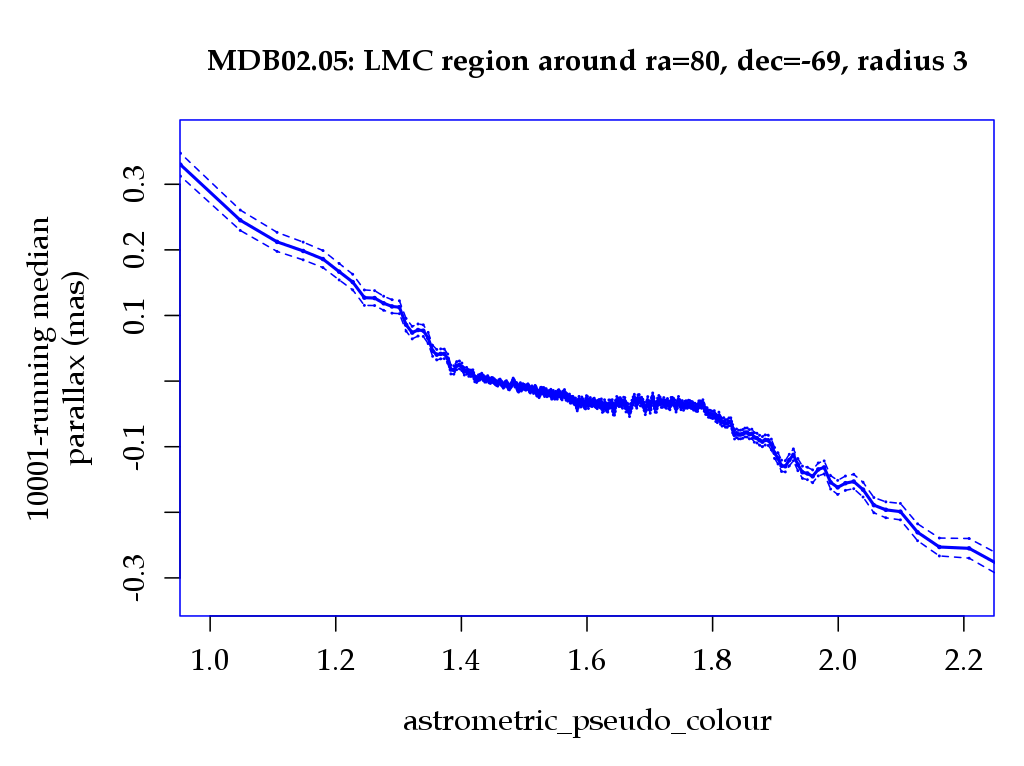}
\caption{Running median of parallaxes (mas) on 10001 points with $\pm 1\sigma$ standard error. There are no large parallax systematics linked to the $(\bpminrp)$ colour neither using QSO (top left), nor for a sample of stars in the LMC direction (bottom left). However, a significant correlation of the parallaxes with the pseudo colour is present for QSOs (top right), and even much larger in the LMC direction (bottom right).
}\label{fig:astrometry_vs_pseudo_colour}\end{center}\end{figure}

Although significant variations of the parallax zero-point with magnitude and colours is probably present, e.g. for the QSO parallaxes versus {\bpminrp} colour in Fig.~\ref{fig:astrometry_vs_pseudo_colour}a, the trend is nowhere as obvious as with the astrometric pseudo colour, Fig.~\ref{fig:astrometry_vs_pseudo_colour}b, about 0.05 mas peak-to-peak for QSO. 
The amplitude is even much larger on a subset of sources in the LMC direction\footnote{A sample of 1.56 million sources in a 3{\degr} radius around
$(\alpha,\delta)=(80\degr, -69\degr)$, keeping only those with null astrometric excess noise and $(\varpi,\mu_\alpha,\mu_\delta)$ within $4\sigma$ of (0.02-0.03, 1.8, 0.2), which accounts both from the average LMC values and the average DR2 parallax zero-point.}, Fig.~\ref{fig:astrometry_vs_pseudo_colour}d. This cannot be due to contamination by foreground sources, as the parallax peak-to-peak variation with \bpminrp\ (0.05 mas, Fig.~\ref{fig:astrometry_vs_pseudo_colour}c) is less than one order of magnitude smaller than with the pseudo colour (0.6 mas, Fig.~\ref{fig:astrometry_vs_pseudo_colour}d).
The pseudo colour has absorbed a fraction of the astrometric systematics, as can be seen from their spatial variations on the LMC, Fig.~\ref{fig:pseudo-colour-LMC}a. Pseudo colour variations, when they are not representative of the colour variations themselves, may then help to detect astrometric systematics.

It could also be tempting to ``correct'' the parallaxes from the global zero-point. Individually, this would be inadvisable, first because its contribution level is generally below the random error; second, as seen above, the local (\secref{ssec:astroacc_small}), regional (\secref{ssec:astroacc_large}) variations, or colour, magnitude effects may actually be larger than the global zero-point. However, for a sample well distributed over the whole sky which is being used for e.g. a luminosity calibration, then the zero-point may be corrected, or, better, solved for, as mentioned by \citet{DR2-DPACP-51}.

\subsection{Uncertainties of the astrometric random errors}\label{ssec:astrometry_precision}

\subsubsection{Distribution of the astrometric errors}\label{ssec:astroacc_6}
\begin{figure}\begin{center}
\includegraphics[width=0.8\columnwidth]{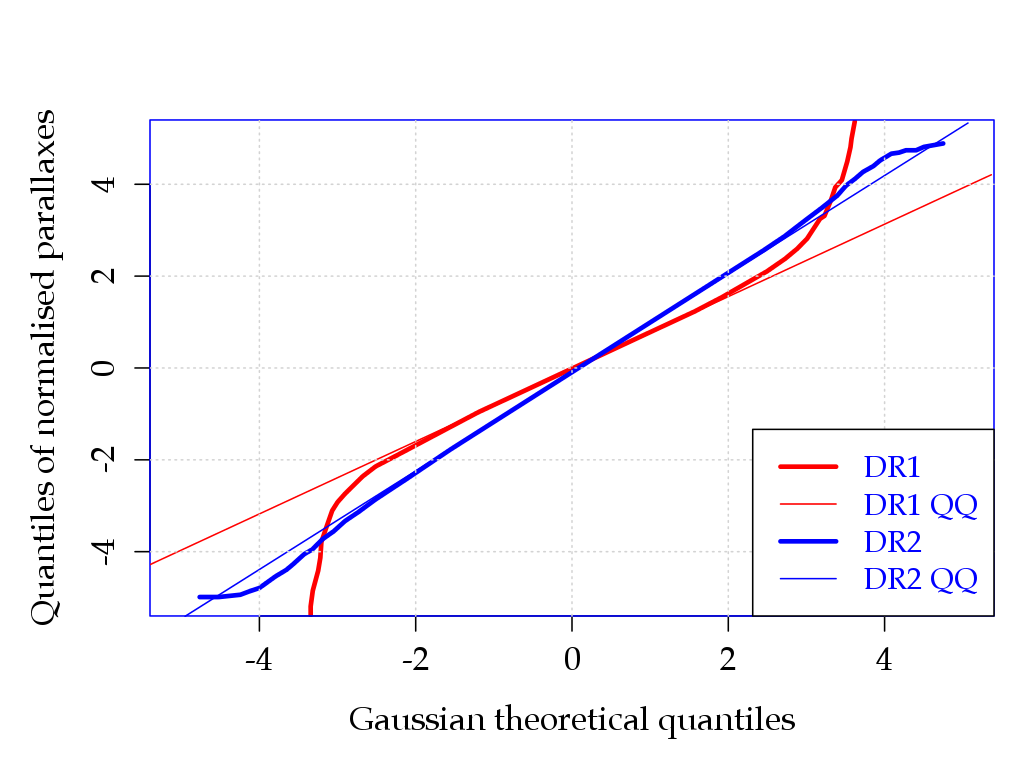}
\caption{Normalised QSO parallaxes truncated to $\pm 5$. As seen using a Gaussian Q-Q plot, the parallax errors of {\gdrtwo} QSOs (blue) are now much closer to the Gaussian(0,1) diagonal than the DR1 ones (red). The thin lines represents the lines passing through the first and third quartiles of the corresponding samples. }\label{fig:MDB02.05-norm-plx-qqplot}
\end{center}\end{figure}

The astrometric error distribution, at least for faint sources, can be studied using the QSO. We used the sources
with \dt{frame\_rotator\_object\_type} equal to 2 or 3, keeping only those with a parallax over error smaller than 5 in absolute value and keeping even those with non-zero excess noise (488\,805 sources).
The statistical distribution of the errors (parallax over uncertainty) can then be directly seen, and the deviation from normality beyond $2\sigma$ which was present in {\gdrone} has now disappeared in {\gdrtwo} (\figref{fig:MDB02.05-norm-plx-qqplot}), the errors
being now much more Gaussian. This legitimates the use of the normal distribution in likelihood functions where the astrometric errors appear.

\subsubsection{Internal comparisons}\label{ssec:astrocal_1}

\begin{figure}\begin{center}
\includegraphics[width=0.8\columnwidth]{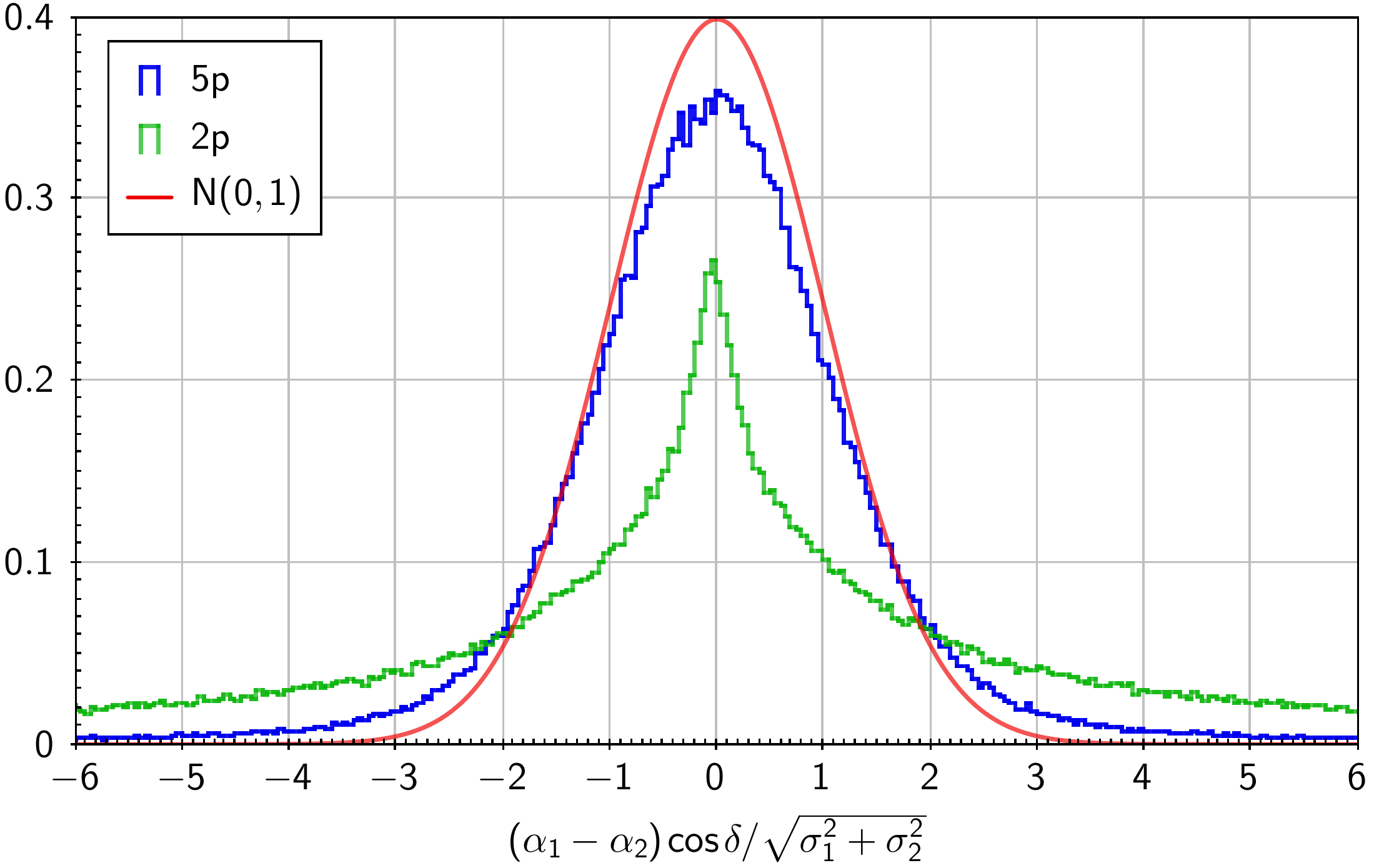}
\includegraphics[width=0.8\columnwidth]{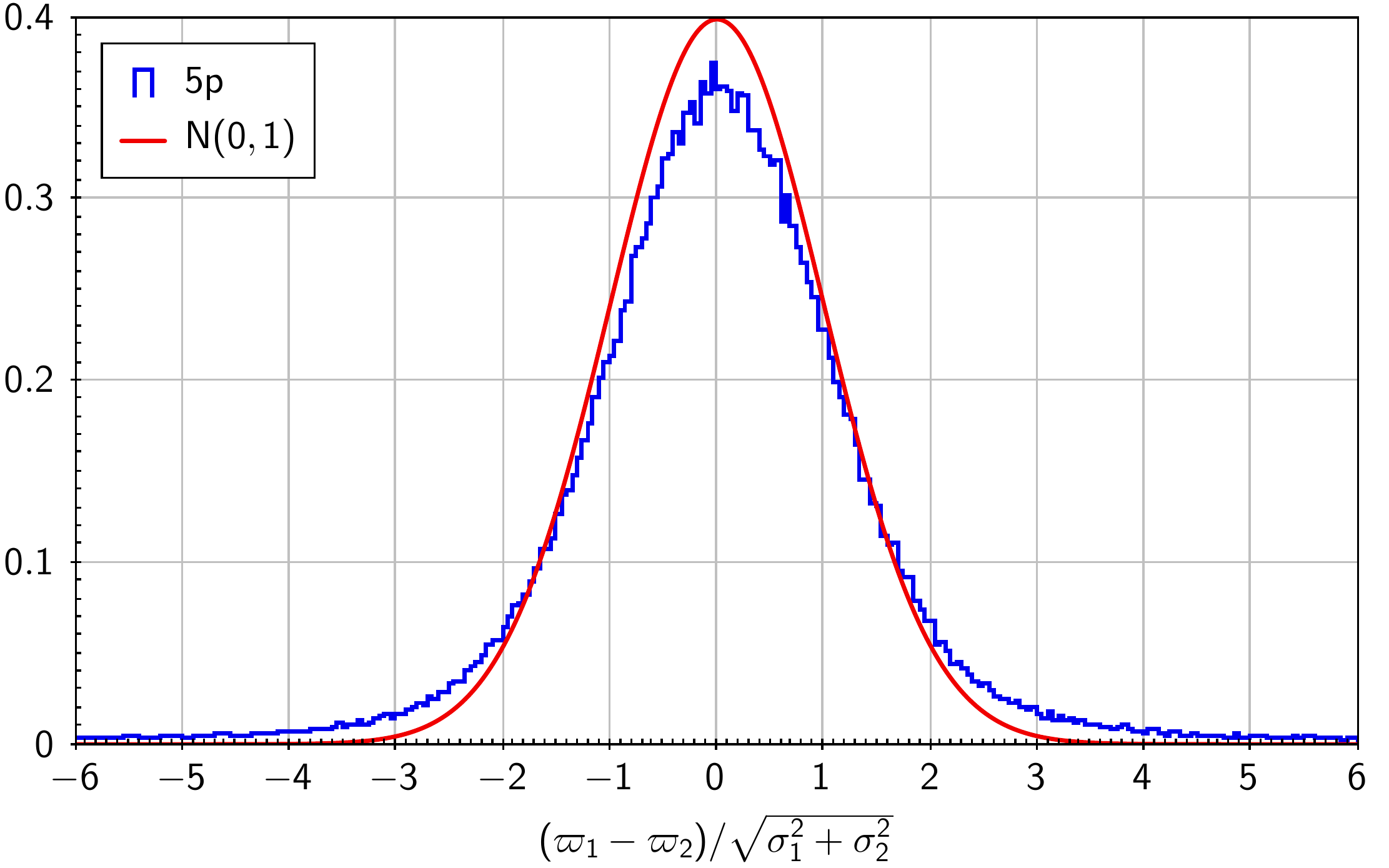}
\caption{\small
Histograms for
the normalised differences for right ascensions (top) and parallaxes (bottom) for duplicate source 
pairs brighter than $G=17$~mag. A normal distribution is overplotted.
}\label{fig:dup_dif}
\end{center}\end{figure}

A simple test on astrometric precision is a comparison of parameters for the
duplicated source pairs mentioned in \secref{sec:dup}. Figure~\ref{fig:dup_dif}
shows histograms for the normalised differences of right ascensions
and parallaxes for sources brighter than 17~mag. The properties for
declinations are similar to the ones for right ascensions and the proper motion
components show features similar to the parallaxes. For sources with the full
five-parameter, astrometric solution for both solutions the comparison in
\figref{fig:dup_dif} only suggests that formal uncertainties are slightly
underestimated, perhaps 10\%.  However, for sources where the full solution for some reason
failed for one or both solutions, the differences are non-Gaussian and show very large wings. 
As discussed above, \secref{astroqual}, these
sources may be binaries or show structure or have only few observations. As
mentioned in \secref{sec:dup}, the duplicated sources are more affected than
the average sources with only two astrometric parameters. This may explain why
the uncertainties are so strongly  underestimated for this specific subset.

\begin{figure}[tbhp]\begin{center}
\includegraphics[width=0.95\columnwidth, height=0.65\columnwidth]{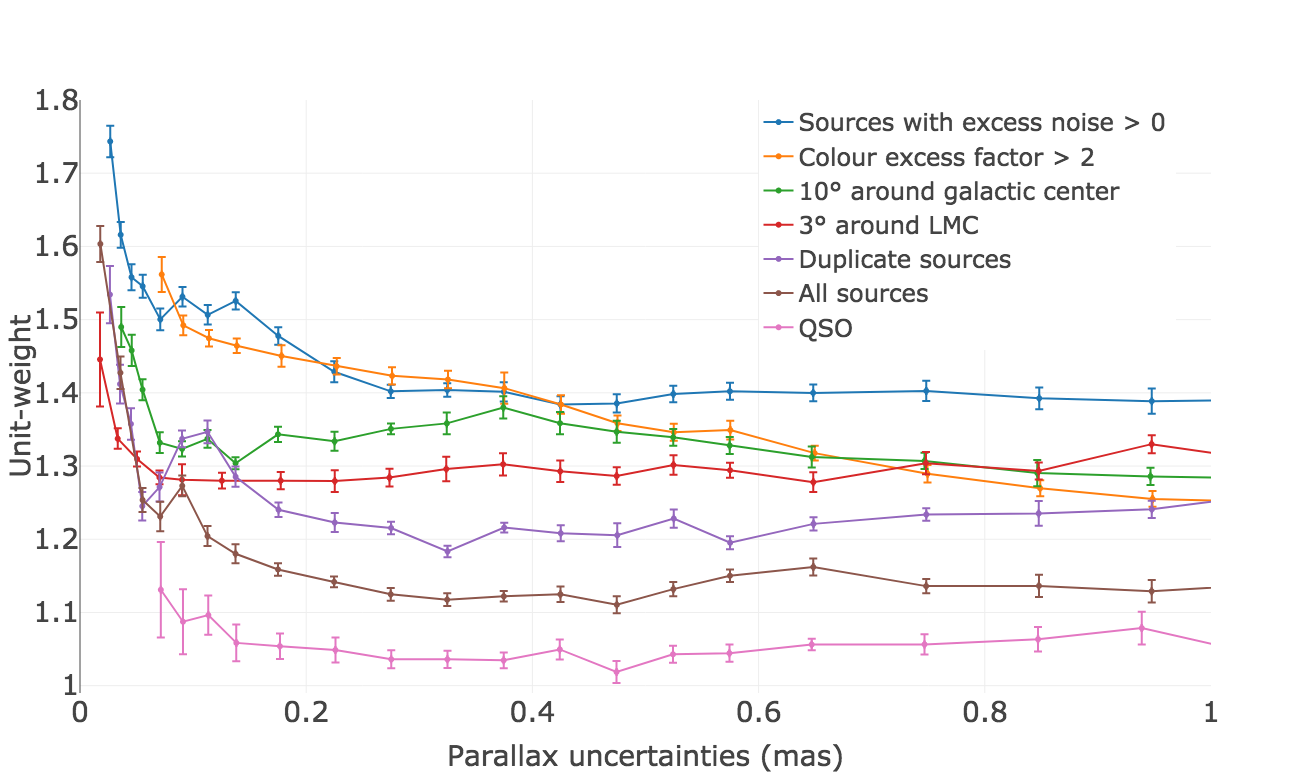}
\caption{Unit-weight uncertainties of parallaxes estimated by deconvolution, versus these uncertainties. From top to bottom, sources with non-zero excess noise (blue), \dt{phot\_bp\_rp\_excess\_factor} larger than 2 (orange), within 10{\degr} of the Galactic center (green), within 3{\degr} towards LMC (red), duplicated sources (violet), all Catalogue sources (brown), QSO (pink). Only sources with more than 8 visibility periods and GoF $<5$ have been kept in all subsets, except for the subset with non-zero astrometric excess noise where no GoF upper limit was applied.}\label{fig:wp942:allparallaxdeconv}
\end{center}\end{figure}

The uncertainties on the parallaxes have also been studied using the dispersion of the negative parallax tail estimated by deconvolution  \citep[see ][Sect.~6.2.1 for details]{2017A&A...599A..50A}. The unit-weight uncertainties\footnote{We call unit-weight uncertainty (uwu) the factor (ideally one) that needs to be applied to the published uncertainties to be equal to the estimated dispersion of the astrometric parameters.} are shown \figref{fig:wp942:allparallaxdeconv} for several illustrative subsets of the {\gdrtwo} Catalogue, as a function of these uncertainties.
Although the formal uncertainties are primarily increasing with magnitude \citep[see e.g.][Fig. 9]{DR2-DPACP-51}, they also increase with the astrometric excess noise. Non-zero excess noise can be either due to the non-single character of the source, or to imperfect calibrations \citep{2012A&A...538A..78L}. Beside, \dt{phot\_bp\_rp\_excess\_factor} is an indication of binarity or duplicity (contamination) in dense fields, or of extended objects \citep{DR2-DPACP-40}. From top to bottom (largest unit-weights to smallest ones), the subsets with non zero excess noise and large colour excess factor most probably represent respectively the contribution from non-single stars and extended objects, showing that the added excess noise was actually not enough to cope with the actual dispersion.
Below, the subsets in the direction of the galactic center and LMC probably show the effect of contaminated sources in dense fields. Then, the duplicated stars, mostly made of single stars for small uncertainties, with a possible contribution of binaries for larger uncertainties. For faint, average stars, the unit-weight is only about 15\% too large. Then the QSO uncertainties look the most realistic, as can also be seen Fig.~\ref{fig:wp944:varfac}. 

For all subsets, the unit-weight is increasing towards small uncertainties (i.e. concerning stars with magnitude between 13 and 15), which could be underestimated by about 40\%. This probably originates from the reweighting which has been applied to the uncertainties \citep[][Appendix A]{DR2-DPACP-51} to correct a bug found lately in the data processing cycle. It was found that this reweighting correctly improved the uncertainty estimates of the stars brighter than $G=13$, but had an adverse effect for stars with \dt{astrometric\_n\_obs\_ac}$=0$, i.e. fainter than 13, and stars $13\lesssim G\lesssim 15$ are those with the smallest uncertainties.

\subsubsection{Comparison to distant external  data}\label{ssec:astrocal_2}
\begin{figure}
\centering
\includegraphics[width=0.8\columnwidth]{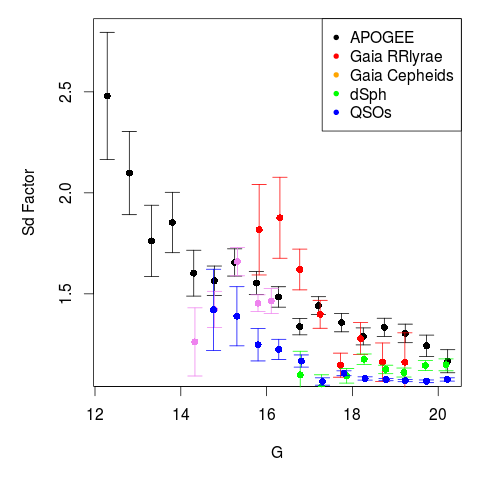}
\caption{Unit-weight uncertainty that would need to be applied to the {\gaia} parallax uncertainties to be consistent with the residual distribution versus APOGEE, {\gaia} RRlyrae and Cepheid distance moduli as well as dSphs and LQRF QSOs.}\label{fig:wp944:varfac}
\end{figure}

The uncertainties have also been tested via the comparison to distant stars or QSOs (Table~\ref{tab:cu9val_wp944_summaryplx}). Those comparisons are complicated by the fact that the uncertainties of the external catalogues may not be accurately determined and by the pollution from wrong identifications for QSOs. Still, the under-estimation of the parallax errors is seen to increase with magnitude in all the tests conducted, as illustrated in Fig.~\ref{fig:wp944:varfac}. This is the same trend as shown in \figref{fig:wp942:allparallaxdeconv}, though of a larger amplitude for the reasons just explained. For bright stars, however, the comparison with {\hip}, Table~\ref{tab:cu9val_wp944_summaryplx}, shows that the parallax uncertainties are unlikely to be much underestimated, as {\hip} parallax uncertainties may well have themselves been slightly underestimated.

\begin{figure}\begin{center}
\includegraphics[width=0.7\columnwidth]{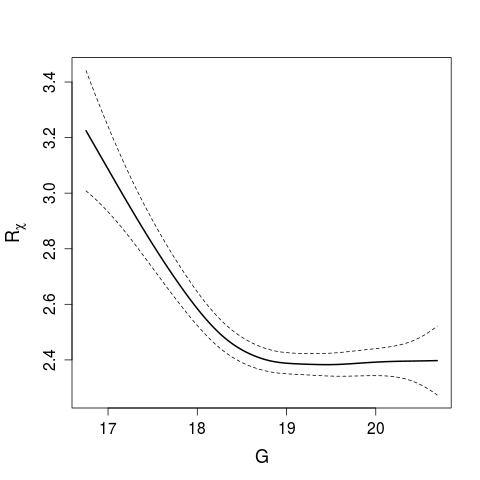}
\caption{$\chi^2$ test of the LQRF QSOs proper motions as a function of $G$ magnitude. The residual $R_\chi$ should follow a $\chi^2$ with 2 degrees of freedom. The dotted lines corresponds to the 1 $\sigma$ confidence interval. The correlation observed here is most likely due to the under-estimation of the uncertainties as a function of magnitude.}\label{fig:wp944:pmchi2}
\end{center}\end{figure}

The variation of the uncertainties with magnitude explains why, depending on how the uncertainties and handled in computing the differences with external data, a correlation between the difference and the magnitude is seen or not. The most striking example is when using a $\chi^2$ test: while no significant correlation of the individual proper motions of QSOs is seen with magnitude, the correlation is significant when combining them through their covariance matrix (Fig.~\ref{fig:wp944:pmchi2}). This is also seen in the comparison with the {\hip} proper motions, similarly to what was found in {\gdrone} \citep{2017A&A...599A..50A}. 

\subsubsection{Parallax and proper motion precision tested using distant clusters}\label{ssec:astroacc_3bis}
\begin{figure}\begin{center}
\includegraphics[width=0.9\columnwidth]{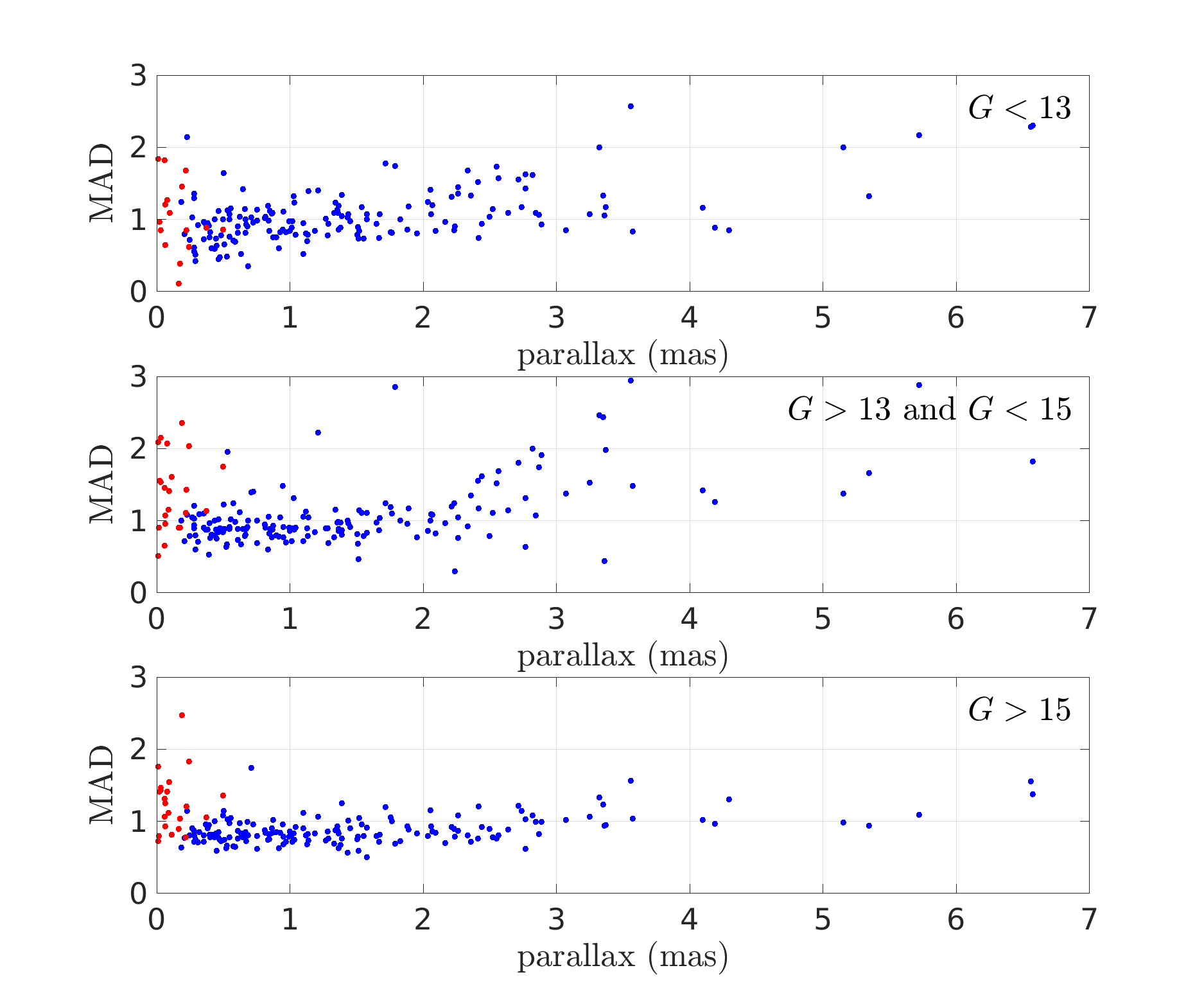}
\caption{MAD of the distribution of the dispersion on the parallaxes normalised by the nominal uncertainties, as a function of the parallaxes (in mas) for open clusters (black dots) and globulars (red crosses). Top to bottom: $G<13$, $13<G<15$, $G>15$.} \label{fig:allmembers_parzMAD}
\end{center}\end{figure}

The astrometric precision has also been estimated using a sample of about 200 OCs and  about 20 GCs. We used as reference values the DAML and MWSC catalogues. 
We calculated for each cluster the dispersion of parallaxes from the median value, after normalising the offsets by the nominal uncertainties, selecting only stars with errors on parallax smaller than 2 mas.
Figure~\ref{fig:allmembers_parzMAD} shows the Median Absolute Deviation (MAD) of the above distribution as a function of the parallaxes for open and globular clusters. For nearby clusters there is a clear internal parallax dispersion. However also for distant clusters the MAD does not really converge to one, as it would be expected if the uncertainties on the parallax are correctly estimated. The results suggest that the uncertainties are underestimated.  This  holds in particularly at the bright end of the star distribution (for $G< 15$). The uncertainties are definitively underestimated for GCs (see Fig.~\ref{fig:allmembers_parzMAD}): clearly the high crowding is responsable for the lower number of observations per star and the degraded astrometric precision.

\begin{figure}\begin{center}
\includegraphics[width=0.9\columnwidth]{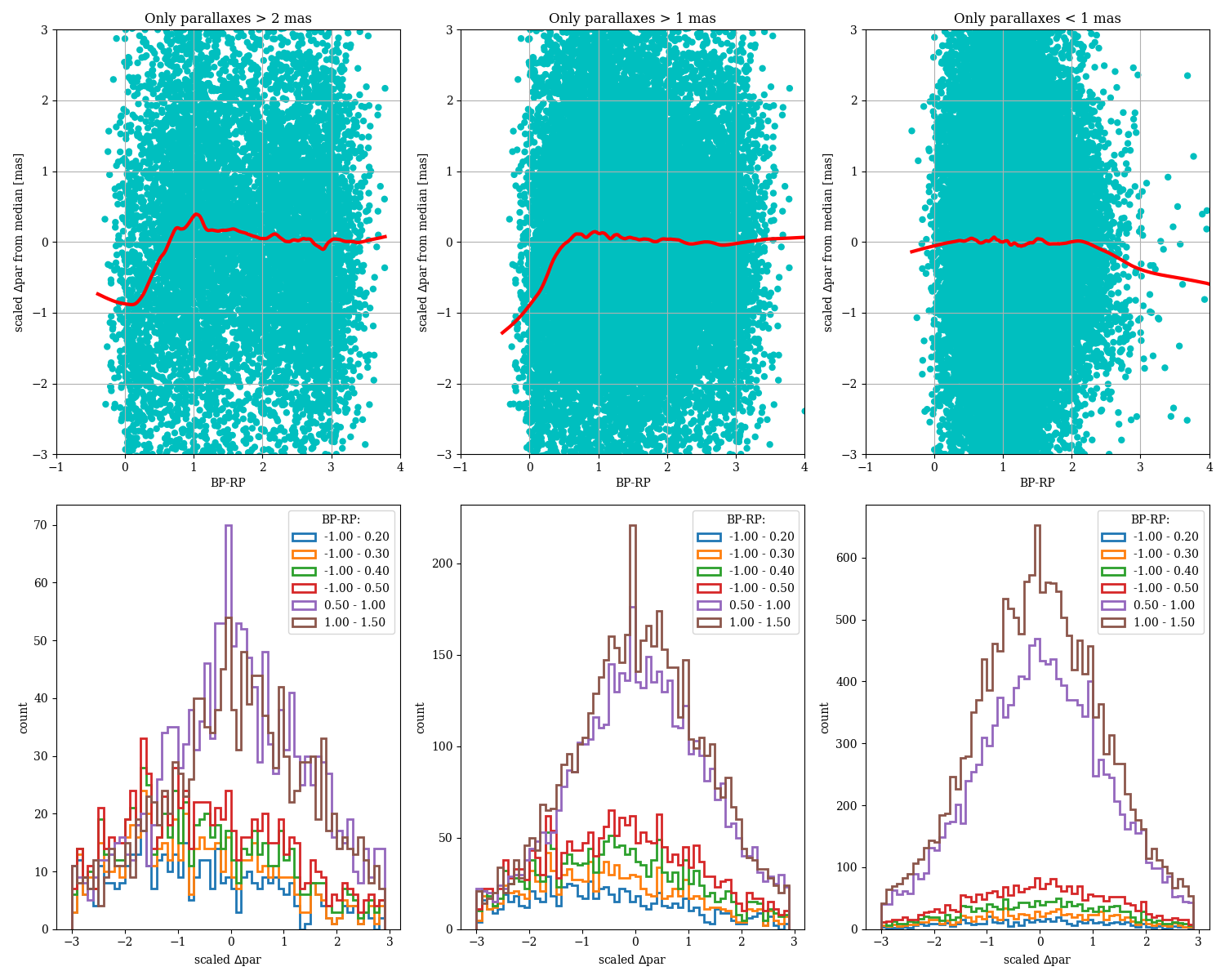}
\caption{Distribution of the differences between the median cluster parallaxes and the single star parallaxes normalised by the nominal uncertainties for stars with $\varpi > 2$mas (upper left panel); $1<\varpi < 2$mas (upper central panel);  $\varpi < 1$mas (upper right panel). The red lines are the smoothed distributions. The lower panels present the histograms of the distributions for different colour ranges.
}\label{fig:allmembers_scaleddeltapar}
\end{center}\end{figure}

A residual parallax trend with colours is shown in Fig.~\ref{fig:allmembers_scaleddeltapar} for all the stars in the cluster sample in the blue edge and possibly in the red edge of the colour domain although in this case with a poor statistics. This could be a consequence of an imperfect chromaticity  correction. However since in our sample the majority of stars are on the main sequence, there is a strong correlation between magnitude and colour, and it is hard to distinguish both effects.

\begin{figure}\begin{center}
\includegraphics[width=0.8\columnwidth]{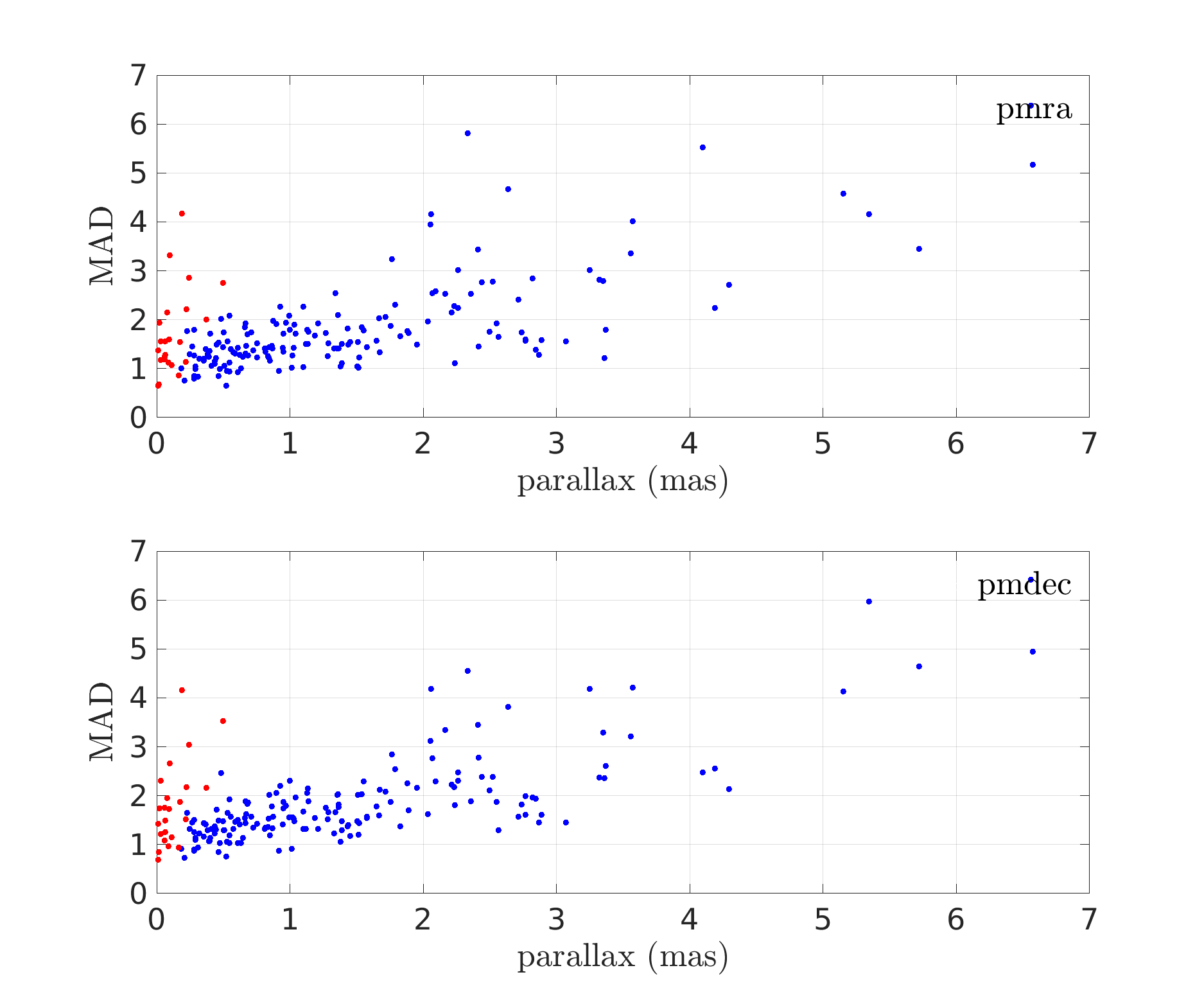}
\caption{MAD of the distribution of the dispersions of normalised differences on the proper motions (\masyr) in right ascension (top panel) and in declination (bottom panel)  as a function of the parallax (in mas) for open clusters (blue dots)  and globulars (red dots). }\label{fig:allmembers_PM_MAD}
\end{center}\end{figure}

Figure~\ref{fig:allmembers_PM_MAD} shows the MAD of the distribution of the dispersion on the proper motions in right ascension  and in declination  normalised by the nominal uncertainties as a function of the parallax. Nearby clusters are affected by intrinsic proper motion dispersion, while distant clusters tend to MAD=1, albeit with a large dispersion. At small parallax, all the objects belonging to the tail having MAD$>1.5$ are globulars, implying that  proper motion uncertainties are also underestimated in the central regions of this type of cluster.

\begin{figure}\begin{center}
\includegraphics[width=0.7\columnwidth]{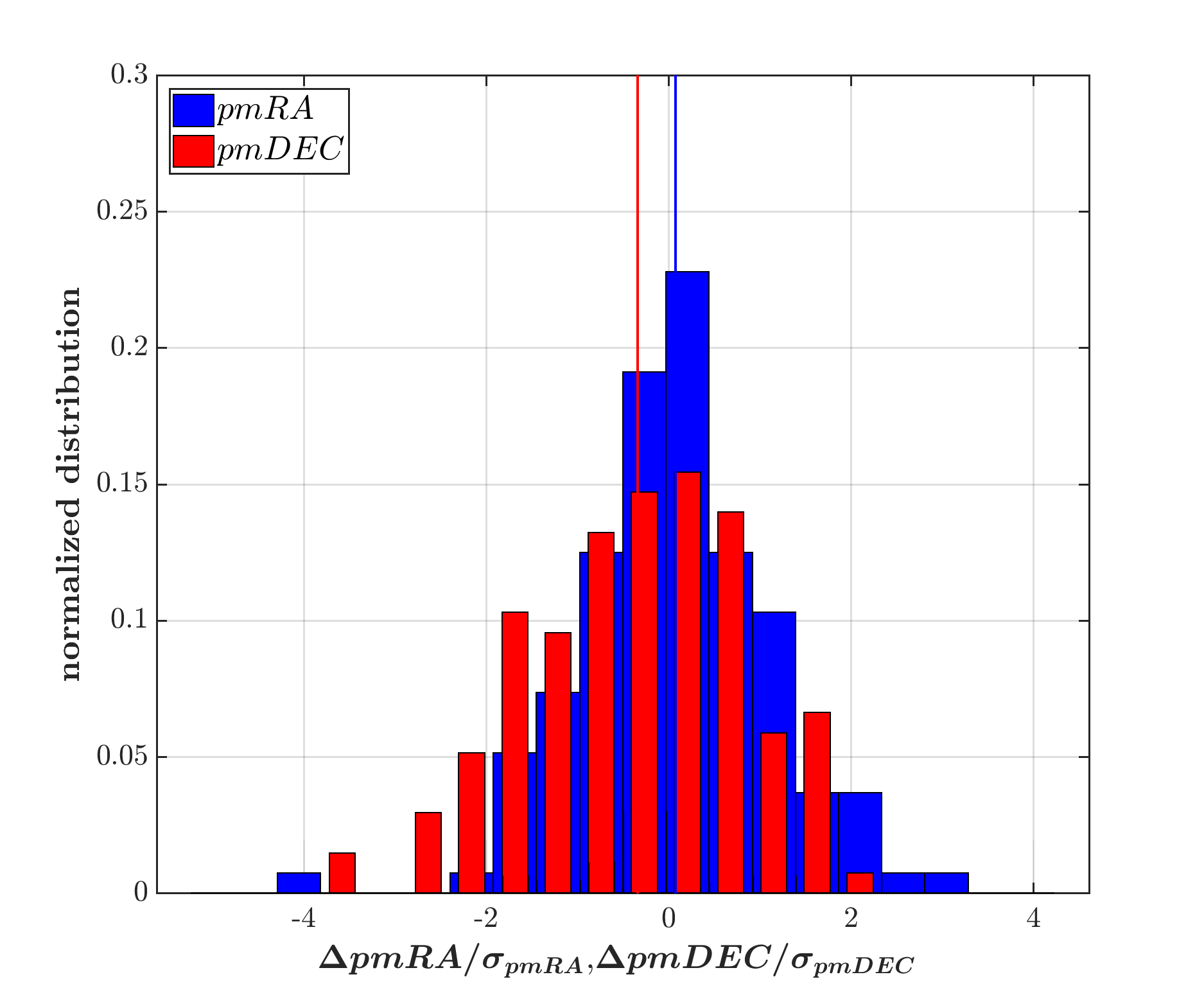}
\caption{Normalised differences in Omega Cen cluster proper motions (ra and dec) between Gaia DR2 data and HST data. The lines represent the mean values of the distributions}\label{fig:omegacenpm}
\end{center}\end{figure}

We compared the proper motions in Omega Cen with external HST data  by \cite{2018ApJ...854...45L}, where relative proper motions are available down to very faint magnitudes, and a proper motion zero point is provided. About 140 stars were found in common.
The studied  field is located at the outskirts of the cluster and it is not very crowded. We compared the normalised dispersion of the differences in proper motions for the stars in both samples.  The normalised dispersion is very close to one both for $\mu_{\alpha  *}$ and $\mu_{\delta}$, implying that the proper motions uncertainties are correctly estimated (see Fig.~\ref{fig:omegacenpm}). 

\begin{figure}\begin{center}
 \includegraphics[width=0.8\columnwidth]{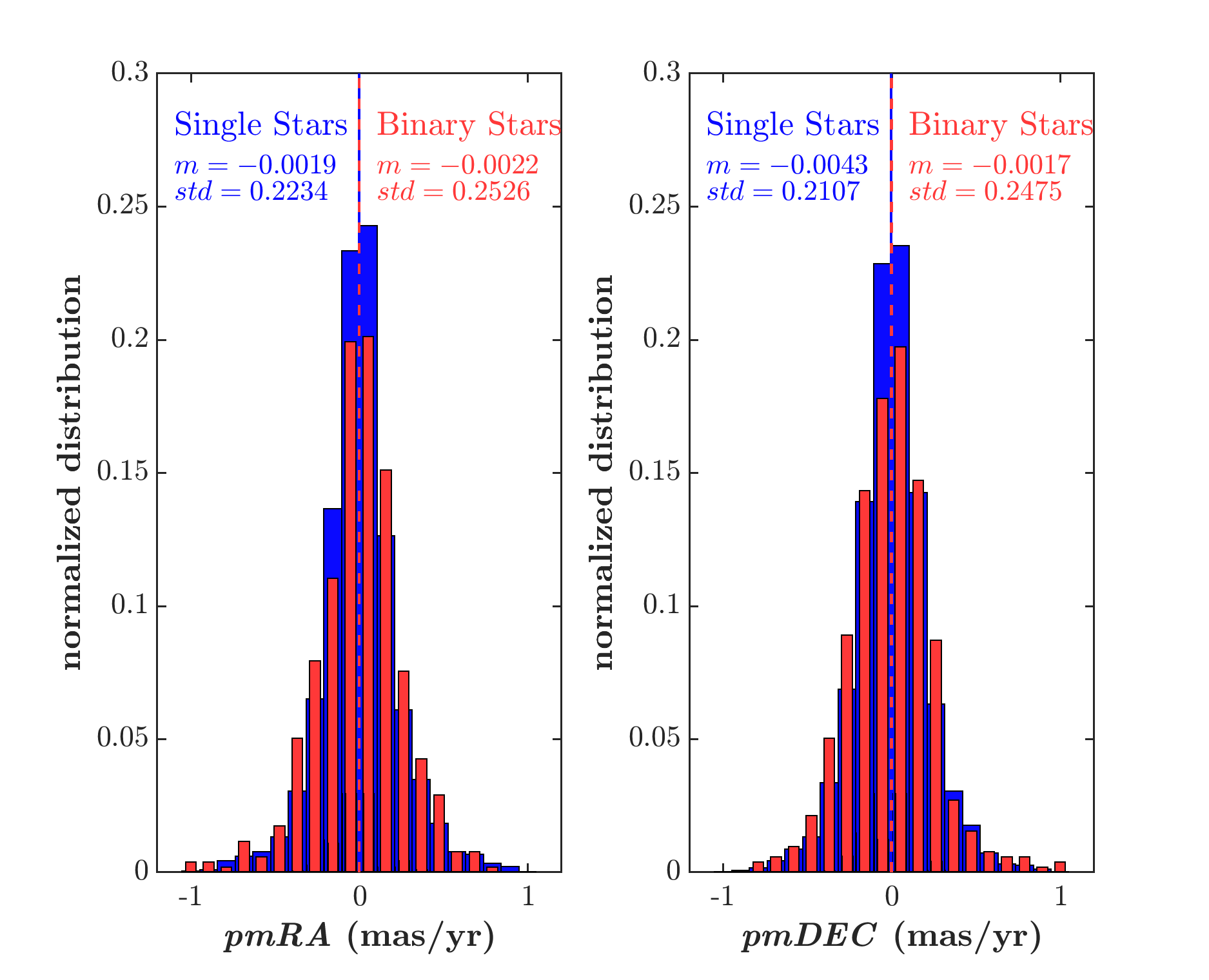}
 \caption{Distribution of the deviation from the cluster median of the proper motions in the main sequence and photometric binaries for a sample of 12 OCs.}\label{fig:binarypm}
 \end{center}\end{figure}

Finally, we check the quality of the astrometry for the unresolved photometric  binary sequence that is clearly visible in the CMDs (see for instance Figure~\ref{fig:NGC2630_teff}) for about 12 OCs selected after visual inspection and located farther than about 400 pc. This would minimize the effect of the internal velocity dispersion and of the mass segregation.  The procedure and a few examples are discussed in Section \ref{photo-prec_ext}. We derived the deviation from the cluster median for every star in the main sequence and in the binary star sequence in the proper motions space. The global distributions are shown in Fig.~\ref{fig:binarypm}.   For all the OCs, the  Kolmogorov-Smirnov test does not reject the null hypothesis that the two samples are drawn from the same distribution at the 5\% significance level for $\mu_{\alpha*}$, and for the parallaxes, while for $\mu_{\delta}$ the null hypothesis is rejected only for two objects with a marginally inconsistent p-value $=4.8$\%, $3.8$\%. This shows that the unresolved binaries with a small magnitude difference have an astrometric quality not significantly different from that of the single stars. 

\subsection{Quality indications of the astrometric solution, outliers}\label{ssec:astrocal_5}
\begin{figure}\begin{center}
\includegraphics[width=0.9\columnwidth]{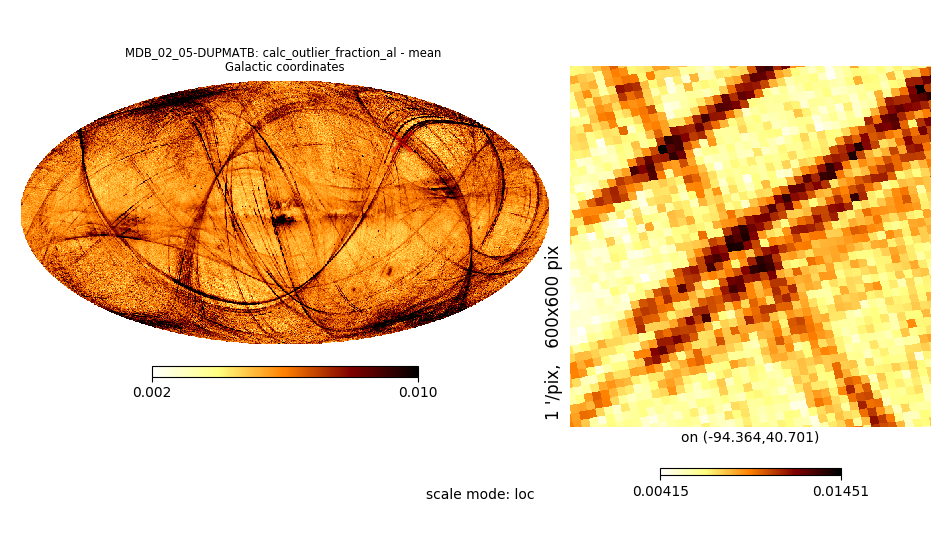}
\caption{Mean fraction of outlying CCD measurements over the whole sky (left) and 10{\degr} size detail near $(l,b)=(-94\degr,41\degr)$. Several bad scans or larger regions have a larger fraction of outliers.}\label{fig:calc_outlier_fraction_al-mean}
\end{center}\end{figure}

\gdrtwo\ includes many quality indicators for the astrometric solution as described in detail in \citet{DR2-DPACP-51}, and we will here just mention a few. An obvious indicator is the number of rejected observations (\dt{astrometric\_n\_bad\_obs\_al}) as compared to the retained ones (\dt{astrometric\_n\_good\_obs\_al}).
Figure~\ref{fig:calc_outlier_fraction_al-mean} shows the fraction of outliers over the whole sky and in a small area. The sky map shows whole great circles with more than average rejections. This points to specific time intervals of maybe half a day with a specific problem for the astrometric calibration. This is also clear in the zoom, where we see examples of two or three consecutive scans, each 0.7{\degr} wide, with some issue.

\begin{figure}\begin{center}
\includegraphics[width=0.9\columnwidth]{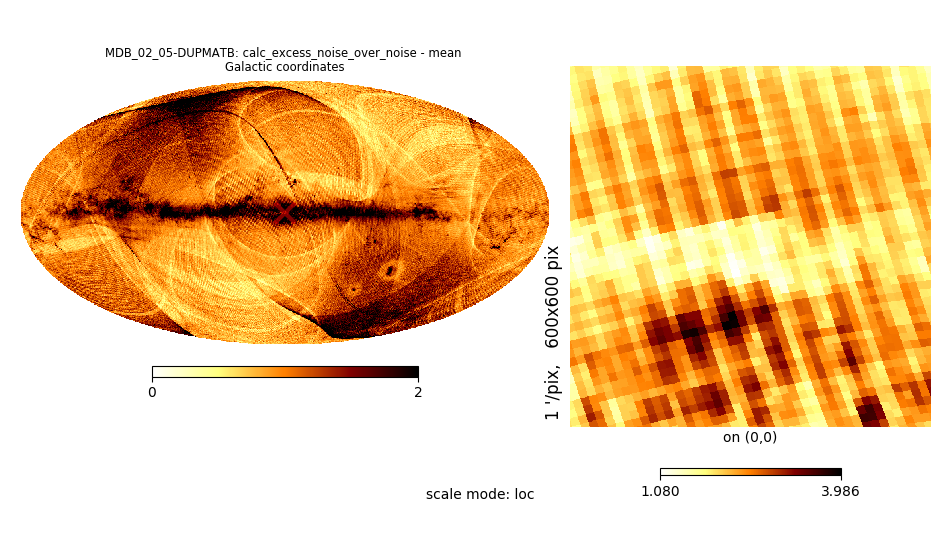}
\caption{Mean value of astrometric excess noise divided by parallax uncertainty. {\em Left:} for the whole sky; {\em Right:} in a 10{\degr} size detail near the Galactic centre. Imprints of the scanning law are present over the whole sky. }\label{fig:calc_excess_noise_over_noise-mean}
\end{center}\end{figure}

Another useful quality parameter is the excess noise, \dt{astrometric\_excess\_noise}, expressing in angular measure the insufficiency of the source model to match the observations. This is illustrated in Fig.~\ref{fig:calc_excess_noise_over_noise-mean} showing the excess noise normalised by the parallax uncertainty. It demonstrates that specific zones have small parallax errors as compared to how well the astrometric solution has behaved. It again points to the scanning pattern, but not to specific time intervals because we do not see problematic great circles.

The presence of scanning patterns in quality maps is not in itself a concern, given that the astrometric solution depends on well-distributed scans. However, note that similar patterns are visible in e.g.\ the parallax itself as shown in \figsref{fig:astro-syst-bulge} and \ref{fig:pseudo-colour-LMC}b.

\section{Photometric quality of {\gdrtwo}}\label{photoqual}

The photometry in {\gdrtwo} consists of three broad bands: a \gmag\ magnitude for all sources and a \gbp\ and \grp\ magnitude for the large majority. The photometry and its main validation is described in \citet{DR2-DPACP-40} and we will here merely present some additional tests. 
As for astrometry, photometry has had very large improvements since {\gdrone} thanks to better calibrations, better image parameter determination and the availability of colours. 

The photometric quality of {\gdrtwo}, accuracy and precision, has been tested using both internal methods 
(using \gaia\ photometry only) and by comparisons to external catalogues.

\subsection{Photometric accuracy}\label{photo-acc}
\subsubsection{Internal comparisons}\label{photo-acc_int}

\begin{figure}\begin{center}
\includegraphics[width=0.7\columnwidth]{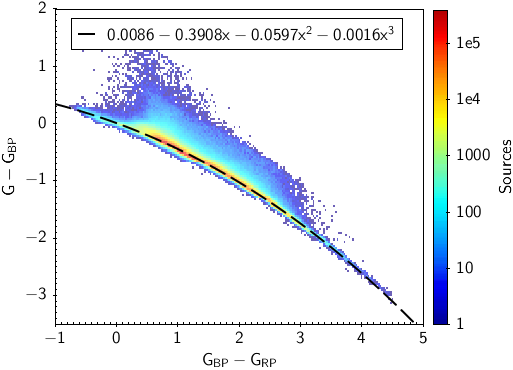}
\includegraphics[width=0.8\columnwidth]{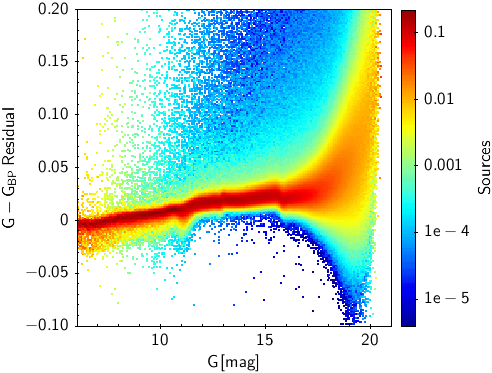}
\caption{Top: {\gminbp} relation vs {\bpminrp}. Bottom: 2D histogram of the {\gminbp} residuals after subtraction of the colour dependent relation shown on top. Only sources at high Galactic latitudes were used and the histogram was reweighted to give the same weight to each magnitude interval.} 
\label{fig:phot_G_G-BP}
\end{center}\end{figure}

Figure~\ref{fig:phot_G_G-BP}b shows a comparison between the {\gmag} magnitude and the {\gbp} magnitude at high Galactic latitudes. The differences depend on the spectral type, and as a first approximation the colour dependence (\figref{fig:phot_G_G-BP}a) was subtracted. The comparison shows a trend with magnitude of a few mmag/mag corresponding to {\gbp} getting relatively brighter for fainter sources. The trend is even stronger for the faintest sources. The small kinks at magnitudes 11, 13 and 15.5 are discussed in \citet{DR2-DPACP-40} and correspond to changes
in the on-board windowing. Although the comparison is presented for {\gmag}$-${\gbp}, we cannot distinguish if the bias comes from {\gmag}, {\gbp} or \grp. A \grp\ comparison would give the same trend and kinks because of the way the colour dependence is subtracted.

\begin{figure}\begin{center}
\includegraphics[width=0.8\columnwidth]{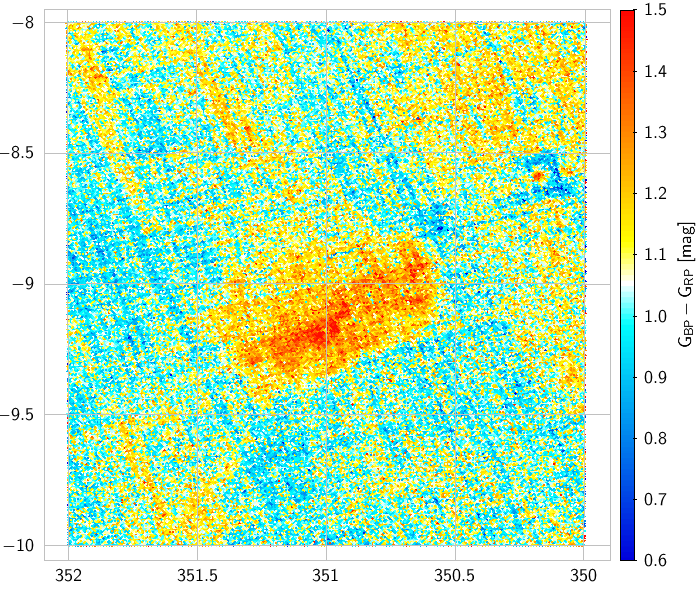}
\caption{Median colours, \bpminrp, in a dense field (Galactic coordinates) showing artefacts from the scan pattern.}\label{fig:gal_cen_col}
\end{center}\end{figure}

\begin{figure}
 \begin{center}
\includegraphics[width=8cm]{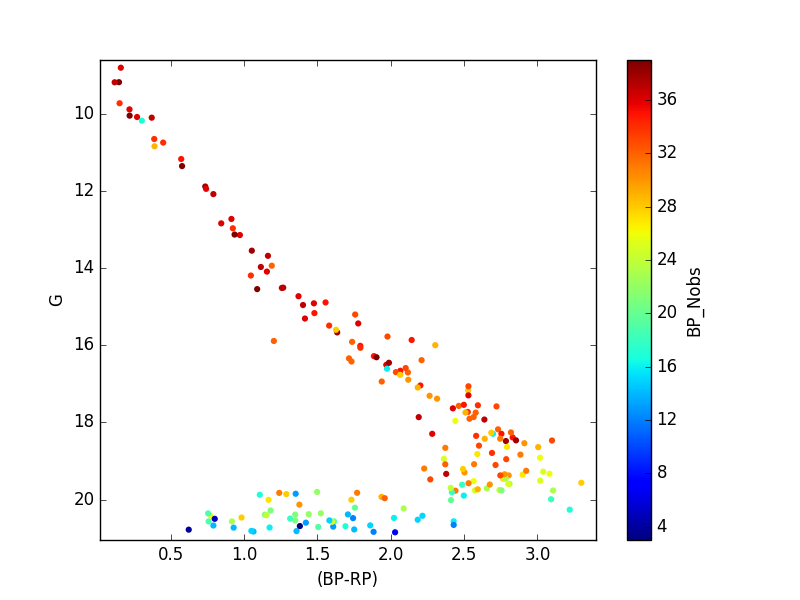}
\caption{CMD of Alessi 10 where the colour map shows the number of {\gbp} observations per CCD in used in the data analysis, $BP_{\rm Nobs}$} \label{fig:Alessi10_excess-flux}
\end{center}\end{figure}

As discussed in \citet{DR2-DPACP-40} and in Sect.~5.5.2 
of the on-line Catalogue documentation the {\gbp} and {\grp} bands may suffer from an uncorrected flux excess as indicated by \dt{phot\_bp\_rp\_excess\_factor}. This occurs especially in dense fields, for binaries, near bright stars, and for the fainter sources. 
This excess is caused partly by an underestimation of the sky background level and partly by the fact that no deblending of overlapping spectra was carried out. 
An example of distorted colours in a dense field is illustrated in \figref{fig:gal_cen_col} which shows artificial patterns originating in the individual scans
(the streaks), while the red blob is an area with few stars and therefore probably a real feature (a cloud). Fig.~\ref{fig:Alessi10_excess-flux} shows as another example the CMD of the cluster Alessi~10. At the faint end, $G> 18 - 19$, the main sequence turns out to be excessively blue; the number of {\gbp} observations for these stars is lower than the average value of the cluster. 

\subsubsection{Comparisons with external catalogues}\label{photo-acc_ext}

\begin{figure*}\begin{center}
\includegraphics[width=0.33\textwidth]{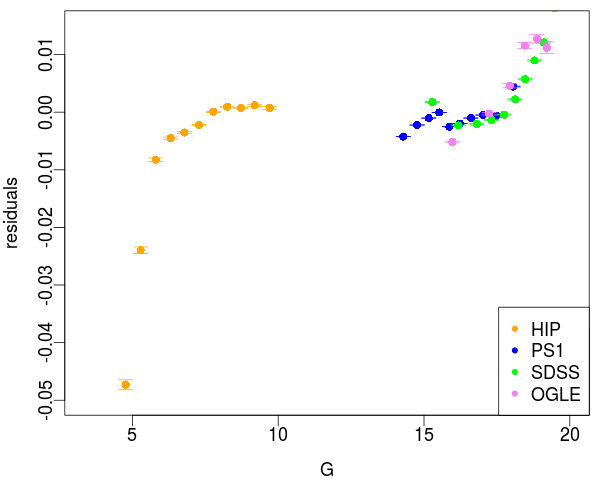}
\includegraphics[width=0.33\textwidth]{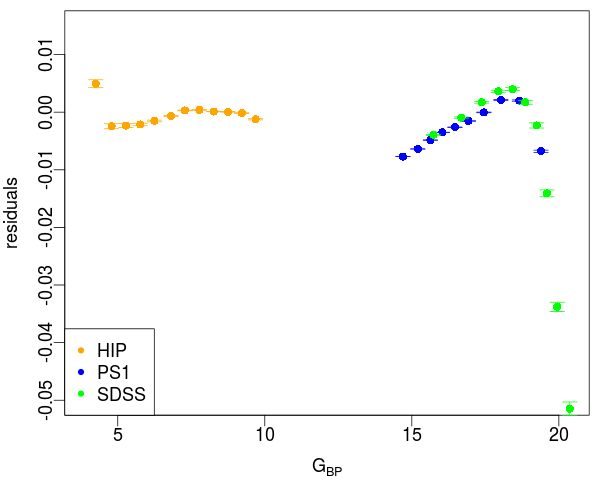}
\includegraphics[width=0.33\textwidth]{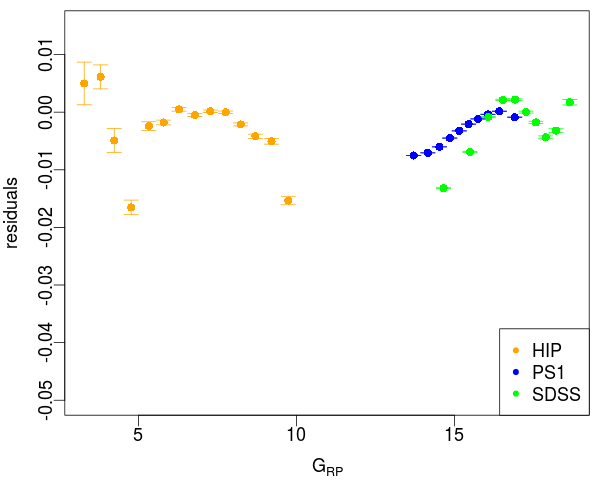}
\caption{From left to right, {\gmag}, {\gbp} and {\grp} photometry versus external photometry: {\hip} (orange), SDSS (green), PS1 (blue), OGLE (magenta). a) $G-r$ residuals of the global $G-r=f(g-i)$ spline for SDSS and PS1, $G-V$ residuals of the global $G-V=f(V-I)$ spline for OGLE. b) {\gbp}$-g$ residuals of the global {\gbp}$-g=f(g-i)$ spline. c) {\grp}$-z$ residuals of the global {\grp}$-z=f(g-i)$ spline. For {\hip} the residuals are computed versus the $X-Hp=f(V-I)$ spline, where $X$ denotes respectively {\gmag}, {\gbp} and {\grp}. The zero point of those different residuals is arbitrary.}\label{fig:wp944_photcomp}
\end{center}\end{figure*}

We compared {\gdrtwo} photometry to the {\hip}, {\tyctwo}, 2MASS \citep{2006AJ....131.1163S}, the SDSS tertiary standard stars of \cite{Betoule13} and Pan-STARRS1 \citep[PS1,][]{2016arXiv161205560C} photometry, selecting low extinction stars only ($E(B-V)<0.015$)  using the 3D extinction map of \cite{2017A&A...606A..65C}. We also compared to OGLE data in regions of relatively homogeneous extinction. An empirical robust spline regression was derived which models the global colour-colour relation. The residuals from those models are plotted as a function of magnitude in Fig.~\ref{fig:wp944_photcomp}. Comparison with 2MASS shows the effect of the 2MASS $J$ band saturation at $J=9$~mag of their "Read 2-Read 1" frames, rather than a possible {\gaia} issue, and is therefore not shown here.  

In the {\gmag} band, a strong saturation effect at $G<6$ is visible in the comparison with {\hip} and {\tyctwo} \citep[see also][]{DR2-DPACP-40}. The strong increase of the residuals for the faint stars, seen with SDSS in \cite{DR2-DPACP-40}, is confirmed here with PS1 as well as with OGLE data. 
A small dip at $G\sim$16, seen in Fig.~\ref{fig:phot_G_G-BP}b, is also present in the comparison with PS1, indicating that it is a feature of the {\gmag} band only. 

\gbp\ starts to deviate at \gbp$\sim$18 in low density regions (Fig.~\ref{fig:wp944_photcomp}b), due to the under-estimation of the sky background level. 
No strong feature is seen in \grp. 

\begin{figure}\begin{center}
\includegraphics[width=0.9\columnwidth,height=0.8\columnwidth]{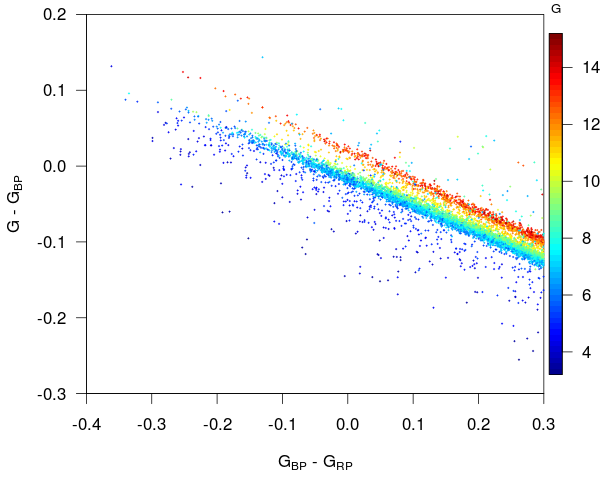}
\caption{Colour-colour relation for hot stars, using low extinction stars ($E(B-V)<0.015$)  with $\varpi/\sigma_\varpi>10$ and $M_G<2.5$,  colour-coded according to the mean {\gmag} magnitude.}\label{fig:hotstarsphotjump}
\end{center}\end{figure}

The small global increase ($\sim$2~mmag/mag) of the residuals with magnitude present in the internal comparison (Fig.~\ref{fig:phot_G_G-BP}) is seen with the external catalogues but is much more difficult to follow due to the relatively small interval coverage of each of the catalogues.
Moreover we applied our internal comparison tests to the external catalogues photometry also found global variations of this order of magnitude for PS1 and larger for SDSS (up to 10~mmag/mag). 

The variation of the residuals with magnitude is much stronger for the blue stars, as illustrated in \afterReferee{Fig.~\ref{fig:hotstarsphotjump}}. The brighter stars have a colour-colour relation more dispersed than the faint stars and the difference versus faint stars decreases with increasing magnitude up to $G\sim11$ where a jump of around 0.02~mag occurs, much larger than seen in the global Fig.~\ref{fig:phot_G_G-BP}. Comparison with 2MASS photometry indicates that the issue lies in the {\gmag} band but its cause is not yet known.

\subsection{Photometric precision}\label{photo-prec}
\subsubsection{Internal comparisons}\label{photo-prec_int}

The duplicated sources, cf.\ \secref{sec:dup}, have been used for a simple test
of the published uncertainties for the three broad-band magnitudes. It was
found that even for pairs of good astrometric quality, i.e.\ two full
astrometric solutions, the uncertainties appear underestimated. The normalised
magnitude differences are best understood if an error floor of 2.3\,mmag is
added in quadrature to the magnitude uncertainties (see the on-line
documentation, Figs.~10.12).  This test was made on a bright subset ($G <
17$\,mag) and as already mentioned in \secref{sec:dup} the duplicated sources
need not be representative for the catalogue as such. The apparent
inconsistencies between the magnitudes are not understood and we therefore
refrain from any recommendation regarding the use of a floor for the magnitude
uncertainty.

\subsubsection{Photometric precision using Clusters}\label{photo-prec_ext}

\begin{figure*}\begin{center}
\includegraphics[width=17cm, height=11cm]{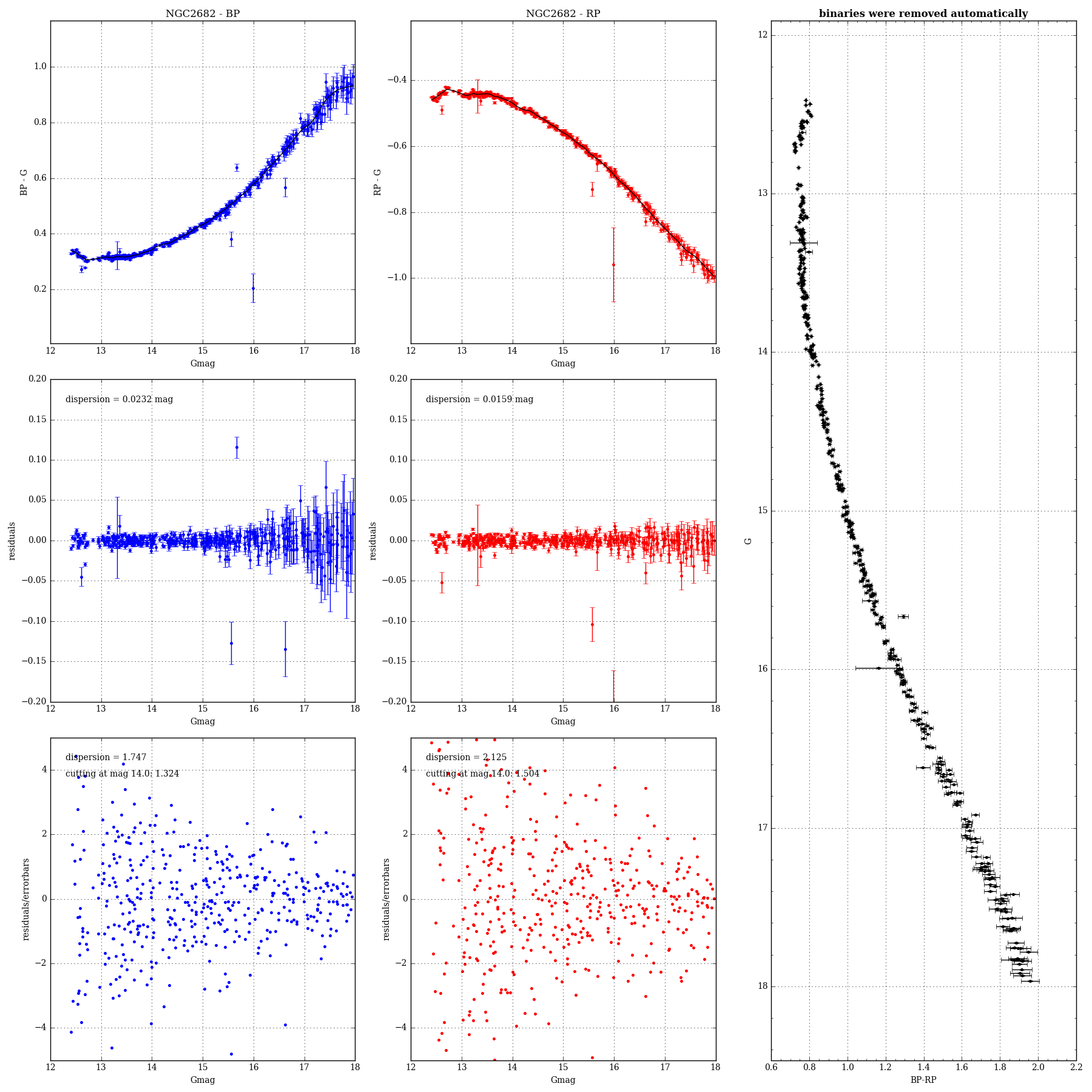}
\caption{CMD of NGC~2682, after binary sequence removal. Left: {\gbp}, middle: {\grp}, right: {\gmag} vs {\bpminrp}. Top: XP-{\gmag} vs {\gmag}, middle: residuals, bottom: normalised residuals.} \label{fig:NGC2682_BP_RP_errorbars}
\end{center}\end{figure*}

The quality of the {\gdrtwo} photometry using open and globular clusters was verified for stars brighter than $G\sim 18$, limiting our diagnostics to clusters with little extinction.
Cluster members were always selected using {\gdrtwo} proper motions and parallaxes. 
We assessed the quality of the {\gbp}/{\grp} photometry by estimating the width of the sequence in the CMD of clusters with secure membership. A downside of having to rely on secure membership is that it is difficult to provide diagnostics for stars fainter than $G\sim18$, as the photometry/astrometry (and thus our ability to discriminate cluster stars from field stars) strongly decreases in quality at that magnitude. Binary stars are first selected and removed.  We used a LOWESS fitting \citep{Cleveland} to follow the sequence, and removed binary star candidates by clipping out sources with {\gbp} fluxes two error bars lower and {\grp} fluxes higher than the fitted relation. Then we derived the dispersions of the relations 
 {\gbp}  and {\grp} vs {\gmag}.  We obtained very clean sequences for 12 OCs. An illustration is shown in Fig.~\ref{fig:NGC2682_BP_RP_errorbars} for NGC~2682.

The typical dispersion in both {\gbp} and {\grp} is of the order of 0.02\,mag.  We restrict  our analysis to stars in the un-evolved part of the main sequence, to avoid evolutionary effects. Scaling the difference to the fitted relation by the individual error bars of each star we find a unit-weight uncertainty of 1.3 on the average for {\gbp} and 1.5 for {\grp}. Because of effects such as rotation, magnetic field, stellar activity, the main sequence has a  natural width that is  difficult to estimate since it may vary from one cluster to another. This means that what we derive is an upper limit to the uncertainties on the photometry. Our result suggests that the errors on magnitudes in both filters are correctly estimated or only slightly underestimated.

In a few cases we detected a wide main sequence 
where a comparison with extinction maps \citep{1998ApJ...500..525S} strongly suggests the presence of differential extinction across the field \citep[see Fig.~A.8 of][for an example]{DR2-DPACP-31}.

\begin{figure}\begin{center}
\includegraphics[width=0.99\columnwidth, height=0.6\columnwidth]{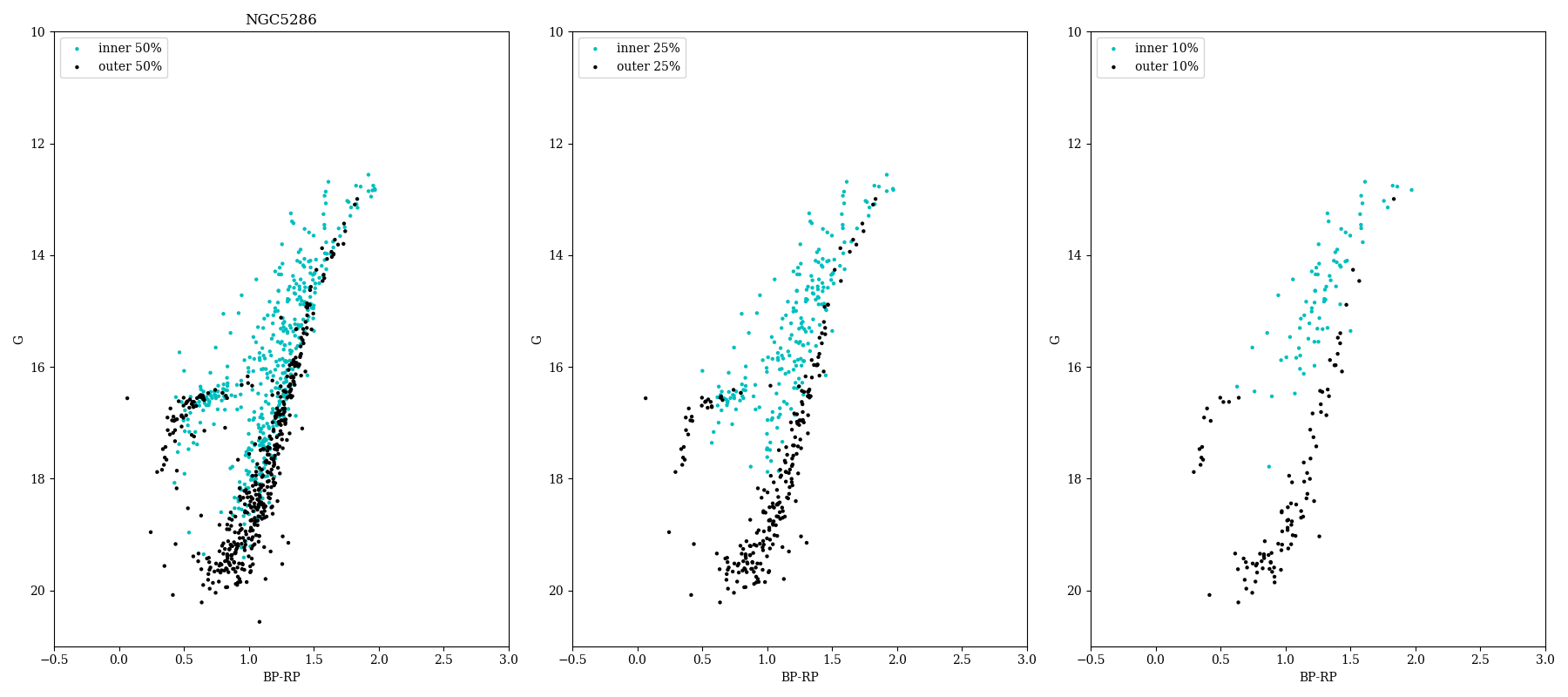}
\caption{ CMD of the globular cluster NGC~5286 inside a radius of 2.2\arcmin. Left panel is inside (cyan)/outside(black) of 1.4\arcmin; central panel plots the data  inside 0.89\arcmin (cyan) and outside 1.75\arcmin (black); right pannel gives the CMDs inside 0.55\arcmin and outside 2.0\arcmin.}  \label{fig:glob_mag1}
\end{center}\end{figure}

The quality of the photometry is substantially degraded in the inner regions of globular clusters (inside the core radius), due to high crowding. This effect is not present in the external regions. Fig.\ref{fig:glob_mag1} 
gives an example of the magnitude/colour shift between the inner and outer regions of the globular cluster NGC~5286.

\subsection{Photometric quality indicators and outliers}\label{photo-outliers}
There is an extensive discussion on the {\gbp}/{\grp} flux excess factor, \dt{phot\_bp\_rp\_excess\_factor}, in Sect.~8 of \cite{DR2-DPACP-40}. As it is sensitive to contamination by close-by sources in dense fields, binarity, background subtraction problems, as well as for extended objects, \citet[][Eq.~1]{DR2-DPACP-40} recommends to use it, with a colour term such as in Eq.~\ref{eq2}, to filter the photometry from outliers. As was seen at \secref{sec:err_ast} above, this has also a beneficial impact for astrometry.  

\begin{figure}\begin{center}
\includegraphics[width=0.49\columnwidth, height=0.6\columnwidth]{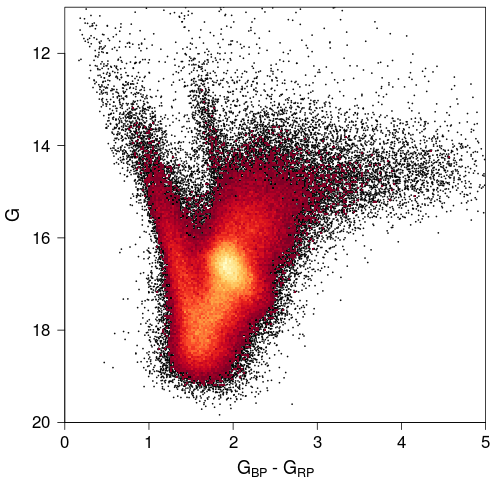}
\includegraphics[width=0.49\columnwidth, height=0.6\columnwidth]{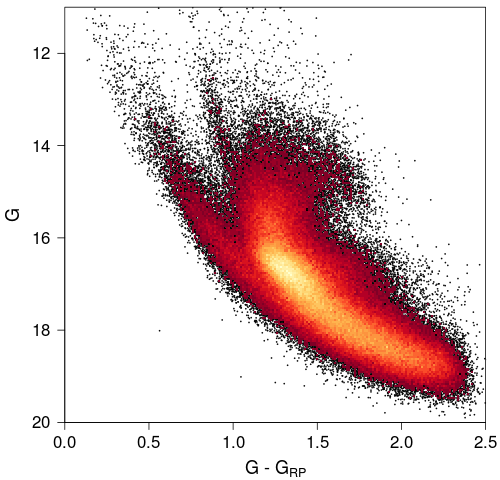}
\caption{Bulge colour magnitude diagram around Sagittarius window ($l=1.6\degr$, $b=-2.65\degr$) using either the a) {\bpminrp}, b) {\gminrp} colour, with photometric precision better than 5\% for {\gbp} and/or {\grp} and 2\% for \gmag, but without the filter \eqref{eq2} applied.}  
\label{fig:bulgeCMD}
\end{center}\end{figure}

To avoid the background issues for the faint {\gbp} stars, one can use the {\gminrp} colour.
However, counter-intuitively, in crowded areas with stars having roughly a similar spectral type (e.g. a selection of distant bulge stars which will consist of mainly red giants), the CMD using {\bpminrp} colour distribution will look reasonable (Fig.~\ref{fig:bulgeCMD}a), but not the CMD using {\gminrp} (Fig.~\ref{fig:bulgeCMD}b). This is due to the fact that the contamination flux will be present in both the {\gbp} and {\grp} bands, as they are integrated over the same spatial scale, averaging out in the {\bpminrp} colour, while it will not be in the {\gmag} band, derived from a narrow image profile-fitting, leading to a strong artificial reddening tail. The filter \eqref{eq2} is needed, especially in crowded area, even if {\gbp} is not used. This filter removes almost all stars fainter than $G>16$ in Fig.~\ref{fig:bulgeCMD}.

\subsection{Variability}\label{photovar}

The occurrence of variability along with the presence of outliers in the time series  photometry can strongly affect the mean magnitudes derived for variable sources by the {\gaia} photometric processing. In order to study this effect we have compared the two independent estimates of the {\gmag} mean magnitude  provided in the {\gdrtwo} archive for variable stars of RR Lyrae and Cepheid types, namely, \dt{phot\_g\_mean\_mag} listed in the gaia\_source table and \dt{int\_average\_g} provided for the same stars in the variability tables. 
The \dt{phot\_g\_mean\_mag} mean magnitudes are the result of the {\gaia} photometric processing which is described in detail in \citet{DR2-DPACP-40}, while the \dt{int\_average\_g} mean magnitudes  
are computed as part of the specific processing of RR Lyrae stars and Cepheids which takes into account the variability of these sources. The \dt{int\_average\_g} mean magnitudes  are derived from the Fourier models best fitting the time series data of the sources \citep[][2018, in preparation]{2016A&A...595A.133C}. Furthermore, the outlier rejection procedures applied in estimation of the  \dt{phot\_g\_mean\_mag} and \dt{int\_average\_g}  mean magnitudes are different. Nevertheless, the two measurements of the mean {\gmag} magnitudes are in good agreement for the large majority of stars. 

\begin{figure}\begin{center}
\includegraphics[trim=20 170 30 50,width=0.8\linewidth]{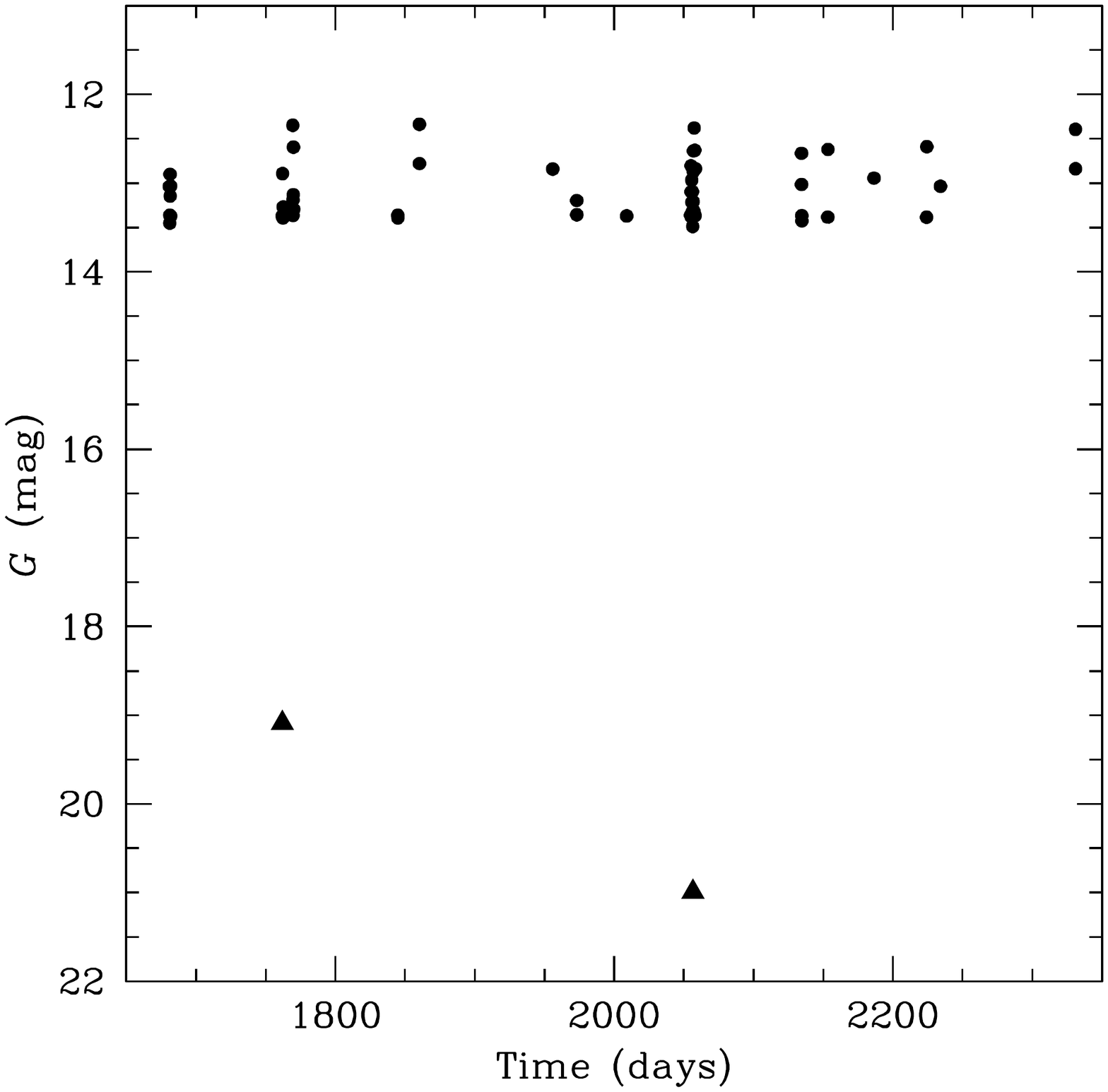}
\includegraphics[trim=20 170 30 50,width=0.8\linewidth]{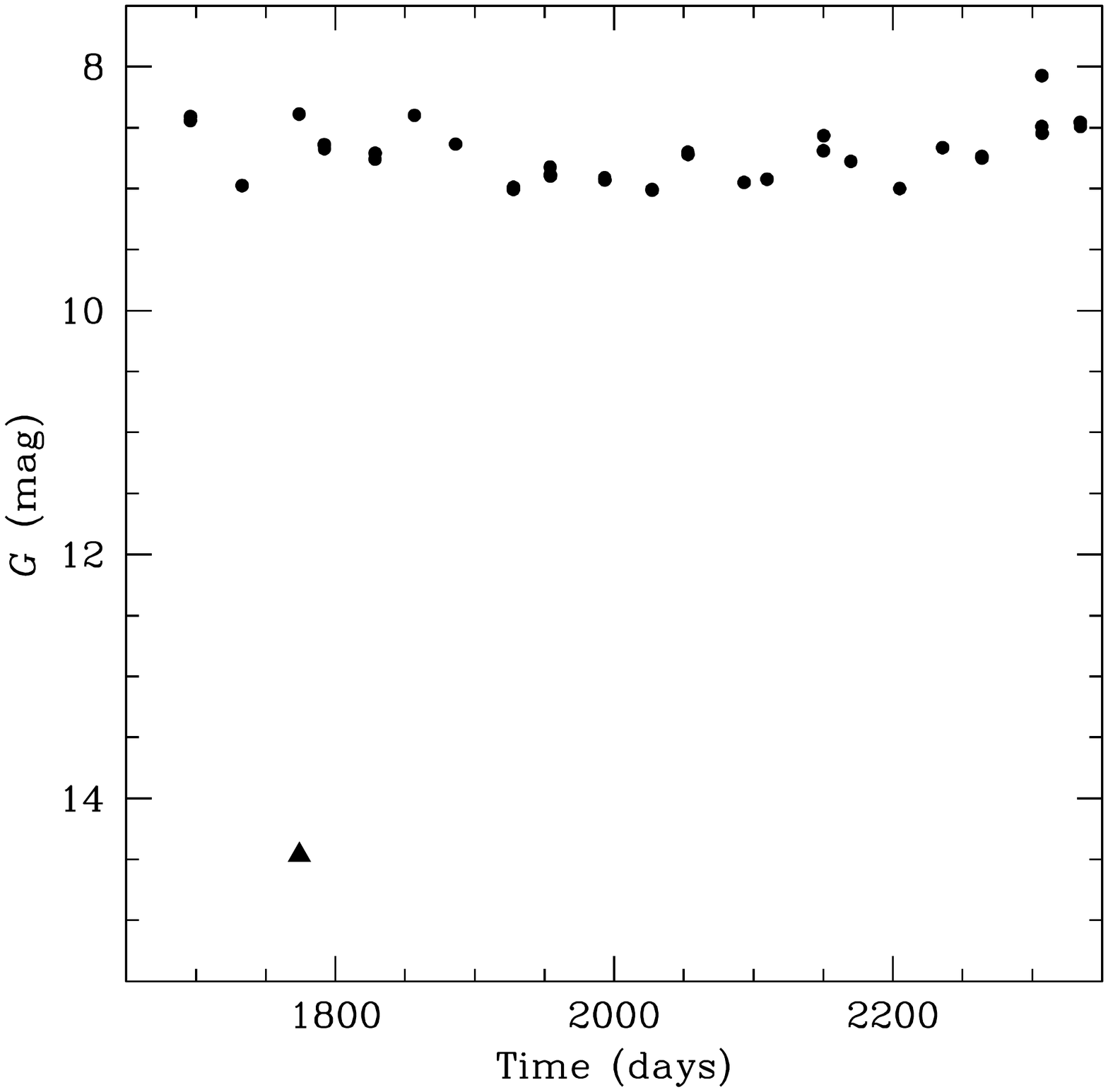}
\caption{Time series {\gmag}-band photometry of the RR Lyrae variable WY~Scl (upper panel) and the Cepheid UY~Car (bottom panel). Black dots and triangles represent measurements  used to calculate the \dt{int\_average\_g} values and rejected outliers, respectively.}
\label{fig:wp946_lc}
\end{center}\end{figure}

For a small fraction of variables: eight RR Lyrae stars and six Cepheids, the two mean {\gmag} magnitudes differ by more than 1~mag. We have visually inspected the time series  photometry  of these 14 variables  and found that their datasets contain faint outliers significantly deviating from the majority of the photometric measurements.  As an example, the time series photometry of the RR Lyrae variable WY~Scl and the Cepheid UY~Car are presented in Fig.~\ref{fig:wp946_lc}. The intensity-averaged mean {\gmag} magnitude of  WY~Scl is $\dt{int\_average\_g}=13.04$~mag, while $\dt{phot\_g\_mean\_mag}=15.25$~mag. The upper panel of Fig.~\ref{fig:wp946_lc} shows that the determination of the \dt{phot\_g\_mean\_mag}  was affected by two faint outliers (triangles) that were instead rejected in the estimation of the  \dt{int\_average\_g} magnitude. The same issue affects  also  UY~Car (bottom panel of Fig.~\ref{fig:wp946_lc}), for which $\dt{phot\_g\_mean\_mag}=14.19$~mag, while $\dt{int\_average\_g}=8.69$ mag. 

The reason for this discrepancy is the estimation of the mean {\gmag} magnitudes \dt{phot\_g\_mean\_mag}: the outlier rejection procedure is still non optimal for variable sources \citep{DR2-DPACP-40},  while it is efficient for constant stars.  The large spread of the  measurements  caused by the variability makes the estimation of the sigma used in the rejection difficult and overestimates its value. Hence, outliers may fall within a few sigma from the median value and, therefore, they are not rejected. Furthermore, the weighted mean value is calculated in the flux space, where the fainter observations have smaller uncertainties and, consequently, higher weights than  brighter values. Thus, if there are faint outliers in the photometric dataset of a variable source, they will most likely drag the estimated weighted mean flux and magnitude towards the faint tail of the distribution. 

This issue has affected the estimation of the mean magnitude of RR Lyrae itself, a relatively bright star ($V\sim 7.12$~mag) that has given its name to the whole class of RR Lyrae variables. 
The mean {\gmag} magnitude of this star provided in gaia\_source table $\dt{phot\_g\_mean\_mag}=17.04\pm1.57$ mag, which is $\sim10$ mag fainter than the true value. The reason is, as in the previously described cases, a faint outlier in the {\gmag}-band  time series  that drags the distribution towards the faint end. Unfortunately,  no \dt{int\_average\_g} mean magnitude is provided  for the star in the variability tables: due to limited number of  measurements available during the variability processing the star was rejected by the algorithm based on the Fourier parameters of the light curve (Clementini et al. 2018, (in preparation)). Incorrect value of \dt{phot\_g\_mean\_mag} for RR Lyrae itself caused  incorrect estimation  of the magnitude-dependent term applied  in the astrometric instrument calibration and, consequently,  wrong estimation of parallax for this star in {\gdrtwo}: $\varpi=-2.61\pm0.61$~mas.

To summarise, the  method to determine  the mean magnitudes \dt{phot\_g\_mean\_mag} of the photometric processing  worked properly for  the large majority of  stars and produced incorrect results only for a small fraction of variables. It will be further improved in {\gaia} Data Release 3.

\section{Radial velocity}\label{radial}

We refer the reader to \cite{DR2-DPACP-54} for a description of the radial velocity data in {\gdrtwo} and their extensive validation. This publication describes the stars
which did not pass the quality filters to be published in {\gdrtwo}, thus affecting the completeness of the radial velocity data.
For instance, only stars with a radial velocity uncertainty $\leq 20 ${\kms} have been published in {\gdrtwo}. Also, the publication has been restricted to stars with effective temperature between 3500 and 7000 K due to a degraded performance of the radial velocity and to the restricted grid of templates, respectively. A sky map of the completeness can be also seen in \cite{DR2-DPACP-54}, showing the expected decrease in dense areas where there are conflicts between acquisition windows. 
The completeness also depends on the initial list of sources observed by the {\gaia} spectrograph to be published for {\gdrtwo}, whose quality was very dependent on the sky region. The reader can also find in the above paper several considerations on the global zero point (see also below), comparison with external data and on the precision and accuracy of the data as a function of magnitude, stellar properties, sky position, etc.

\subsection{Accuracy}\label{radial-accuracy}

The radial velocities have been compared to external catalogues: GALAH DR1 \citep{2017MNRAS.465.3203M}, RAVE DR5 \citep{2017AJ....153...75K}, APOGEE DR14 \citep{2015AJ....150..148H}, GES DR3 \citep{2012Msngr.147...25G}, SIM \citep{2015MNRAS.446.2055M} and a home made compilation of several smaller catalogues UMMSV composed of \cite{DR2-DPACP-48}, \cite{2005A&A...430..165F},  \cite{2008A&A...485..303M,2009A&A...498..949M}, \cite{2002ApJS..141..503N},  \cite{2004A&A...418..989N}, \cite{2012A&A...542A..48W} and \cite{2012arXiv1207.6212C}. The results are summarised in Table~\ref{tab:cu9val_wp944_summaryvr}. The overall differences can be due to either {\gaia} and/or the external catalogue. Similar comparisons are presented in more details in \cite{DR2-DPACP-54}. A global zero point offset between 0.1 and 0.3~{\kms} is found with respect to all catalogues, including a global increase of this offset with magnitude. The other correlations are catalogue-dependent and therefore not discussed further.

\begin{table*}
\caption[Comparison between Gaia DR2 radial velocities and the external catalogues]{Summary of the comparison between the radial velocities and the external catalogues.} \label{tab:cu9val_wp944_summaryvr}
\begin{center}
\begin{tabular}{lcccccl} 
\hline\hline
{ Catalogue} & Nb & { Outliers} & $<G>$ & {RV~difference} & {RV~uwu} & Correlations \\
\hline
GALAH & 571 & 1\% & 11.9 & \textcolor{purple}{$0.16 \pm 0.02$} & \textcolor{purple}{$1.29\pm0.04$} &  \\
SIM &  1927 & 4\% & 9.4 & \textcolor{purple}{$0.24 \pm 0.006$} & \textcolor{purple}{$1.12\pm0.02$} & \bpminrp \\
APOGEE & 60282 & 2\% & 12.3 & \textcolor{purple}{$0.24 \pm 0.002$} & \textcolor{purple}{$1.285\pm0.004$} & \gmag, \bpminrp, Teff, logg \\
RAVE &  373755 & 3\% & 11.4 & \textcolor{purple}{$0.27 \pm 0.002$} & \textcolor{purple}{$1.480\pm0.002$} & \gmag, \bpminrp, Teff, logg, [Fe/H] \\
GES & 2201 & 3\% & 12.7 & \textcolor{purple}{$0.13 \pm 0.02$} & \textcolor{purple}{$1.33 \pm 0.02$} & \gmag \\
UMMSV & 6843 & 4\% & 7.5 & \textcolor{purple}{$0.15 \pm 0.003$} & \textcolor{purple}{$1.38 \pm 0.01$ } & \gmag, \bpminrp \\
\hline
\end{tabular}
\tablefoot{The total number of stars used in the comparison (Nb) as well as the percentage of outliers (at 5$\sigma$) excluded as well as the median {\gmag} of the sample are presented.  The radial velocity difference (Gaia-Ext, in \kms) and the unit-weight uncertainty (uwu) that need to be applied to the data to adjust the differences are indicated in purple when they are significant (p-value limit: 0.01). Significant correlations of the differences with other parameters are indicated in the last column and may as well originate from the external catalogue. 
}
\end{center}
\end{table*}

The data for duplicate sources have been removed from {\gdrtwo} but have also been used beforehand for validation purposes. Similar to previous sections, here we look at the RV related data for the duplicate sources to test for internal consistency. However, we have to bear in mind that this sample may not be representative of all the RV dataset. We found 100\,406 pairs of duplicated sources with RV data (see Sect.~\ref{sec:err_dup}). When looking at the templates used for each component of the pair of duplicates, we see that the same template has been used in 40\% of the cases for \logg, 86\% of the cases for \feh, and 41\% for \teff.

\begin{figure}\begin{center}
\includegraphics[width=8cm]{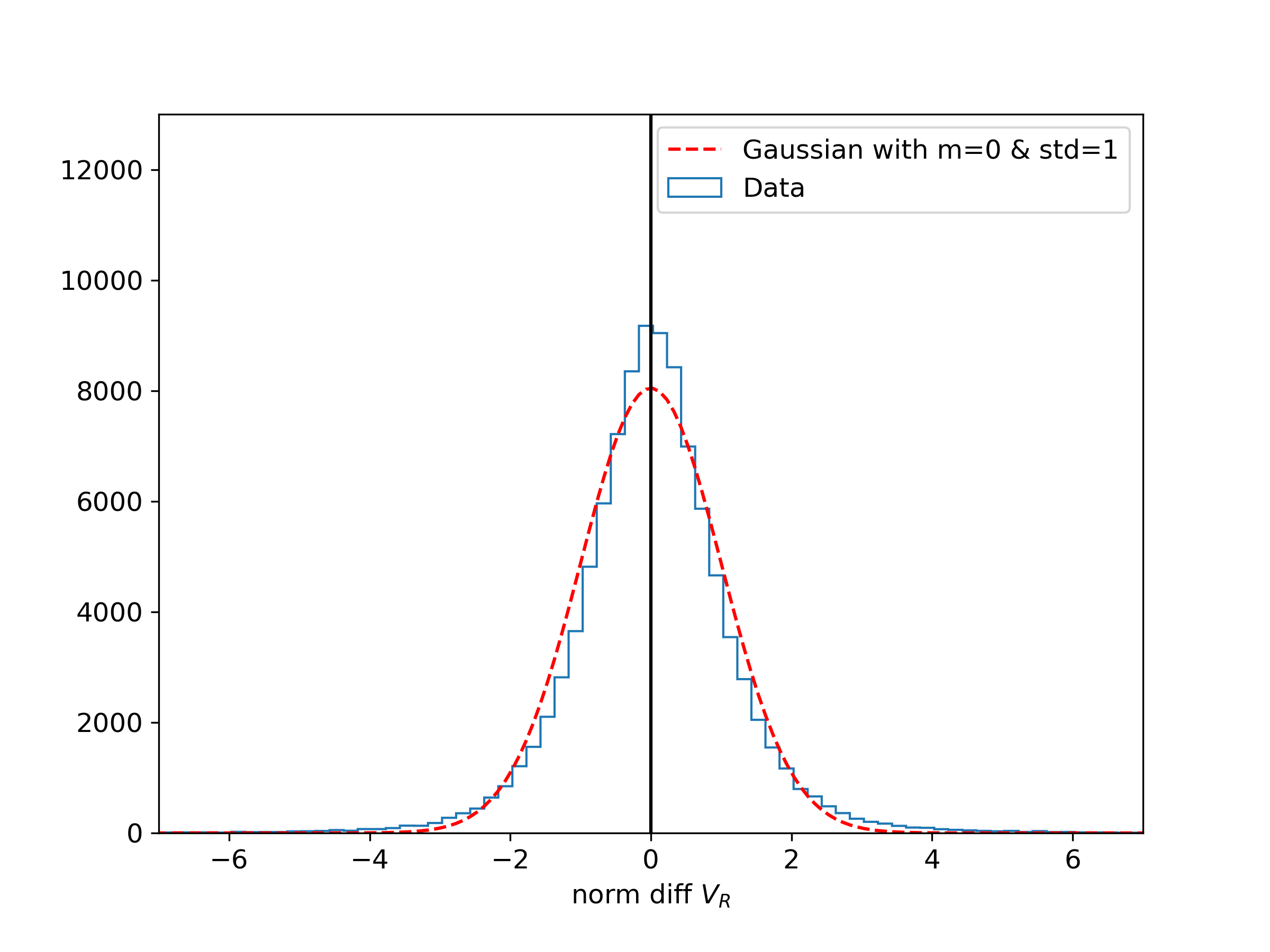}
\caption{Blue histogram: differences between the radial velocity within pairs of
duplicate sources, normalised by their uncertainty. The red dashed curve is a 
Gaussian fit forced to have mean=0 and standard deviation = 1 but free height.}
\label{fig:cu6duprvsnorm2}
\end{center}\end{figure}

Overall, the RV data coming from duplicate sources is consistent. Under the assumption of Gaussian errors, the differences between the measurements of radial velocity for the components of the pairs, when normalised by the errors, should yield a Gaussian distribution centred at 0 and with dispersion equal to 1. We see in Fig.~\ref{fig:cu6duprvsnorm2} that the data (blue histogram) follows well the expected distribution (red curve). The normalised median and robust dispersion of the data are $0.023\pm 0.0036$  (in the sense that the eliminated sources show slightly larger radial velocities) and 0.91 respectively.  

This suggests a very small bias, again significantly below the random errors. It does not seem to correspond to differences in the templates used, and may rather originate from stars with a low number of observations, which is usually the case for one of the components of duplicate pairs. However, another estimation of the mean and dispersion can be obtained by fitting a Gaussian to the histogram, giving a mean of -0.08 and standard deviation of 0.87, indicating again, overall, the good internal consistency of the data.

\subsection{Precision of radial velocities}\label{radial-prec}

As indicated in the section above, the core of the distribution of the normalised differences of duplicates has a 0.87 dispersion consistent with the 0.91 value obtained using a robust estimate of the full distribution. The robustness is actually needed to mitigate the effect of the few binaries which must be present within the duplicates and produce outliers. This internal comparison thus points to a pessimistic estimate of the uncertainties.

On the contrary, all the comparisons with external catalogues (Table~\ref{tab:cu9val_wp944_summaryvr}) indicate some underestimation of the radial velocity uncertainties. However those are a combination of the {\gaia} and the external catalogue errors and of the intrinsic radial velocity variation due to binarity or duplicity in dense fields. The comparisons with the cleanest catalogues (e.g. with multi-epoch radial velocity measurements) indicate the smallest underestimation, and suggest that the {\gaia} RV uncertainties are probably not significantly underestimated.

\section{Astrophysical parameters}\label{astropar}
In this section we review some of the key features found during the validation of the astrophysical parameters (AP) using different approaches, namely open clusters and internal or external data. Part of these features are also reported in \cite{DR2-DPACP-43} which devotes a large part to the AP validation. In \secref{astropar-temp}, we show the results found for the effective temperature {\teff}. In \secref{astropar-ext}, we focus on the extinction $A_G$ and reddening {\ebpminrp}, while in \secref{astropar-lum}, we give details of the validation of the radius and luminosity. Finally, in \secref{astropar-dup}, we use the duplicate sources present in internal releases to validate the astrophysical parameters.

\subsection{Temperature}\label{astropar-temp}

\begin{figure}\begin{center}
\includegraphics[width=0.8\columnwidth]{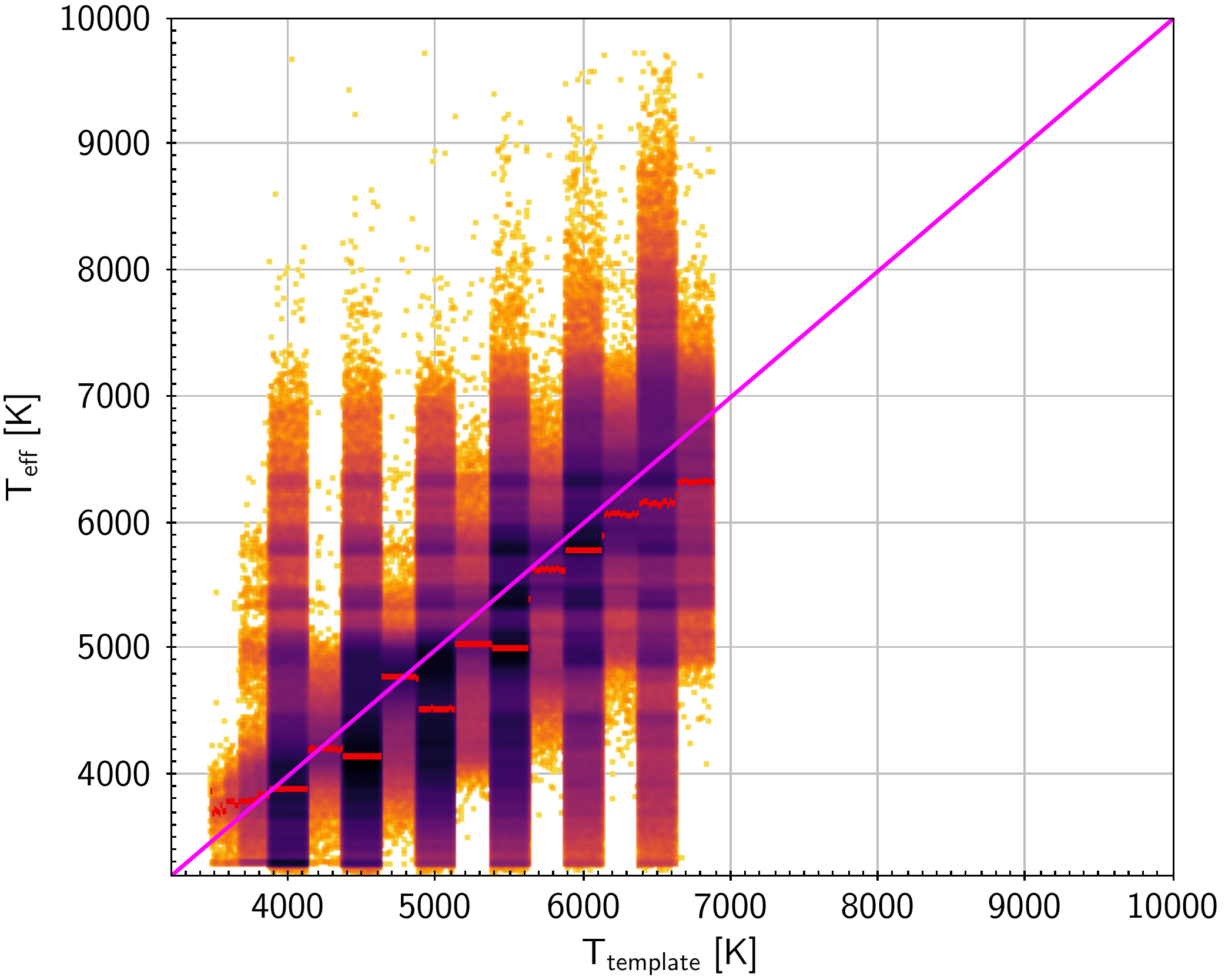}
\includegraphics[width=0.8\columnwidth]{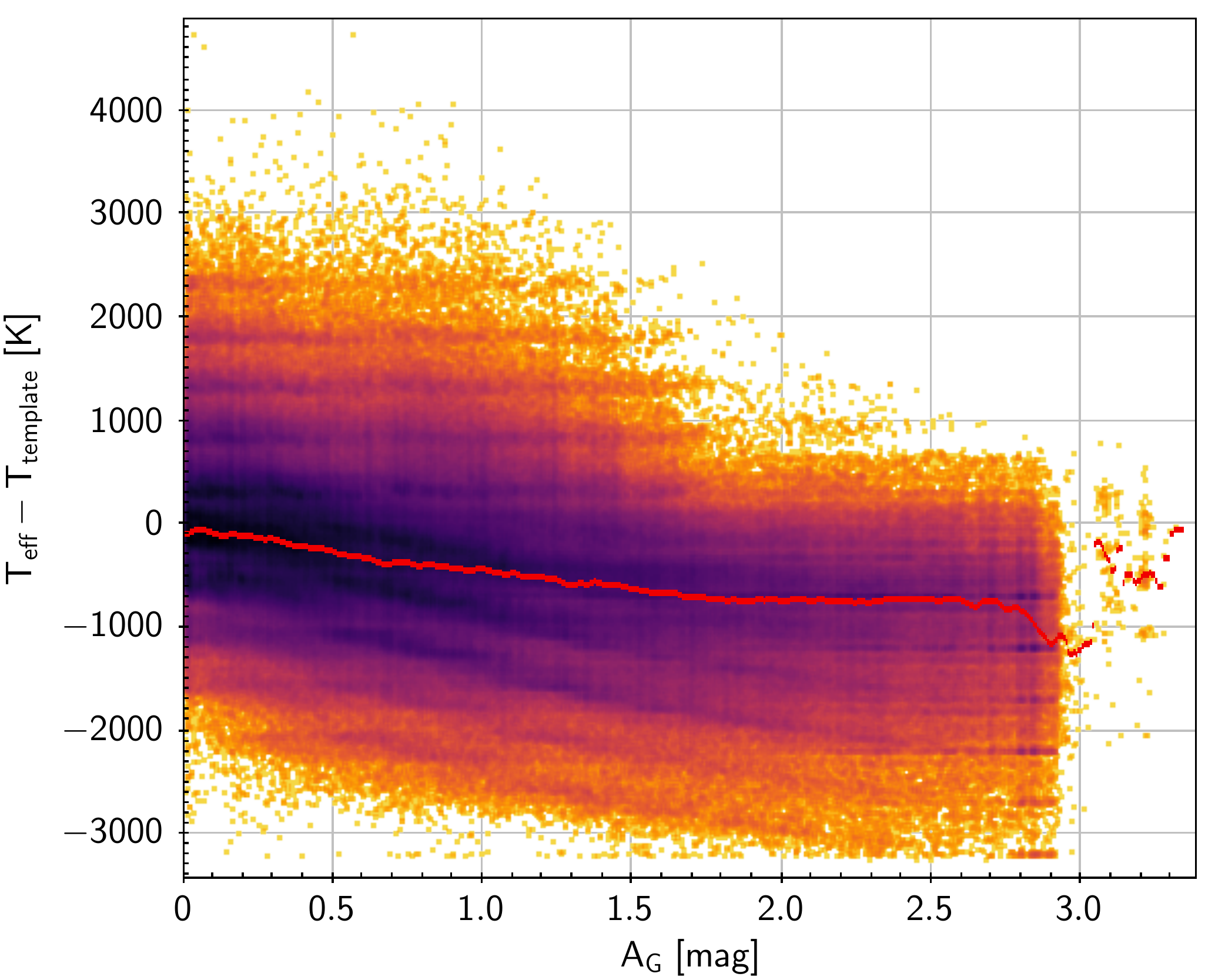}
\caption{{Top:} {\teff} estimated from the photometry versus {\teff} of the radial velocity template. The red lines show median values, while the diagonal line is the unity line. The template {\teff} have been randomly smeared $\pm250$\,K for clarity. {Bottom:} difference between the two temperatures versus the extinction. The red line shows the median.
}\label{fig:teff_comp}
\end{center}\end{figure}

As a first internal consistency test, in \figref{fig:teff_comp}, we plot the comparison between the {\teff} provided in {\gdrtwo} and the effective temperature template used to derive the radial velocity of the star, by comparing the linear fit to the data and the 1:1 correspondence line.  Taking into account that, first, the effective temperature template is largely unaffected by the extinction and should not be used as an estimation of the effective temperature of the star, and, second, the extinction could not be used when deriving the effective temperature \citep{DR2-DPACP-43}, we see that the effective temperature in {\gdrtwo} is underestimated.

\begin{figure}\begin{center}
\includegraphics[width=0.9\columnwidth]{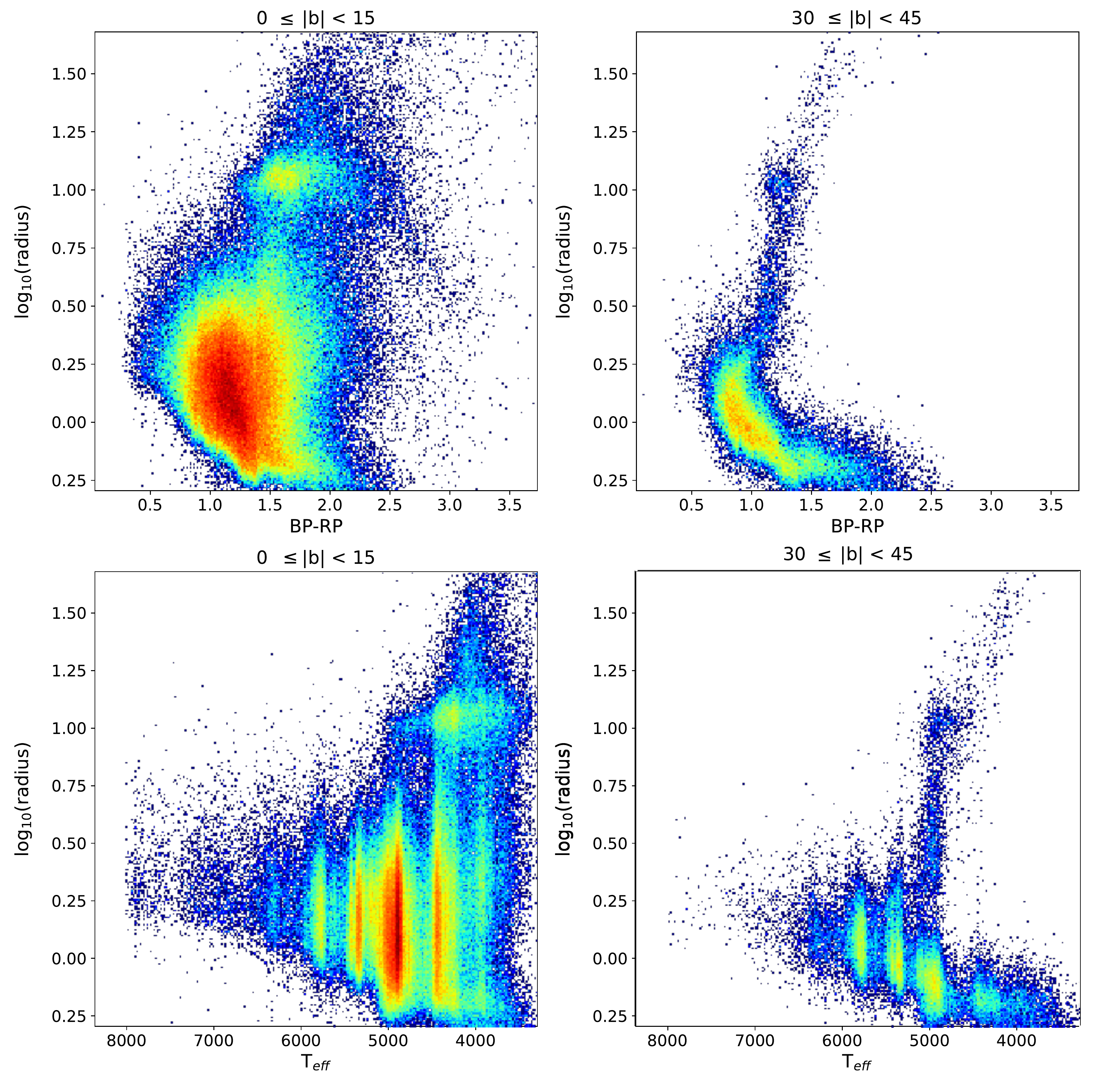}
\caption{Log of radius vs colour (top) and temperature (bottom), in the Galactic plane (left) and intermediate latitude (representative of the behaviour at all latitudes $b \ge 15$\degr, right). The temperature of the red clump appears to follow the reddening, and cold stars on the main sequence have too large radii for $|b| \le 15$\degr.}\label{fig:s90int6-color_teff-vs-radius_log}
\end{center}\end{figure}

This is also apparent from
\figref{fig:s90int6-color_teff-vs-radius_log} where we explore the
relation between colour/ temperature and radius as a function of
Galactic latitude for a thin slice at Galactic longitude
$l=90\degr$. When comparing the distributions for low and intermediate
latitude bins, we see for example that the cloud of points with $\log
R \sim 1$, which has colour $G_{\rm BP} - G_{RP} \sim 1.2 - 1.4$ and
effective temperature {\teff}$ \sim 5000$~K for $30\degr < |b| < 45\degr$,
because of reddening its colour becomes {\ebpminrp} $ > 1.5$
for $|b| < 15\degr$. As a result the derived temperature is
artificially shifted to below 4500~K. Similarly, for this bin we see
that for the lowest temperatures, the stars have too large radii
compared to stars at higher galactic latitudes.

\begin{figure}\begin{center}
\includegraphics[width=0.8\columnwidth]{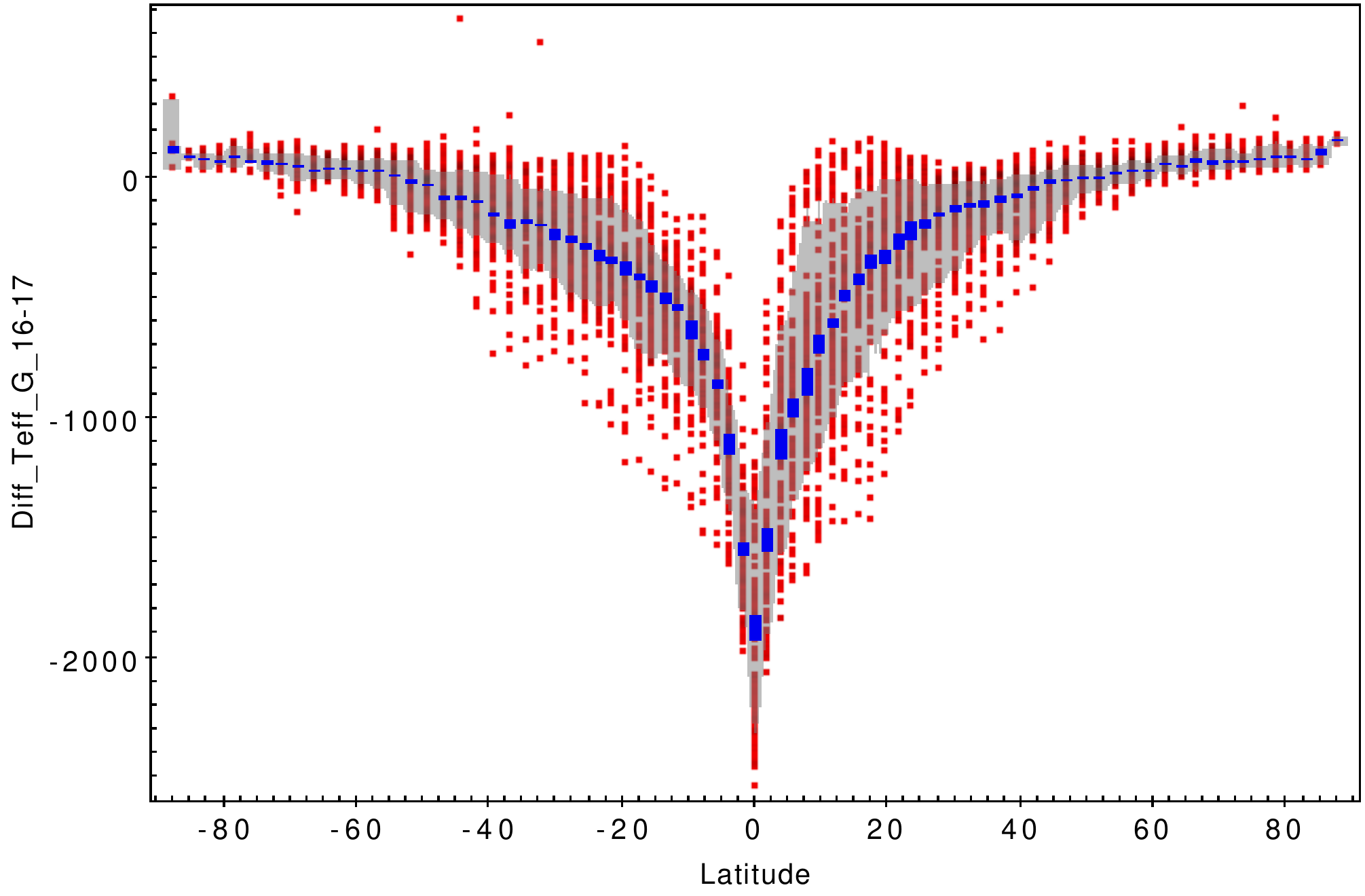}
\caption{Difference of the mean {\teff} (K) between {\gdrtwo} data and GOG simulation. Each red dot corresponds to the mean difference in a healpix bin at a given latitude (abscissa, in degrees) limited to the magnitude range $16 < G < 17$. The grey area indicates the $1\sigma$ quantiles ($Q = 0.15$ to 0.85) and the blue one the central values ($Q = 0.46$ to 0.54).}\label{fig:Diff_Teff_G_16-17_Lat}
\end{center}\end{figure}

Consistent with these findings, using GOG \citep{2014A&A...566A.119L}, a {\gaia}-based simulation based on the Besan\c{c}on Galaxy model \citep{2012A&A...543A.100R}, the influence of extinction suggests that the temperature may have an large bias in the Galactic plane (\figref{fig:Diff_Teff_G_16-17_Lat}), while it would be correctly estimated over the plane. 

The same feature is observed when comparing to the APOGEE DR14 temperatures, the systematic offset being larger than the uncertainties at galactic latitude smaller than $\vert b \vert<20\degr$, as presented in Fig.~12 of \cite{DR2-DPACP-43}. 

\begin{figure}\begin{center}
\includegraphics[width=8cm]{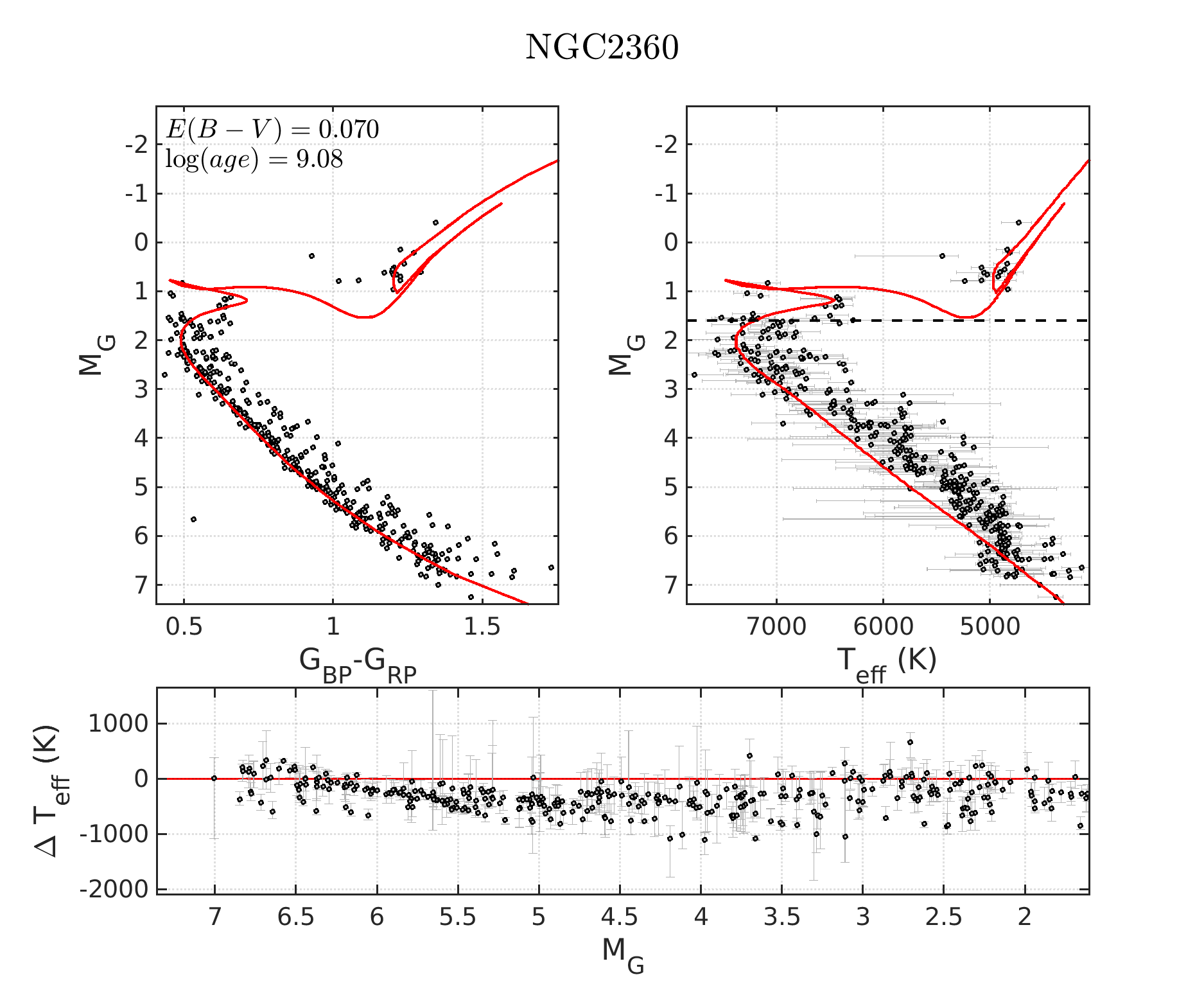}
\caption {{\gdrtwo} {\teff} (blue points) compared with the expectations from a PARSEC isochrone (red line)  for NGC~2360 on the M$_{\rm G}$ absolute magnitude vs {\bpminrp} plane (left upper panel); on the M$_{\rm G}$ vs {\teff} plane (right upper panel), and $\Delta$ {\teff} vs M$_{\rm G}$ (lower panel)}\label{fig:NGC2630_teff}
\end{center}\end{figure}

\begin{figure}\begin{center}
\includegraphics[width=8cm]{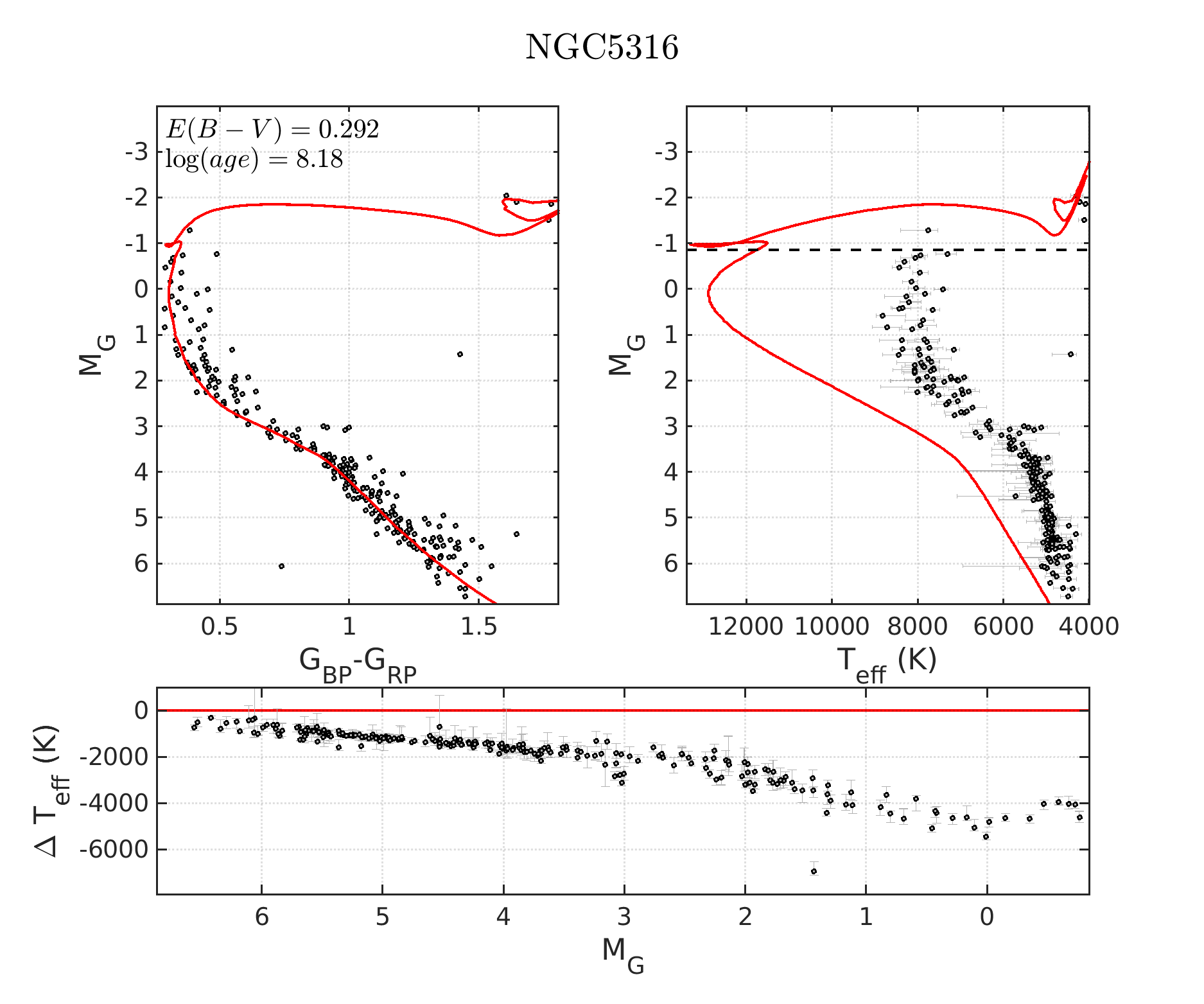}
\caption{Same as \figref{fig:NGC2630_teff} for NGC~5316.}\label{fig:NG5316}
\end{center}\end{figure}

Then, for about 180 open clusters, we compared the value of {\teff} with the expectations from  PARSEC isochrones by \cite{2014MNRAS.444.2525C} where magnitudes are calculated with the {\gaia} passbands \citep[see][for details]{DR2-DPACP-31}. We used literature values for the age of each cluster \citep{2013A&A...558A..53K} and solar metallicity for NGC~2156 and NGC~5316, while the information about NGC~2516 are taken from \cite{2016A&A...586A..52J}.  In general there is a reasonable agreement for the clusters
located in  regions of low extinction until the expected temperature is below {\teff} $\sim 7000-8000$ K. In  \figref{fig:NGC2630_teff} we present the distribution of the temperatures for NGC~2630 having  $E(B-V)=0.07$. This result is consistent with the fact that the temperature {\teff} was derived under the assumption of $A_G=0$. Significant deviations of the temperature are expected in moderate/high extinction regions, as in the case of NGC~5316 having $E(B-V)=0.29$, \figref{fig:NG5316}.

\begin{figure}\begin{center}
\includegraphics[width=8cm]{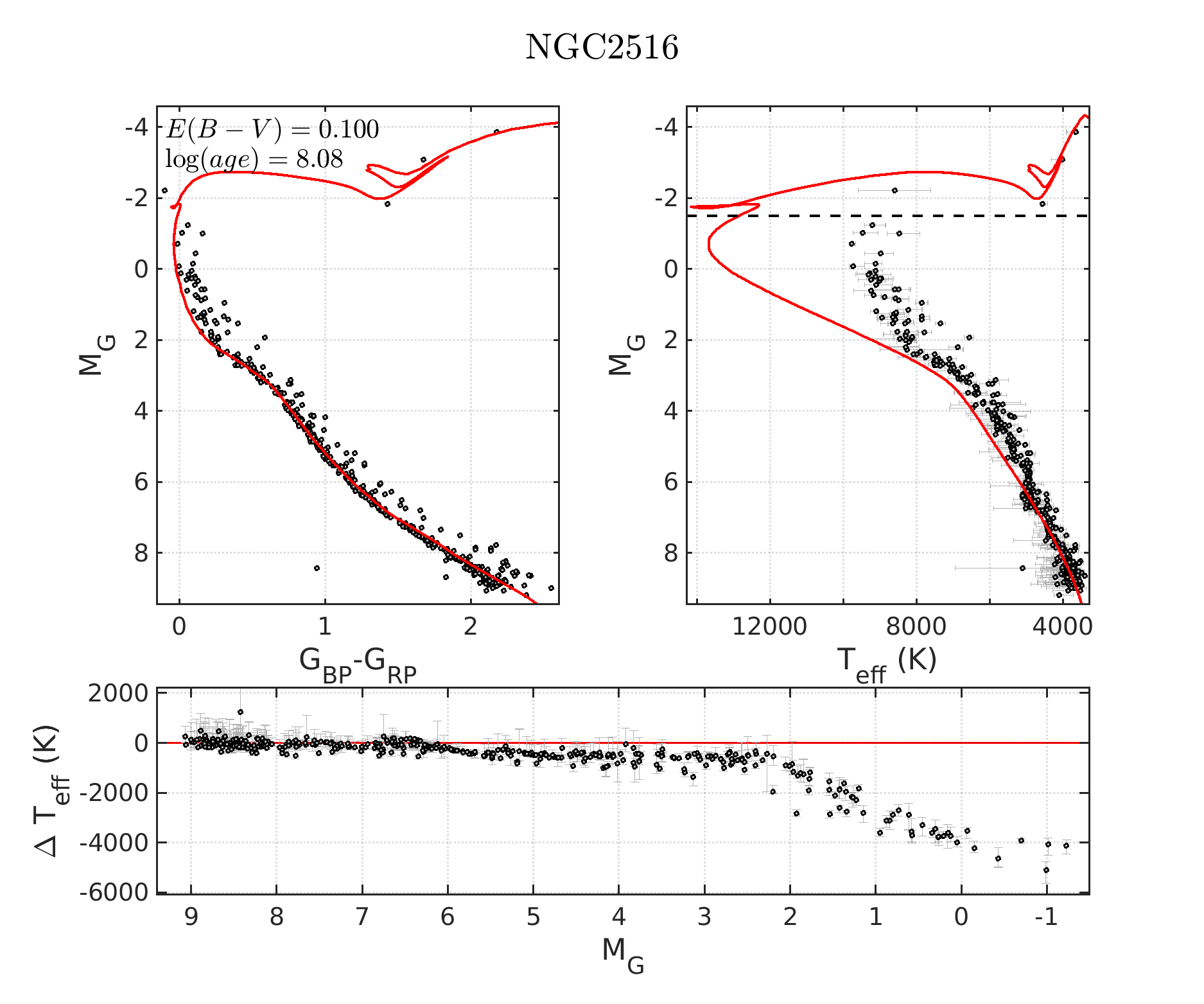}
\caption{Same as  \figref{fig:NGC2630_teff} for NGC~2516}\label{fig:NG2516_sat}
\end{center}\end{figure}

{\gdrtwo} {\teff} was derived training the regression algorithms with observational templates in the range 3000 K$<${\teff}$< 10000$ K \citep{DR2-DPACP-43}. This has produced  a saturation effect for all stars hotter/cooler  than these limits. However a significant deviation from the expected values is detected already at {\teff}$= 8000$ K (see \figref{fig:NG2516_sat} for an example). 
A spurious effect of granularity on the {\teff} distribution is present. This is understood as coming from the inhomogeneities in the {\teff} training data distribution \citep[see Fig.18 in ]{DR2-DPACP-43}.   

\subsection{Extinction and reddening}\label{astropar-ext}

\subsubsection{Extinction using Open and Globular Clusters}\label{ssec:exticluster}

\begin{figure}\begin{center}
\includegraphics[width=8cm]{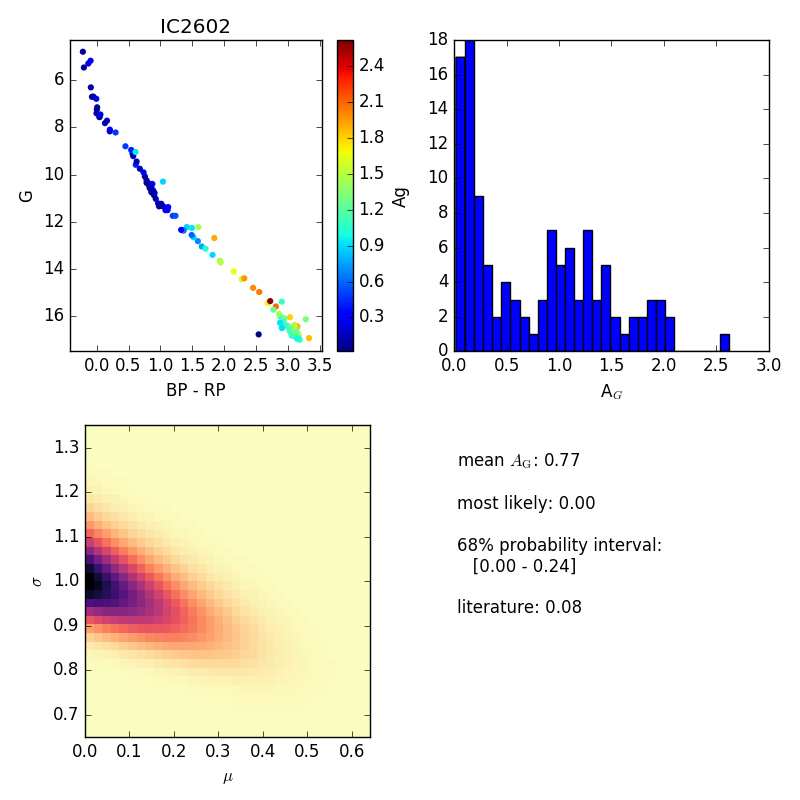}
\caption{CMD of IC~2602 where the colours represent the value of  the extinction $A_G$;(top left); $A_G$ distribution in the cluster (top right); distribution of $\mu$, the extinction value obtained using the ML method vs the uncertainty $\sigma$ (bottom panel)}\label{fig:IC2602ext}
\end{center}\end{figure}

As explained in \cite{DR2-DPACP-43}, deriving the extinction and {\teff} from the $G$, {\gbp}, and {\grp} magnitudes has to face the fact that the system is degenerate and these degeneracies  lead to large random errors. The consequence is that the $A_G$ values presented in the {\gdrtwo} Catalogue cannot be easily used on star-by-star basis. An illustration of this is presented in Fig.~\figref{fig:IC2602ext} for the cluster IC~2602.  Indeed $A_G$ presents a large spread inside the cluster, depending on the colour of the stars. Redder stars, either faint main sequence or binaries turn out to have always attributed an higher extinction.  Similar analysis on globular clusters show that the majority of stars have extinction values higher than 1. Halo Globular clusters are expected to have extinctions in the range $A_G$=0.05-0.09. It should be pointed out that what is expected to show no variation with the temperature is the extinction parameter $A_0$, while $A_G$ has a dependence on \teff. When $A_0=0.3$ we have $(A_G-A_O) >-0.11$ for   $(G_{BP}-G_{RP})< 3$ \citep{2018arXiv180201670D}. However the variations presented here cannot be entirely ascribed to this effect.

The non-negativity constraint imposed on extinction and the noise level (with typical uncertainties of about 0.6\,mag on the extinction of individual sources) makes the probability distribution function for the members of a cluster highly asymmetrical. As a consequence, their uncertainty distribution is far from being Gaussian. \cite{DR2-DPACP-43}  recommend to use a maximum-likelihood method (ML) to combine the data and  derive the most likely value inside the area, while simple (or weighted) mean overestimate the extinction. If we apply this method and filter the best measured stars following the flags in \cite{DR2-DPACP-43},  we derive  for IC~2602 the most likely value for $A_G$ as $0.0$ with a 68\% probability interval in the range $0 \leq A_G <0.24$. Applying a simple average gives instead $A_{G,\text{mean}}=0.77$. \cite{2013A&A...558A..53K} quote an extinction of $E(B-V)=0.031$, corresponding to $A_G=0.08$ (see \figref{fig:IC2602ext} for an example).

\begin{figure}\begin{center}
\includegraphics[width=8cm]{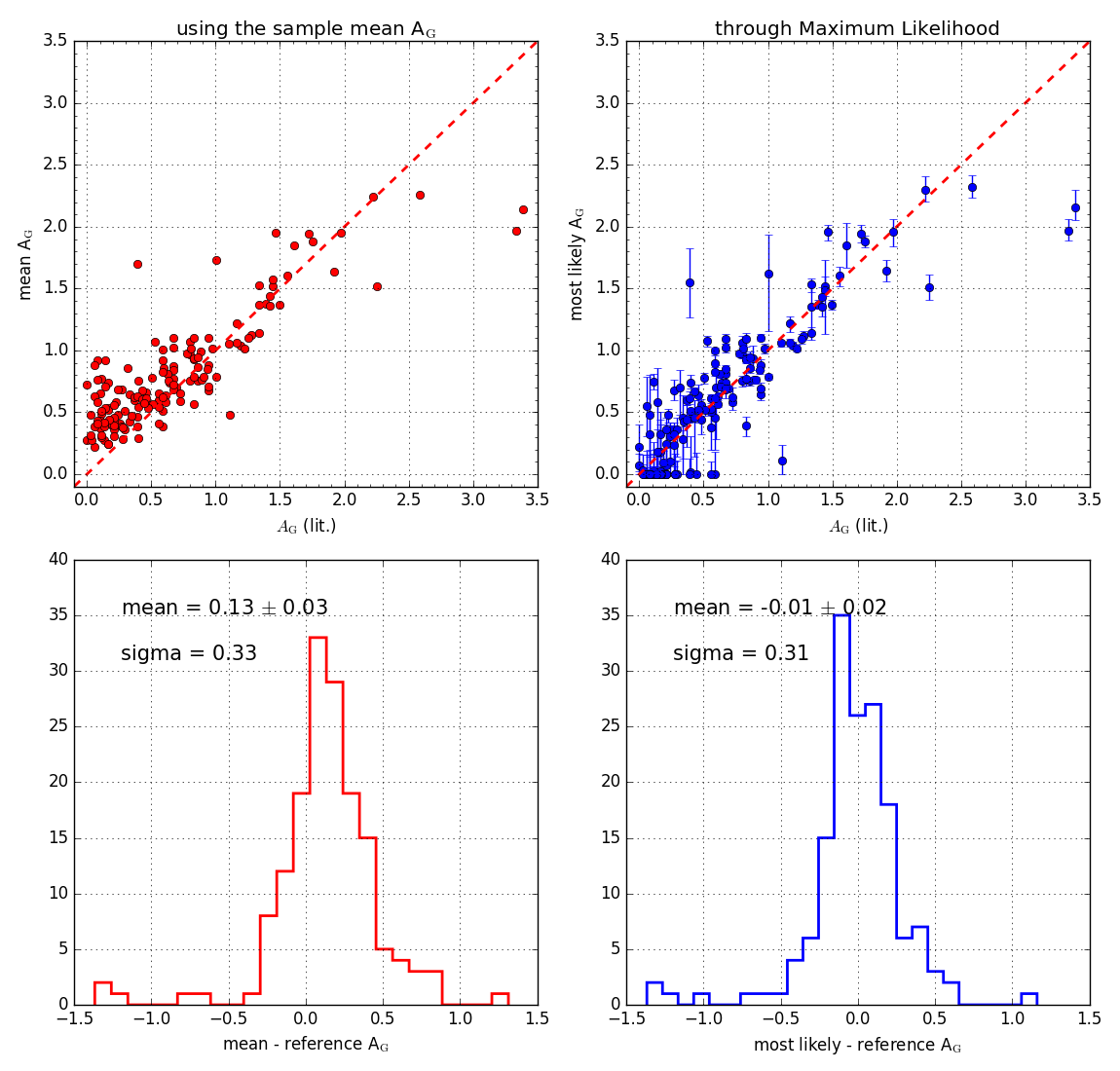}
\caption{ Mean {\gaia} $A_\mathrm{G}$ values vs literature values (upper left panel) and distribution of the differences (lower left panel) for OCs. On the right, the analogous plot using the recommended maximum likelihood values of $A_\mathrm{G}$ instead of the mean values.}\label{fig:ag_lit_all}
\end{center}\end{figure}

\begin{figure}\begin{center}
\includegraphics[width=8cm]{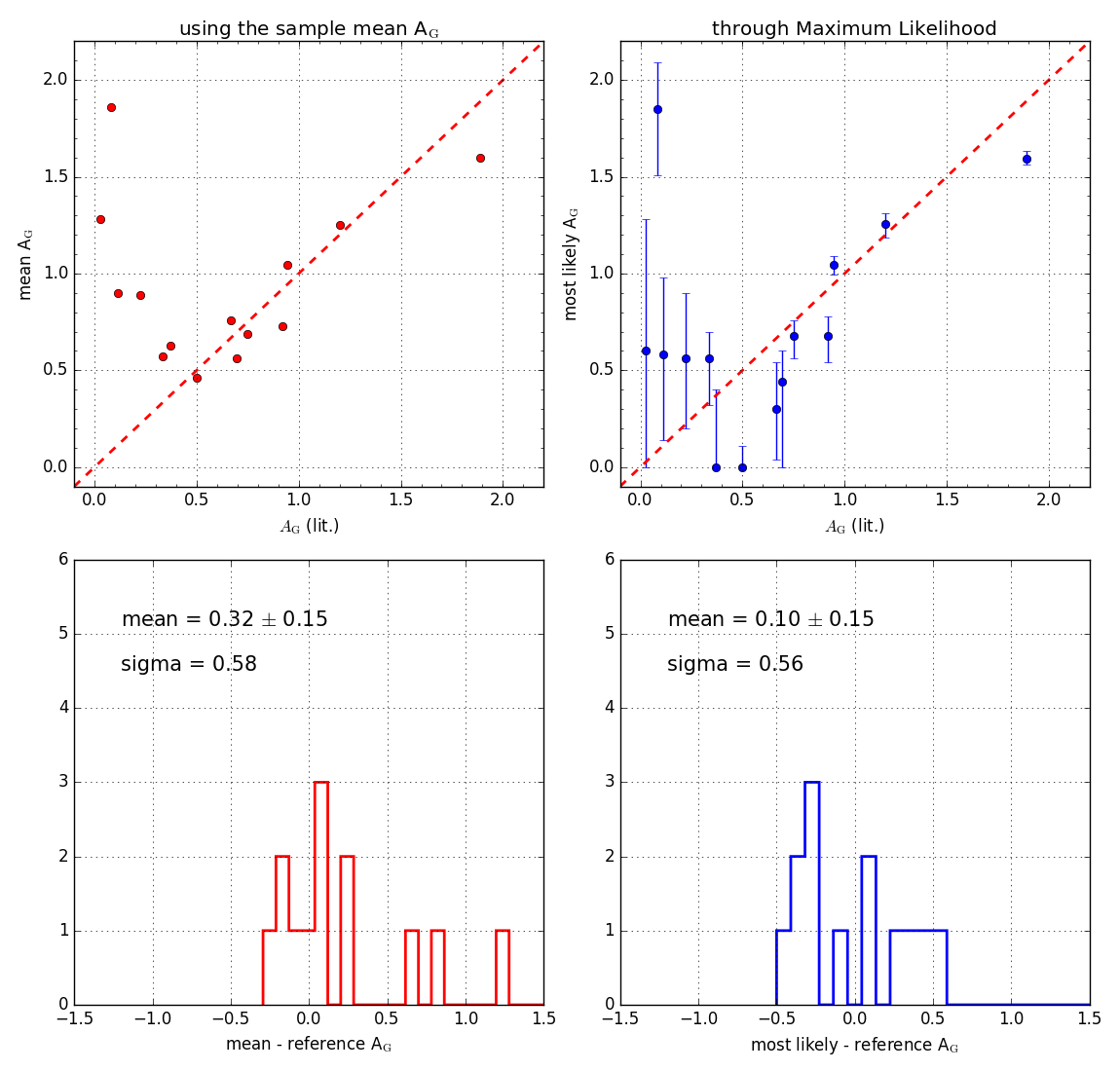}
\caption{Same as \figref{fig:ag_lit_all} for halo Globulars.}\label{fig:ag_lit_glob}
\end{center}\end{figure}

We compared the $A_G$ values derived using the ML method with reference literature data \citep{2013A&A...558A..53K} for a sample of 100 clusters, including a few halo globulars. 
Figures~\ref{fig:ag_lit_glob} and \ref{fig:ag_lit_all} present the results for the disk and the halo sub-sample respectively. 

For the disk stars, we found a general reasonable agreement with literature values, $A_{G,{Gaia}}-A_{G,{\rm ref}}=-0.01 \pm 0.02$ albeit with a high dispersion ($\sigma$=0.31). In the case of metal poor populations such as halo globulars the agreement is less good:  $A_{G,{Gaia}}-A_{G,{\rm ref}}=0.10 \pm 0.15$ with $\sigma=0.56$.  We emphasize that comparing literature values with the arithmetic mean $A_G$ of cluster members leads to a general overestimation of about 0.2\,mag (in our sample), with differences reaching up to 0.3 -- 0.4\,mag in the case of clusters with $A_G<0.1$\,mag, so indeed, the estimation method recommended in \cite{DR2-DPACP-43} is preferable.

\subsubsection{Internal validation of the reddening and extinction}\label{astropar-int-ext}
\begin{figure}\begin{center}
\includegraphics[width=0.88\columnwidth]{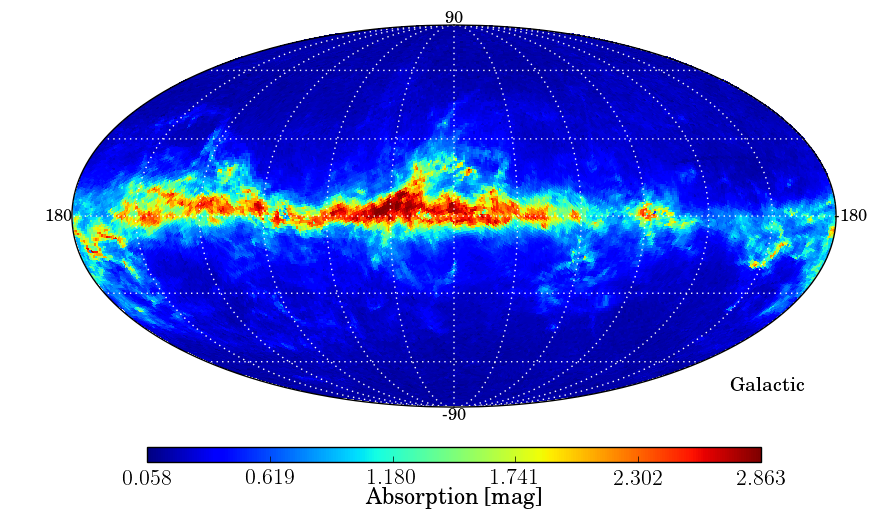}
\includegraphics[width=0.9\columnwidth]{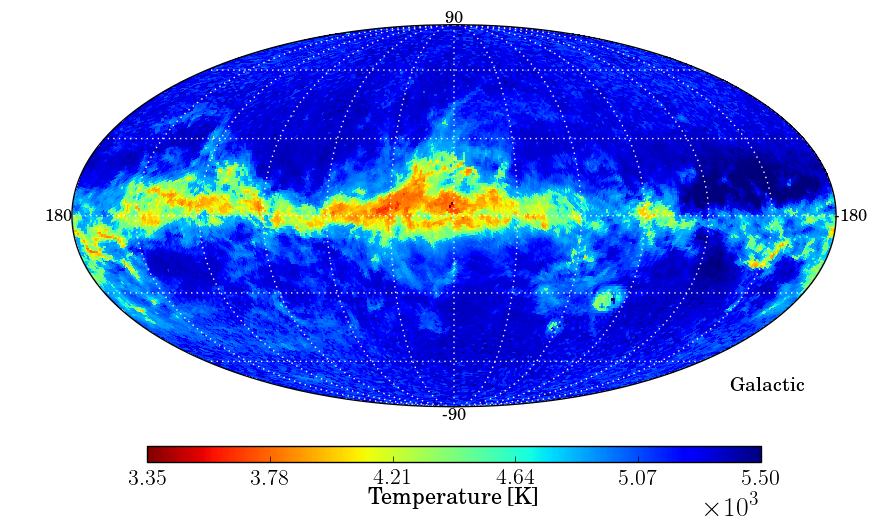}
\caption{Healpix maps (level 6, i.e.\ resolution $\sim$0.9\degr) of median of extinction $A_G$ (top panel) and effective temperature (bottom panel) in Galactic coordinates.}\label{fig:aitoff}
\end{center}\end{figure}

In \figref{fig:aitoff}, we plot Healpix maps for the extinction in $G$, $A_G$, and effective temperature, {\teff}, given in {\gdrtwo} (top and bottom panels, respectively) in Galactic coordinates and with a resolution $\sim 0.9\degr$. As expected, the extinction map traces the large-scale dust structure seen in the Galaxy, decreasing towards large latitudes. The temperature map shows a big tendency towards cooler stars, where lower temperatures seem confined to the Galactic disk, the anticenter and the Magellanic clouds. We emphasise here that while there certainly are differences in the stellar populations at different latitudes, these are unlikely to lead to mean temperature differences as large as these ones.

\begin{figure}\begin{center}
\includegraphics[width=0.8\columnwidth]{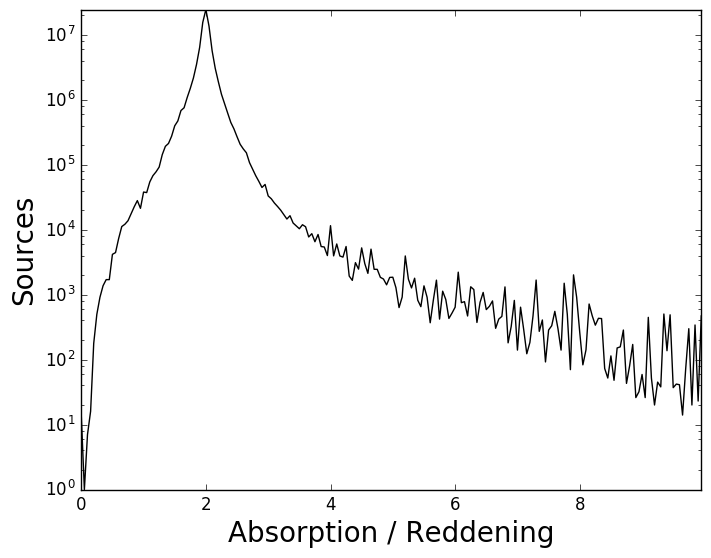}
\caption{Histogram of the ratio between the extinction in $G$, $A_G$, and the reddening, {\ebpminrp}.}\label{fig:A/Edist}
\end{center}\end{figure}

Figure~\ref{fig:A/Edist} shows the histogram of the ratio between the extinction in $G$, $A_G$, and the reddening, {\ebpminrp}. This ratio peaks around $2$ as expected from \cite{2010A&A...523A..48J}. Note, however, a large dispersion towards larger values. Since $A_G$ and {\ebpminrp} are estimated independently of each other, if both are low then random noise can let {\ebpminrp} come very close to zero such that the ratio $A_G$/{\ebpminrp} becomes very large. Therefore, caution is necessary when using this ratio, specially if it reaches unrealistically large values (i.e. $A_G/${\ebpminrp}> 3).
 
\begin{figure}\begin{center}
\includegraphics[width=0.8\columnwidth]{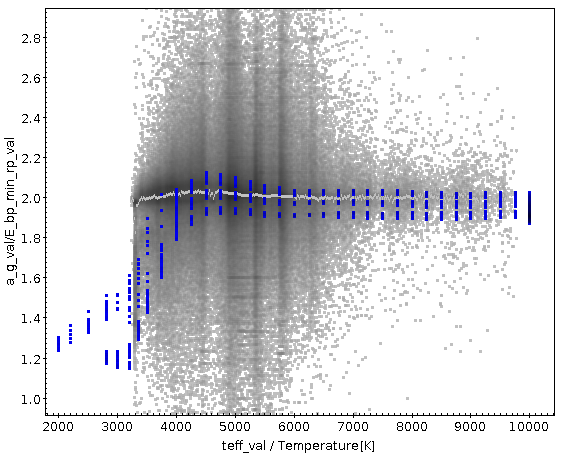}
\caption{Ratio $A_G/${\ebpminrp} as a function of {\teff} as derived in DR2 (grey points). Grey line is a running median. Blue dots show the prediction using stellar energy distribution of solar metallicity stars and the DR2 passbands \citep{DR2-DPACP-40}.}\label{fig:A/Etemp}
\end{center}\end{figure}

The ratios $A_G/${\ebpminrp} have been compared with predictions using spectral energy distributions of solar metallicity stars and the DR2 passbands \citep{DR2-DPACP-40}
in \figref{fig:A/Etemp}. It can be seen that the predicted ratios are about 2 for temperatures larger than about 4000~K and decrease to about 1.2 at $2000-3000$~K. Similar trends were present with the nominal passbands in \cite{2010A&A...523A..48J}. Instead, the computed ratios do not show the decrease at temperatures below 4000~K, demonstrating an issue with the extinction parameter. As can be seen from Fig.~7 in \cite{DR2-DPACP-43}, the training set does not have enough models for which $A_G/${\ebpminrp} could become $\sim$1.5 or less.

\subsubsection{Extinction using external catalogues}\label{astropar-extinction}

\begin{figure}\begin{center}
\includegraphics[width=0.8\columnwidth]{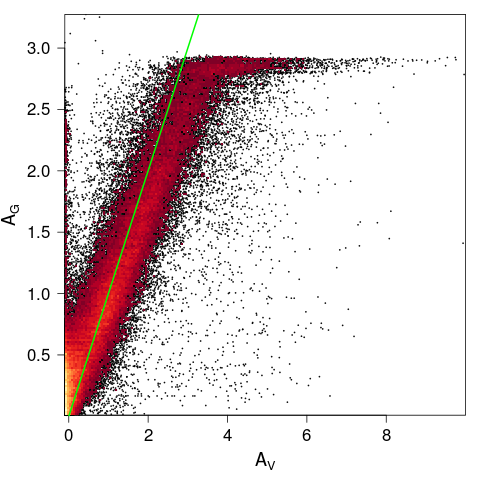}
\caption{Comparison between {\gdrtwo} $A_G$ and $A_V$ determined on APOGEE DR14 stars by  \cite{2018MNRAS.tmp..326Q}. In green the one-to-one relation. An over density of stars with over-estimated extinction is seen at low extinction due to the positivity constraint and saturation at high extinction is seen due to the training grid boundaries.}\label{fig:apogee_bestext_A0}
\end{center}\end{figure}

The comparison to external data is complicated at the low end by the non-negativity constraint, and at the upper end by saturation due to the training grid boundaries. Those effects can be seen in \figref{fig:apogee_bestext_A0} comparing $A_G$ to $A_V$  determined for the APOGEE DR14 by \cite{2018MNRAS.tmp..326Q}. Note that the few outliers which remain at $A_V\sim0$ with $A_G>1.5$, indicate that the outliers filtration detailed in \cite{DR2-DPACP-43} is imperfect.

The uncertainties, provided as percentiles, are difficult to use on those highly skewed uncertainties. This is illustrated by a sample of low-extinction stars at high Galactic latitudes or within the local bubble (within 50~pc). Not only does the estimated extinction reach large values in those samples \citep[][Sect. 6.5]{DR2-DPACP-43}, but even the 16th percentiles exceeds 0.05~mag for 60\% of the stars. This shows that the percentiles are not accurate enough to be used as estimates of the individual errors; still \cite{DR2-DPACP-43} showed that they are useful for outliers filtration. 

Selection of stellar types using the extinction and colour excess is to be done with caution. For example when attempting to select OB stars, the global over-estimation of the extinction due to the non-negativity constraint moves many cool stars into the hot stars colour range. Such a selection cannot be done with the {\gaia} data alone and external photometry is required. We tested on low extinction stars that even after applying a colour-colour cut based on 2MASS photometry, as done in \cite{DR2-DPACP-33}, which removes the coolest stars from an OB sample, 20 times more stars than expected are found, simply due to the large errors of the DR2 extinctions combined to the fact that the hottest stars are less numerous than the cooler ones (see the online documentation, Sect.~10.2.7.2). 

\begin{figure}\begin{center}
\includegraphics[width=0.8\columnwidth]{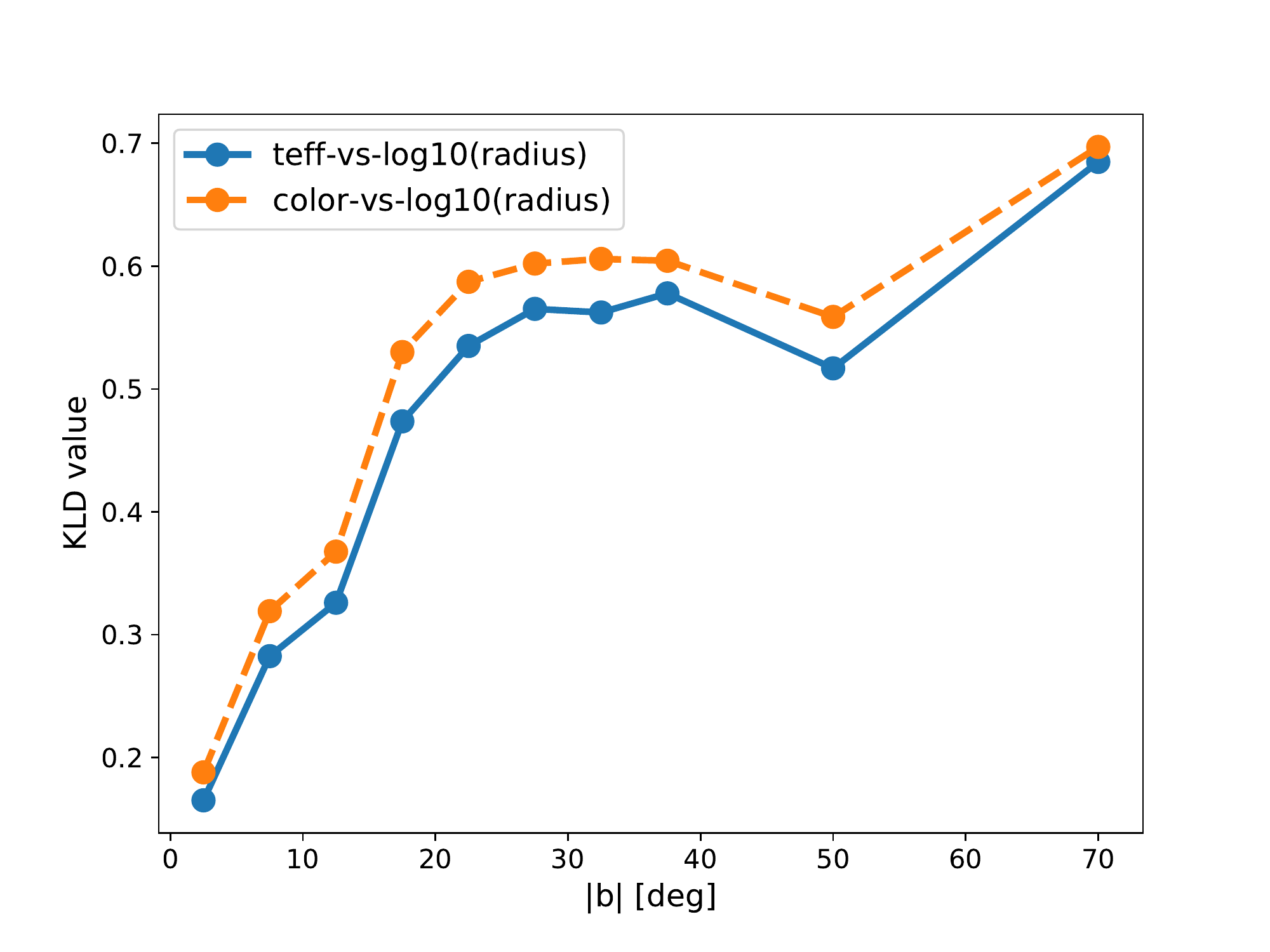}
\caption{Quantitative comparison of the distributions of $\log R$ vs {\teff} (and {\bpminrp} colour) as function of Galactic latitude, as measured by the  KLD statistic. The data is consistent with the expectation that the stellar populations are similar for all Galactic latitudes, except for $|b| \lesssim 20$\degr where the behaviour is markedly different, revealing the systematic issues with the determinations of radii and temperatures at low latitudes.}\label{fig:s90-bins_kld-1}
\end{center}\end{figure}

\subsection{Luminosities and radii}\label{astropar-lum}

The radii and luminosities are computed using the temperatures, with $A_G$ set to 0.0 mag, they therefore suffer from the same issues as described above. The radius may however be recalculated from any estimation of $A_G$ using Eq.~6 given in \cite{DR2-DPACP-43}.

Given the expectation that stars at different latitudes will not vary
dramatically in their intrinsic properties, we expect the 2D
distributions in $\log R$ and {\teff} to be roughly independent of
Galactic latitude. The degree of similarity of distributions can be
quantified using the Kullback-Leibler divergence \citep[Kullback \&
Leibler 1959, see also][Sect.~5.1]{2017A&A...599A..50A}.  This is
shown in \figref{fig:s90-bins_kld-1} which clearly indicates that the
distributions in $\log R$ vs {\teff} for $|b| \lesssim 20\degr$ are
significantly different from those at higher $|b|$, where they resemble
each other (i.e. the KLD value remains approximately constant). On the
other hand, the KLD values obtained when computing the distribution of
stars in the space of $\log R$ vs {\bpminrp} vary with
latitude especially strongly for low $|b|$ at least partly as expected
because of reddening (see also \figref{fig:s90int6-color_teff-vs-radius_log}). 

\begin{figure}\begin{center}
\includegraphics[width=0.8\columnwidth]{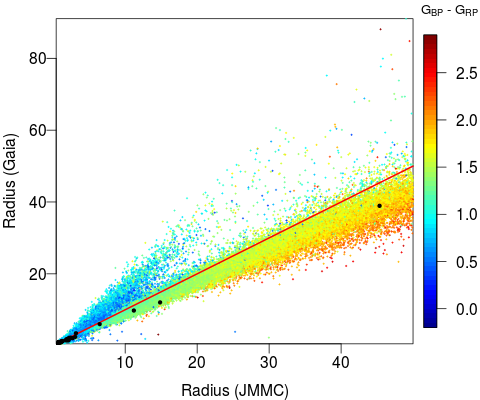}
\caption{Radius comparison with the JMMC stellar diameter catalogues for stars with relative parallax error smaller than 10\% and $G>6$. JSDC v2 is colour coded with $\bpminrp$ colour, JMDC in black. The red line corresponds to the one-to-one relation.}\label{fig:wp944_jmmc}
\end{center}\end{figure}

Tests using asteroseismic targets are presented in \cite{DR2-DPACP-43} and not repeated here. We present in \figref{fig:wp944_jmmc} a comparison with the JMMC Stellar Diameters Catalogue \citep[JSDC v2 from][selecting stars with $\chi^2<2$]{2017yCat.2346....0B} and the JMMC Measured Stellar Diameters Catalogue  \citep[JMDC,][]{2016yCat.2345....0D} for stars with relative parallax errors smaller than 10\% with $G>6$. The trail of stars for which {\gaia} is over-estimating the radius corresponds to hot stars, outside the APSIS {\teff} data training range, suffering strongly from extinction and adopting a cool temperature instead. A global underestimation of the radius is seen, as presented in \cite{DR2-DPACP-43}. We checked that it is still present when selecting only low extinction stars. The relative underestimation increases with increasing radius.

\subsection{Precision on the AP using the duplicate sources}\label{astropar-dup}

Regarding the duplicate sources mentioned in \secref{sec:dup}, we checked, as in \secref{radial}, whether the astrophysical parameters of the two components of duplicated sources are consistent. {\gdrtwo} provides the 16th percentile and 84th percentile of the probability density function for each of the astrophysical parameters. We adopted as the uncertainty for each of the parameter half of the difference between the upper and lower percentiles, although it is known that the extinction errors are far from normal. We have used in the tests other uncertainty estimators and the following results are equivalent.

\begin{figure}\begin{center}
\includegraphics[width=\columnwidth]{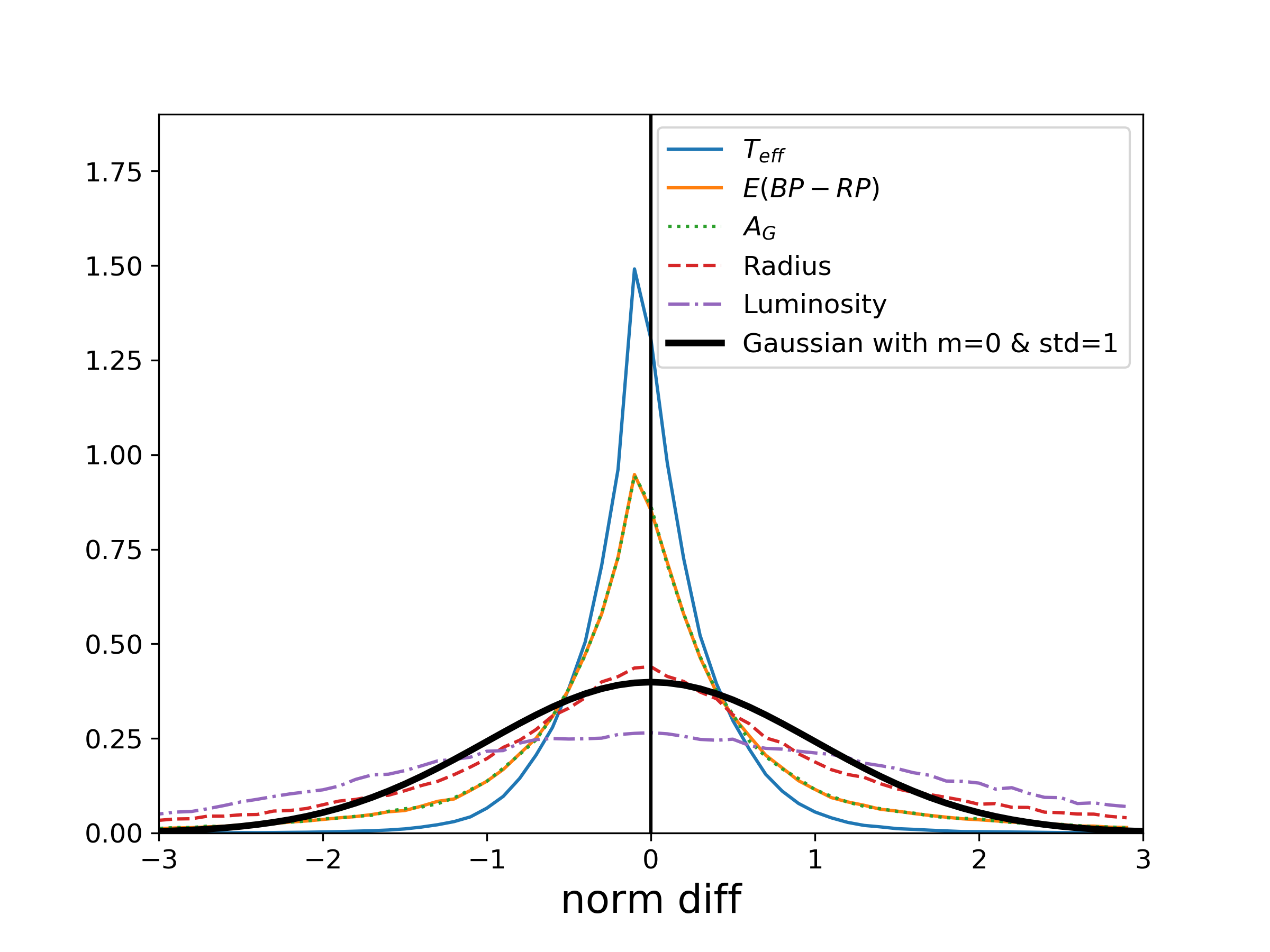}
\caption{Density of the normalised differences of pairs of duplicate
sources, as in Fig.~\ref{fig:cu6duprvsnorm2}, for the effective temperature (blue solid), extinction
(green dotted), $E(G_{BP}-G_{RP})$ (orange solid), radius (red dashed) and luminosity (purple dot dashed).
The Normal distribution with 0. mean (shown in a black vertical line)
and dispersion 1 is in black thick solid line. Note that the curves 
for the extinction and reddening perfectly overlap.}\label{fig:dup}
\end{center}\end{figure}

In \figref{fig:dup} we plot the differences for duplicate sources of the five astrophysical parameters provided in {\gdrtwo}, normalised by their uncertainty. The data look very symmetric, while it is known that e.g. the non-negativity constraint on $A_G$ or $\ebpminrp$ make their error asymmetric. Most probably, the errors for each component of a duplicate pair are little correlated so that the differences of the errors between pairs can be randomly positive or negative. The normalised distribution then appears leptokurtic (due to the lower errors), but with a long tail (due to the upper errors), because the adopted uncertainty was the difference between the upper and lower percentiles; indeed, a robust width for the normalised $A_G$ gives about 0.5, but the standard deviation is close to 1 as expected. For the various APs, {\teff} seem to have uncertainties overestimated, while they seem underestimated for the luminosity; for the other parameters, the uncertainties look as expected. 


\section{Solar System Objects}\label{sso}

\subsection{Data}\label{chap:sso_1}
{\gdrtwo} contains the astrometry for 1\,977\,702 CCD observations of 14\,099 Solar System Objects (SSOs), and it also provides, as additional information, asteroid magnitudes in the {\gmag} band for a selected 52\% of the observations, obtained as a result of the validation process described in~\cite{DR2-DPACP-32}.

The main goal of the validation of SSOs has been achieved, that is to show that the asteroid astrometry is very close to the expected performances, especially in the optimal range of brightness G$\sim$12-17, where the typical accuracy per CCD observation is at a sub-mas level.

The validation approach to asteroid astrometry has been based on an orbit determination process used to assess the quality of the data. The orbit determination is a set of procedures to compute the orbit of an object: it uses an orbit as initial guess (well known or computed with different procedures) and then it fits an orbit on the available observations. We used the least square method and the differential corrections algorithm (the core of the orbit determination) to fit orbits on 22 months of \gaia\ observations, starting from the already well-known orbits of each object.

For {\gdrtwo} we have selected an initial sample of $14\,124$ objects, which covers all the various categories of Solar System Objects. To assess the quality of the data we employ, in the orbit determination process, a high precision dynamical model, we added the contribution of $16$ massive asteroids and Pluto and we use a relativistic force model including the contribution of the Sun, the planets and the Moon. While all these precautions are sufficient in the usual orbit determination process, they were not enough to handle \gaia\ observations. To properly deal with \gaia\ asteroid astrometry, it is fundamental to appropriately take into account that:
\begin{itemize}
    \item \gaia\ astrometry is given in Barycentric Coordinate Time (TCB).
    \item The error model contains the correlations in $\alpha \cos{\delta}$ and $\delta$, which are strong in the epoch astrometry and crucial in the orbit determination.
    \item The positions ($\alpha$ and $\delta$) of the asteroids given in {\gdrtwo} have been corrected with a full relativistic model, but the light deflection assumes that the object were at infinite distance. In the validation process we also apply a further correction to take into account the finite distance.
\end{itemize}

The entire process is described in~\citep{DR2-DPACP-32}, including the computation of the residuals on the equatorial reference frame $(\alpha \cos{\delta}, \delta)$ and on the (AL, AC) plane, and their use for outlier rejection.

As a result of this procedure we discarded 25 objects and $1\%$ of the observations, obtaining the sample published in {\gdrtwo}. 

\subsection{Orbits}\label{chap:sso_2}

\begin{figure}\begin{center}
\includegraphics[width=0.9\columnwidth]{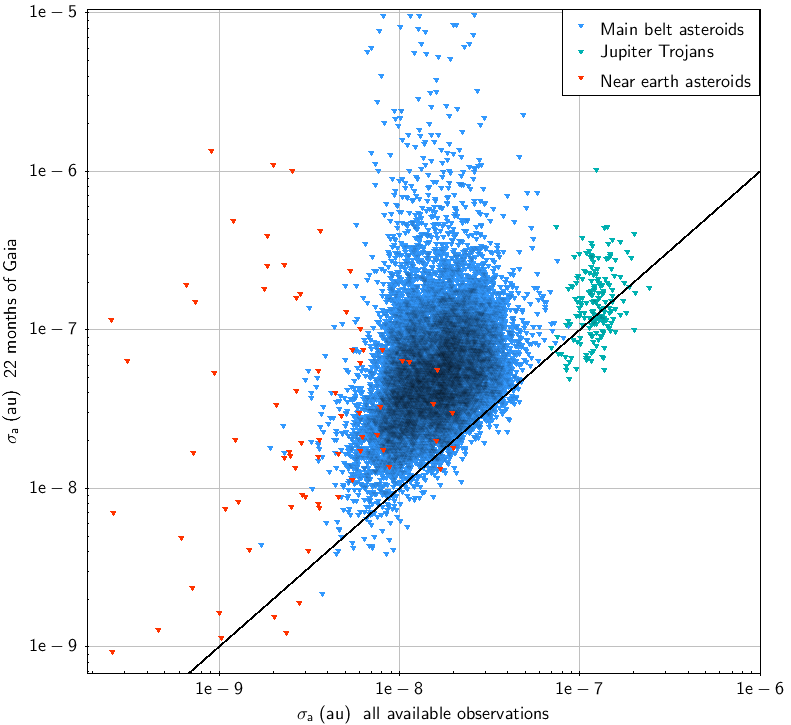}
\caption{Quality of the orbit determination measured by the post-fit uncertainty of the semi-major axis (au) for the objects contained in {\gdrtwo} with respect to the currently available uncertainty. The colours represent the different categories included in {\gdrtwo}: near-Earth and  main-belt asteroid and Jupiter trojans. } 
\label{fig:sigma_a_all_Gaia}
\end{center}\end{figure}

The orbits are of course a secondary product of the validation process. They represent the final outcome of the entire procedure. Since the time span covered by the {\gdrtwo} observations is quite short (compared to the time span of the hundreds of observations available today), we expect that the quality of the orbits should be limited on the average. Nevertheless, as shown in Fig.~\ref{fig:sigma_a_all_Gaia}, there are some asteroids in {\gdrtwo} that have already reached a quality in their orbits equivalent to ground-based data (and in 350 cases the orbit is even better determined using Gaia observations only).

\section{Conclusions and recommendations}\label{sec:conclusions}
We have described the results of the validation tests applied to the
second {\gaia} data release as an indirect quality control of the Catalogue before its publication.

With such a complex mission and so large diversity of sources, less than 2 years of observations and not yet optimal calibrations, the astrometry cannot be perfect. While the overall quality of the data is excellent, the user should consider the following recommendations, depending on the specific application, in order to make optimum use of the data:
\begin{itemize}
\item First, in applications requiring the cleanest possible dataset, spurious solutions need to be filtered.
Section~\ref{sec:err_ast} summarizes the filters suggested for this purpose. 
\item Because only a single-star model has been used and calibrations are sub-optimal, the quality indicators of the solutions (\dt{astrometric\_excess\_noise}, \dt{astrometric\_gof\_al}, \dt{astrometric\_n\_bad\_obs\_al}) may be used to discard other potential outliers.
\item In terms of astrometric systematics, correcting individual parallaxes from the global parallax zero-point (\secref{ssec:astroacc_large}) is discouraged. For applications where the zero-point matters, however, and if the samples are well distributed over the sky, in colour and in magnitude, parallaxes may need to be corrected from (or solved for) the global zero-point.
\item In the special case of samples in small ($< 1\degr$) or intermediate ($< 20\degr$) regions, the contribution of systematics to the error budget has to be taken into account, cf. \secref{ssec:astroacc_small} and \secref{ssec:astroacc_large}.
\item Some substantial underestimation of the formal uncertainties has to be taken into account (\secref{ssec:astrometry_precision}) e.g. when doing sample selections based on astrometric precision, or for likelihood methods. Reweighting the most precise uncertainties using Eq.~A.6 of \cite{DR2-DPACP-51} may be useful for $13\lesssim G\lesssim 15$ stars, or more generally when \dt{astrometric\_n\_obs\_ac}$=0$.
\item The full covariance information between astrometric parameters should always be taken into account.
\end{itemize}

Concerning the photometry, the colours of faint stars, in the neighbourhood of bright stars, or contaminated in dense regions, should be taken with care and sources with large \dt{phot\_bp\_rp\_excess\_factor} may be removed, as recommended by \cite{DR2-DPACP-40} and applied at Eq.~\eqref{eq2}.

For the variable stars present in this release, we recommend the users to adopt mean magnitudes calculated by the variability processing  (\dt{int\_average\_g}), when they are available, in preference to the values (\dt{phot\_g\_mean\_mag}) in the main catalogue. 

For the astrophysical parameters, the extinctions cannot easily be used individually, due to their large uncertainties; when trying to obtain an average sample value, it is important to follow the maximum-likelihood method proposed in \cite{DR2-DPACP-43} to derive the most probable value, as a simple average would overestimate the extinction. As both $A_G$ and {\ebpminrp} are SED dependent \citep{2010A&A...523A..48J,2018arXiv180201670D}, their combination on very different spectral types would increase the uncertainties.
Concerning temperatures and radii, it is preferable to use them in low extinction regions to avoid biases.
In moderate to high extinction regions external photometry combined with the {\gaia} one, using e.g. the cross-matches available directly within the {\gaia} Archive \citep{DR2-DPACP-41}, may help to disentangle extinction and temperature. 

\medskip
In summary, the wealth of data provided in {\gdrtwo} will represent beyond any doubt a landmark in the history of the astronomical catalogues. However, completely avoiding mistakes and shortcomings in the astrometric, photometric, spectroscopic or classification data in a 1.7 billion sources catalogue, with many intricate data for each, was an impossible task, given the short time since the observations were made. 
Nothing coming for free, the data cannot then be used blindly and any serious scientific exploitation of the {\gaia} data must understand and take into account the various limitations and caveats attached to the various {\gdrtwo} Catalogue content.

\begin{acknowledgements}

Funding for the DPAC has been provided by national institutions, in
particular the institutions participating in the Gaia Multilateral Agreement: 
the Centre National d'Etudes Spatiales (CNES), 
the European Space Agency in the framework of the Gaia project.

This research has made an extensive use of Aladin and the SIMBAD, VizieR databases 
operated at the Centre de Donn\'ees 
Astronomiques (Strasbourg) in France 
and of the software TOPCAT \citep{2005ASPC..347...29T}.
This work was supported by the MINECO (Spanish Ministry of Economy) through grant ESP2016-80079-C2-1-R (MINECO/FEDER, UE) and ESP2014-55996-C2-1-R (MINECO/FEDER, UE) and MDM-2014-0369 of ICCUB (Unidad de Excelencia 'María de Maeztu') and the European Community's Seventh Framework Programme (FP7/2007-2013) under grant agreement GENIUS FP7 - 606740.
We acknowledge the computer resources, technical expertise and assistance provided by the Red Espa\~nola de Supercomputaci\'on and specially the MareNostrum supercomputer at the Barcelona Supercomputing Center.
AH, MB and JV acknowledge financial support from NOVA (Netherlands Research School for Astronomy), and from NWO in the form of a Vici grant. 
\end{acknowledgements}

\bibliographystyle{aa} 
\bibliography{biblio} 

\appendix

\section{The {\gdrtwo} general completeness}\label{chap:selectFunc}

One of the most important properties of a catalogue is its completeness. Although in this respect {\gdrtwo} represents a major advance compared to {\gdrone} (as shown by the significant increase in the number of sources) it is still nevertheless an intermediate release and during its processing a variety of truncations and filters have been applied to the different types of data, limiting its completeness. Therefore, the selection function of the {\gdrtwo} Catalogue is difficult to define, and  significantly depends on the type of data.

The {\gdrtwo} truncations and filters are discussed in detail in \citet{DR2-DPACP-36} but we present here a short summary for the convenience of the reader. An overall summary of the properties of {\gdrtwo} can be found at the web page \url{https://www.cosmos.esa.int/web/gaia/dr2}. 

\begin{figure}\begin{center}
\includegraphics[width=0.9\columnwidth]{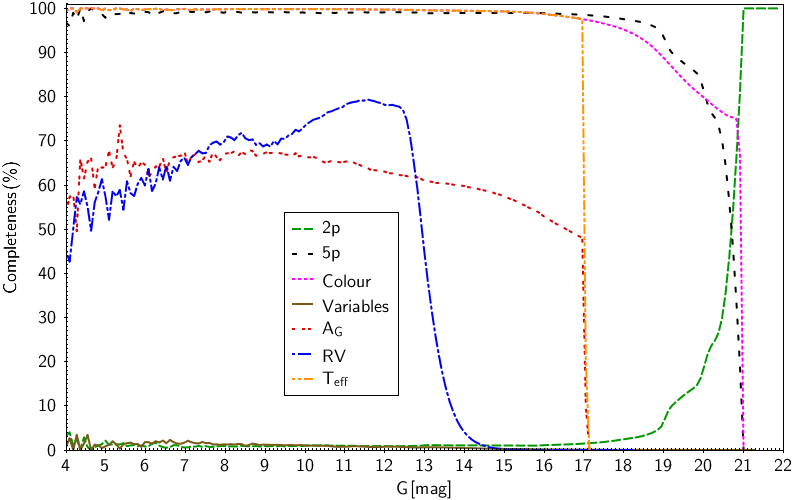}
\caption{Fraction of \gdrtwo\ sources having two (green) or five astrometric parameters (black); having colour (pink), temperature (orange), absorption (red), radial velocity (blue); or being flagged as variable (brown).} \label{fig:completeness}\end{center}\end{figure}

Figure~\ref{fig:completeness} illustrates the completeness, when it comes to astrometric, photometric and astrophysical parameters, for the sources in \gdrtwo. Especially clear are the, slightly arbitrary, magnitude limits imposed in different processes.

\subsection{Payload limits}\label{sec:selectFuncPayload}
The basic limitation for the {\gaia} data is the on-board detection of sources: only sources detected in the sky mapper are tracked on the focal plane and their data stored to be sent to ground. \afterReferee{The on-board capabilities are described in \cite{2015A&A...576A..74D} to which we refer as it describes extensively how the optimisation of the on-board processing has been done, and the resulting selection function for various type of objects.} We may simply note that the detection algorithm has been configured to a limiting sky-mapper magnitude of $G\simeq 20.7$, thus setting the essential limit for the {\gaia} sources \citep[see][]{2016A&A...595A...1G}. However, at these faint magnitudes the detection is not 100\% efficient so only up to $G\simeq 20$ the detection can be approximately complete. 

Furthermore, due to limitations on the on-board data handling resources, in crowded regions the capability to observe all stars is reduced. In combination with the still limited data treatment in crowded areas this means that the survey limit in regions with densities above a few hundred thousand stars per square degree can be as bright as $G = 18$.

\subsection{The full Catalogue}\label{sec:selectFuncCat}
The above described payload limits, combined with some  additional restrictions introduced by the Initial Data Treatment, define the dataset available on ground to the DPAC. This dataset is then processed through several pipelines to produce the final products in the Catalogue. The minimum requirement for a source to be published is the availability of a valid position and a $G$ magnitude. The objects with these parameters constitute the full {\gdrtwo} Catalogue, with a total of 1\,692\,919\,135 sources, a catalogue essentially complete between $G=12$ and $G=18$ but still incomplete at the bright end with an ill-defined faint magnitude limit, which depends on celestial position. 
For binaries or double stars, the current resolution limit is about 0\farcs4 (cf. \figref{fig:pair_stat}).

In the next sections the details on the astrometric and photometric selection leading to these numbers are described.

\subsection{Astrometry}\label{sec:selectFuncAstr}
The astrometric dataset ({\gaia} observed objects with an astrometric solution) is the result of the AGIS processing of the downloaded data \citep[see][]{DR2-DPACP-51}. For this data set the results were filtered by requiring that a source had been observed by {\gaia} at least five times (five focal
plane transits), and that the astrometric excess noise and the semi-major axis of the position uncertainty ellipse are less than $20$ and $100$ mas, respectively. The visibility of a source depends on the position of the sky and is tied to the {\gaia} scanning law; therefore these limitations have a complex effect on the completeness that depends on the sky region. 

Even if all the published sources have at least a position, the parallax and proper motions are determined only for sources satisfying the requirement that they are brighter than $G = 21$, that the number of visibility periods used is at least 6, and that the semi-major axis of the 5-dimensional uncertainty ellipse is below a magnitude dependent threshold. Therefore, the {\gdrtwo} subset of objects with five-parameter astrometry is significantly smaller than the full dataset and is composed of 1 331\,909\,727 sources. This five-parameter astrometry dataset is not complete at any magnitude, and the relative completeness (with respect to the full Catalogue) varies slightly depending on the magnitude limit, as can be seen from Figure 2 in \citet{DR2-DPACP-36}.

An additional limitation is applicable to high proper motion sources. The completeness for these objects has significantly improved with respect to {\gdrone}, but about 20\% of stars with proper motion $>0.6$ arcsec yr$^{-1}$ may still be missing.

\subsection{Photometry}\label{sec:selectFuncPhot}
On top of the selection set by the astrometric solution, sources without a well-determined value for $G$ do not appear in {\gdrtwo}. The photometry in the $G$, $G_{BP}$, or $G_{RP}$ bands is only reported if the source was observed at least twice by {\gaia} in the respective bands, as described in \citet{DR2-DPACP-44}. As in the previous case, since this is tied to the source visibility, these limitations have a complex effect on the completeness that depends on the sky region. 

Furthermore, due to limitations of the current photometric processing, leading to the so-called ``flux-excess factor'' \citep[see again][]{DR2-DPACP-44}, there is a significant fraction of the Catalogue ($\approx 300$ million) with missing values of $G_{BP}$ and/or $G_{RP}$.

\subsection{Spectroscopy}\label{sec:selectFuncSpec}
Mean radial velocities are available for a subset of 7\,224\,631 sources, already a much larger amount than presently available from ground-based observations. Objects without radial velocities are those fainter than $G_{RVS} = 12$ \citep[see][]{DR2-DPACP-47}, as estimated from the magnitudes in the IGSL \citep[Initial Gaia Source List, ][]{2014A&A...570A..87S}, so, roughly corresponding to $G\simeq 13$, plus brighter objects for which some quality or conditions or limits on effective temperature were not fulfilled  \citep[see][]{DR2-DPACP-36}. In addition radial velocity values are not listed for a number of sources with $|v_{\rm rad}| > 500$ {\kms} for which the value was clearly dubious. 

As a result the radial velocities sample shows the distribution depicted in Figure 1 of \citet{DR2-DPACP-36}, incomplete at bright magnitudes, slightly incomplete to $G\simeq 13$ and more incomplete for fainter sources. The completeness also depends on the sky position, showing traces of the distribution of the IGSL Catalogue, that has been used in the spectroscopic processing.

\subsection{Astrophysical parameters}\label{sec:selectFuncAstroPar}
In {\gdrtwo} the astrophysical parameter results are only available for sources brighter than $G = 17$, and among these
only for sources for which $G$, $G_{BP}$ and $G_{RP}$ are available. Further filtering was applied based on the quality of the various inputs to the astrophysical parameter estimation, where particularly strict criteria were applied to the extinction and reddening estimations. We refer to \citet{DR2-DPACP-43} for a detailed description
of the filters applied. Essentially, $T_{\rm eff}$ is available for practically all sources at $G < 17$ in the temperature range 3000--10\,000 K, while estimates of the other astrophysical parameters are published for about 50\% of these sources.

\subsection{Variability data}\label{sec:selectFuncVari}
During the variability analysis a strict internal filtering was applied to the quality of the photometric time series, thus reducing the number of sources flagged as variable,  followed by several additional filters to reduce the contamination due to data processing artefacts, confusion with other variable and to remove sources for which
the results of the light curve analysis were not deemed reliable enough. For the details we refer to \citet{DR2-DPACP-49} where estimations of the completeness of the global variability sample and the subsamples of different types of variables are provided.

\subsection{Solar System Object data}\label{sec:selectFuncSSO}
{\gdrtwo} includes epoch astrometry and photometry for a pre-selected list of 14\,099 known minor bodies in the solar system, primarily main belt asteroids. Thus, in this case the objects are taken from an input list and the filtering applied involves only the removal
of some observations for which the relative flux uncertainty in the $G$ band was larger than 0.1 (this mainly removes observations of the very ``fast'' objects). In addition a selection of the observations
was removed as well as some individual sources. We refer the reader to \citet{DR2-DPACP-32} for details.

\section{The open and globular cluster sample}\label{chap:appendixCluster}

\begin{table*}
\begin{center}
	\caption{Completeness level (in percentage with respect to HST fields) in various magnitude ranges, in the inner and outer regions of 26 globular clusters}\label{tab:completeness26gcs} 
	\small\addtolength{\tabcolsep}{-2pt}
	\begin{tabular}{l l c c c c c c c c c c}
	\hline
	\hline
   Name	& Region &                                    \multicolumn{10}{c}{$G$ magnitudes}		                                       \\
       	&        & 11--13 & 12--14 & 13--15 & 14--16 & 15--17 & 16--18 & 17--19 & 18--20 & 19--21 & 20--22 \\
	\hline
LYN07 & inner	 & -- & -- & -- & -- & -- & 56 & 33 & 18 & 5 & 1 \\ 
LYN07 & outer	 & -- & -- & -- & -- & 54 & 50 & 35 & 24 & 9 & 2 \\ 
\hline
NGC0104 & inner	 & 23 & 2 & 0 & 0 & 0 & 0 & 0 & 0 & 0 & 0 \\ 
NGC0104 & outer	 & 85 & 55 & 43 & 21 & 7 & 1 & 0 & 0 & 0 & 0 \\ 
\hline
NGC0288 & inner	 & -- & -- & -- & -- & -- & -- & 60 & 40 & 15 & 1 \\ 
NGC0288 & outer	 & -- & -- & -- & 100 & 85 & 79 & 70 & 54 & 28 & 6 \\ 
\hline
NGC1261 & inner	 & -- & -- & -- & 77 & 55 & 37 & 10 & 1 & 0 & 0 \\ 
NGC1261 & outer	 & -- & -- & -- & 100 & 96 & 89 & 62 & 30 & 11 & 2 \\ 
\hline
NGC1851 & inner	 & -- & -- & 40 & 29 & 14 & 3 & 0 & 0 & 0 & 0 \\ 
NGC1851 & outer	 & -- & -- & -- & 100 & 86 & 61 & 34 & 14 & 5 & 1 \\ 
\hline
NGC2298 & inner	 & -- & -- & -- & -- & 86 & 67 & 30 & 10 & 3 & 0 \\ 
NGC2298 & outer	 & -- & -- & -- & -- & 97 & 90 & 83 & 62 & 36 & 10 \\ 
\hline
NGC4147 & inner	 & -- & -- & -- & -- & -- & 52 & 29 & 12 & 3 & 1 \\ 
NGC4147 & outer	 & -- & -- & -- & -- & -- & 94 & 78 & 63 & 27 & 7 \\ 
\hline
NGC5053 & inner	 & -- & -- & -- & -- & -- & -- & -- & 78 & 44 & 13 \\ 
NGC5053 & outer	 & -- & -- & -- & -- & -- & 100 & 94 & 82 & 46 & 13 \\ 
\hline
NGC5139 & inner	 & -- & -- & 1 & 1 & 0 & 0 & 0 & 0 & 0 & 0 \\ 
NGC5139 & outer	 & 43 & 11 & 5 & 2 & 1 & 0 & 0 & 0 & 0 & 0 \\ 
\hline
NGC5272 & inner	 & -- & -- & 69 & 44 & 27 & 6 & 0 & 0 & 0 & 0 \\ 
NGC5272 & outer	 & -- & 100 & 99 & 86 & 75 & 45 & 17 & 6 & 1 & 0 \\ 
\hline
NGC5286 & inner	 & -- & -- & 52 & 33 & 13 & 4 & 0 & 0 & 0 & 0 \\ 
NGC5286 & outer	 & -- & -- & 86 & 83 & 69 & 51 & 25 & 6 & 1 & 0 \\ 
\hline
NGC5466 & inner	 & -- & -- & -- & -- & -- & -- & -- & 69 & 31 & 7 \\ 
NGC5466 & outer	 & -- & -- & -- & -- & 100 & 99 & 100 & 86 & 48 & 12 \\ 
\hline
NGC5927 & inner	 & -- & -- & -- & 52 & 37 & 24 & 2 & 0 & 0 & 0 \\ 
NGC5927 & outer	 & -- & -- & 81 & 75 & 75 & 60 & 28 & 6 & 1 & 0 \\ 
\hline
NGC5986 & inner	 & -- & -- & -- & 59 & 34 & 14 & 2 & 0 & 0 & 0 \\ 
NGC5986 & outer	 & -- & -- & 90 & 88 & 81 & 61 & 34 & 10 & 2 & 0 \\ 
\hline
NGC6121 & inner	 & -- & -- & -- & 66 & 54 & 38 & 20 & 5 & 0 & 0 \\ 
NGC6121 & outer	 & -- & 95 & 92 & 85 & 79 & 66 & 48 & 25 & 7 & 0 \\ 
\hline
NGC6205 & inner	 & -- & -- & 68 & 42 & 15 & 1 & 0 & 0 & 0 & 0 \\ 
NGC6205 & outer	 & 89 & 92 & 92 & 80 & 56 & 25 & 7 & 2 & 0 & 0 \\ 
\hline
NGC6366 & inner	 & -- & -- & -- & -- & -- & -- & 69 & 55 & 32 & 9 \\ 
NGC6366 & outer	 & -- & -- & -- & 91 & 90 & 81 & 79 & 69 & 42 & 11 \\ 
\hline
NGC6397 & inner	 & -- & -- & 62 & 49 & 28 & 11 & 2 & 0 & 0 & 0 \\ 
NGC6397 & outer	 & -- & 95 & 89 & 82 & 72 & 56 & 33 & 12 & 2 & 0 \\ 
\hline
NGC6656 & inner	 & -- & -- & 53 & 38 & 11 & 1 & 0 & 0 & 0 & 0 \\ 
NGC6656 & outer	 & 75 & 75 & 69 & 61 & 41 & 10 & 1 & 0 & 0 & 0 \\ 
\hline
NGC6752 & inner	 & -- & 62 & 37 & 18 & 6 & 1 & 0 & 0 & 0 & 0 \\ 
NGC6752 & outer	 & -- & 98 & 94 & 78 & 57 & 32 & 13 & 3 & 0 & 0 \\ 
\hline
NGC6779 & inner	 & -- & -- & -- & -- & 63 & 43 & 16 & 2 & 0 & 0 \\ 
NGC6779 & outer	 & -- & -- & 94 & 87 & 82 & 76 & 56 & 30 & 13 & 2 \\ 
\hline
NGC6809 & inner	 & -- & -- & -- & -- & -- & 48 & 20 & 4 & 0 & 0 \\ 
NGC6809 & outer	 & -- & -- & 100 & 96 & 84 & 66 & 37 & 11 & 1 & 0 \\ 
\hline
NGC6838 & inner	 & -- & -- & -- & -- & -- & 67 & 54 & 31 & 8 & 1 \\ 
NGC6838 & outer	 & -- & -- & 91 & 89 & 72 & 80 & 74 & 56 & 28 & 5 \\ 
\hline
NGC7099 & inner	 & -- & -- & -- & 47 & 34 & 11 & 2 & 0 & 0 & 0 \\ 
NGC7099 & outer	 & -- & -- & -- & 94 & 80 & 66 & 38 & 17 & 5 & 1 \\ 
\hline
PAL01 & inner	 & -- & -- & -- & -- & -- & -- & -- & -- & 73 & 39 \\ 
PAL01 & outer	 & -- & -- & -- & -- & -- & -- & -- & -- & 67 & 21 \\ 
\hline
PAL02 & inner	 & -- & -- & -- & -- & -- & -- & 84 & 62 & 24 & 7 \\ 
PAL02 & outer	 & -- & -- & -- & -- & -- & -- & 92 & 89 & 47 & 14 \\ 
\hline
\end{tabular}
\tablefoot{inner: within 0.5 arcmin; outer: 0.5 to 2.2 arcmin }
\end{center}
\end{table*}

\section{Acronyms}\label{sec:acronyms}
{\small
\begin{tabular}{ll}
\hline\hline
\textbf{Acronym}  &  \textbf{Description}  \\ \hline
2MASS & Two-Micron All Sky Survey \\
AC & {\gaia} ACross scan (direction) \\
ACS & Advanced Camera for Surveys (HST) \\
AGIS & {\gaia} Astrometric Global Iterative Solution \\
AL & {\gaia} ALong scan (direction) \\
AP & Astrophysical Parameters \\
APSIS & {\gaia} Astrophysical Parameters Inference System \\
BP & {\gaia} Blue Photometer \\
CCD & Charge-Coupled Device \\
CMD & Colour Magnitude Diagram \\
DAML & New catalog of Optically Visible Open Clusters and Candidates\\
& Dias et al., 2014 \\ 
DPAC & Data Processing and Analysis Consortium \\
DR1 & {\gaia} Data Release 1 \\
DR2 & {\gaia} Data Release 2 \\
EPSL & Ecliptic Pole Scanning Law \\
GC & Globular cluster \\
GoF & Goodness of Fit \\
HIP & Hipparcos catalogue \\
HPM & High Proper Motion \\
HST & Hubble Space Telescope \\
HealPix & Hierarchical Equal Area isoLatitude Pixelisation \\
IGSL & Initial Gaia Source List \\
JMMC & Jean-Marie Mariotti Center \\
JSDC & JMMC Stellar diameters Catalogue, \citet{2017yCat.2346....0B} \\ 
JMDC & JMMC Measured Stellar diameters Catalogue, \citet{2016yCat.2345....0D} \\
KLD & Kullback-Leibler Divergence \\
LMC & Large Magellanic Cloud \\
MAD & Median Absolute Deviation \\
ML & Maximum-Likelihood method \\
MWSC & Milky Way Star Clusters, Kharchenko et al. (2013) \\ 
OC & Open Cluster \\
OGLE & Optical Gravitational Lensing Experiment \\
PSF & Point Spread Function \\
Q-Q & Quantile-quantile plot \\
RAVE & RAdial Velocity Experiment \\
RECONS & REsearch Consortium On Nearby Stars, \citet{2015IAUGA..2253773H}\\
RP & {\gaia}  Red Photometer \\
RV & Radial Velocity \\
SDSS & Sloan Digital Sky Survey \\
SED & Spectral Energy Distribution \\
SMC & Small Magellanic Cloud \\
SOS & Specific Object Studies of the {\gaia} variability pipeline \\
TDSC & {\tyc} Double Star Catalogue \\
UMMSV & local RV Catalogue using Soubiran et al. (2018),\\ & Famaey et al. (2005), Mermilliod et al. (2008, 2009),\\& Nidever et al. (2002), Nordstr\"om et al. (2004), \\& Worley et al. (2012) and Chubak et al. (2012). \\
uwu & unit-weight uncertainty (ratio of external over internal errors) \\
WDS & Washington Visual Double Star Catalogue, \cite{WDS}\\
\hline
\end{tabular} 
}

\end{document}